\documentclass[defended,openany,11pt,twoside]{new_cit_thesis}

\usepackage{amsmath}
\usepackage{amssymb}
\usepackage{graphicx}
\usepackage{epsfig}
\usepackage{bm}
\usepackage{feynmf}
\usepackage{epic}
\usepackage{eepic}
\usepackage{epigraph}
\usepackage{subfigure}

\newcommand{\nc}{\newcommand}  

\nc\beq{\begin{equation}}
\nc\eeq{\end{equation}}
\nc\beqa{\begin{eqnarray}}
\nc\eeqa{\end{eqnarray}}
\nc\bi{\begin{itemize}}
\nc\ei{\end{itemize}}
\def\nn{\nonumber \\ }

\long\def\sl#1{\hbox{\tiny \it #1}}			
\long\def\teq#1{\hbox{$#1$}}				

\def\vh#1{\hat{\bm{#1}}}					
\nc{\vv}{\boldsymbol}					
\nc\La{{\cal L}}
\nc\Ha{{\cal H}}
\nc{\bert}{\raise-0.45mm\hbox{\Large$\Box$}}	
\nc\A{A^\mu}
\nc{\PgP}{\bar\psi\gamma^\mu\psi}			
\nc{\tensorbilinear}{\bar\psi\frac{i}{2}(\gamma_\mu
\buildrel\rightarrow\over\partial_\nu -
\gamma_\mu\buildrel\leftarrow\over\partial_\nu)\psi}

\renewcommand{\slash}[1]{#1\!\!\!/}
\def\abs#1{ \left| #1 \right| }
\def\ket#1{ \left| #1 \right\rangle }
\def\bra#1{\left\langle #1\right|}
\def\vev#1{ \left\langle #1 \right\rangle }
\def\gsim{\mathrel{\rlap{\lower4pt\hbox{$\sim$}}
    \raise1pt\hbox{$>$}}}					
\def\lsim{\mathrel{\rlap{\lower4pt\hbox{$\sim$}}
    \raise1pt\hbox{$<$}}}					

\def\Dsl{\,\raise.15ex \hbox{/}\mkern-12.8mu D}

\def\OMIT#1{}

\newcommand{\ba}{\begin{array}}
\newcommand{\ea}{\end{array}}

\title{Topics in Theoretical Particle Physics and \\ Cosmology Beyond the Standard Model}
\author{Alejandro Jenkins}
\date{26 May, 2006}

\begin{document}

\degreeaward{Doctor of Philosophy}                 
\university{California Institute of Technology}    
\address{Pasadena, California}                     
\unilogo{cit_logo}                                 
\copyyear{\the\year}                               
\pubnum{CALT-68-}                                          
\dedication{\sc \bigskip Para Viviana, quien, sin nada de esto, hubiera sido posible ({\it sic})}                                      

\maketitle

\begin{frontmatter}
      \makecopyright            
      \makededication          

\vfill
\pagebreak

\raisebox{-2in}{
\begin{minipage}{5.5in}
\small\singlespace\raggedright
If I have not seen as far as others, it is because giants were standing on my shoulders. \\ \flushright{--- Prof. Hal Abelson, MIT}
\end{minipage}}

      \begin{acknowledgments}  
	\begin{epigraphs}
	\qitem{How small the cosmos (a kangaroo's pouch would hold it), how paltry and puny in comparison to human consciousness, to a single individual recollection, and its expression in words!}
	{Vladimir V. Nabokov, {\it Speak, Memory}}	
	\qitem{What men are poets who can speak of Jupiter if he were like a man, but if he is an immense spinning sphere of methane and ammonia must be silent?}
	{Richard P. Feynman, ``The Relation of Physics to Other Sciences''}
	\end{epigraphs}
	
	I thank Mark Wise, my advisor, for teaching me quantum field theory, as well as a great deal about physics in general and about the professional practice of theoretical physics.  I have been honored to have been Mark's student and collaborator, and I only regret that, on account of my own limitations, I don't have more to show for it.  I thank him also for many free dinners with the Monday seminar speakers, for his patience, and for his sense of humor.
	
	I thank Steve Hsu, my collaborator, who, during his visit to Caltech in 2004, took me under his wing and from whom I learned much cosmology (and with whom I had interesting conversations about both the physics and the business worlds).
		
	I thank Michael Graesser, my other collaborator, with whom I have had many opportunities to talk about physics (and, among other things, about the intelligence of corvids) and whose extraordinary patience and gentlemanliness made it relatively painless to expose to him my confusion on many subjects.
	
 	I accuse D\'onal O'Connell of innumerable discussions about physics and about such topics as teleological suspension, Japanese ritual suicide, and the difference between white wine and red.  Also, of reading and commenting on the draft of Chapter \ref{chap:massless}, and of quackery.
	
	I thank Kris Sigurdson for many similarly interesting discussions, both professional and unprofessional, for setting a ridiculously high standard of success for the members of our class, and for his kind and immensely enjoyable invitation to visit him at the IAS.
	
	I thank Disa El\'iasd\'ottir for many pleasant social occasions and for loudly and colorfully supporting the Costa Rican national team during the 2002 World Cup.  {\it !`Ticos, ticos!}  I also apologize to her again for the unfortunate beer spilling incident when I visited her in Copenhagen last summer.
	
	I thank my officemate, Matt Dorsten, for patiently putting up with my outspoken fondness for animals in human roles, for clearing up my confusion about a point of physics on countless occasions, and for repeated assistance on computer matters.
	
	I thank Ilya Mandel, my long-time roommate, for his forbearance regarding my poor housekeeping abilities and tendency to consume his supplies, as well as for many interesting conversations and a memorable roadtrip from Pasadena to San Jos\'e, Costa Rica.
		
	I thank Jie Yang, with whom I worked as a teaching assistant for two years, for her superhuman efficiency, sunny disposition, and willingness to take on more than her share of the work.
		
	I thank various Irishmen for arguments, and Anura Abeyesinghe, Lotty Ackermann, Christian Bauer, Xavier Calmet, Chris Lee, Sonny Mantry, Michael Salem, Graeme Smith, Ben Toner, Lisa Tracy, and other members of my class and my research group whom I was privileged to know personally.
	
	I thank Jacob Bourjaily, Oleg Evnin, Jernej Kamenik, David Maybury, Brian Murray, Jon Pritchard, Ketan Vyas, and other students with whom I had occassion to discuss physics.
	
	I thank physicists Nima Arkani-Hamed, J.~D.~Bjorken, Roman Buniy, Andy Frey, Jaume Garriga, Holger Gies, Walter Goldberger, Jim Isenberg, Ted Jacobson, Marc Kamionkowski, Alan Kosteleck\'y, Anton Kapustin, Eric Linder, Juan Maldacena, Eugene Lim, Ian Low, Guy Moore, Lubos Motl, Yoichiru Nambu, Hiroshi Ooguri, Krishna Rajagopal, Michael Ramsey-Musolf, John Schwarz, Matthew Schwartz, Guy de T\'eramond, Kip Thorne, and Alex Vilenkin, for questions, comments, and discussions.
	
	I thank Richard Berg, David Berman, Ed Creutz, John Dlugosz, Lars Falk, Monwhea Jeng, Lewis Mammel, Carl Mungan, Frederick Ross, Wolf Rueckner, Tom Snyder, and the other professional and amateur physicists who commented on the work in Chapter \ref{chap:sprinkler}.
		
	I thank the professors with whom I worked as a teaching assistant, David Goodstein, Marc Kamionkowski, Bob McKeown, and Mark Wise, for their patience and understanding.
	
	I thank my father, mother, and brother for their support and advice.

	I thank Caltech for sustaining me as a Robert A. Millikan graduate fellow (2001--2004) and teaching assistant (2004--2006).  I was also supported during the summer of 2005 as a graduate research associate under the Department of Energy contract DE-FG03-92ER40701.
	
	\end{acknowledgments}

	\begin{abstract}

	We begin by reviewing our current understanding of massless particles with spin 1 and spin 2 as mediators of long-range forces in relativistic quantum field theory.  We discuss how a description of such particles that is compatible with Lorentz covariance naturally leads to a redundancy in the mathematical description of the physics, which in the spin-1 case is local gauge invariance and in the spin-2 case is the diffeomorphism invariance of General Relativity.  We then discuss the Weinberg-Witten theorem, which further underlines the need for local invariance in relativistic theories with massless interacting particles that have spin greater than 1/2.
	
	This discussion leads us to consider a possible class of models in which long-range interactions are mediated by the Goldstone bosons of spontaneous Lorentz violation.  Since the Lorentz symmetry is realized non-linearly in the Goldstones, these models evade the Weinberg-Witten theorem and could potentially also evade the need for local gauge invariance in our description of fundamental physics.  In the case of gravity, the broken symmetry would protect the theory from having non-zero cosmological constant, while the compositeness of the graviton could provide a solution to the perturbative non-renormalizability of linear gravity.
	
	This leads us to consider the phenomenology of spontaneous Lorentz violation and the experimental limits thereon.  We find the general low-energy effective action of the Goldstones of this kind of symmetry breaking minimally coupled to the usual Einstein gravity and we consider observational limits resulting from modifications to Newton's law and from gravitational \v{C}erenkov radiation of the highest-energy cosmic rays.  We compare this effective theory with the ``ghost condensate'' mechanism, which has been proposed in the literature as a model for gravity in a Higgs phase.
	
	Next, we summarize the cosmological constant problem and consider some issues related to it.  We show that models in which a scalar field causes the super-acceleration of the universe generally exhibit instabilities that can be more broadly connected to the violation of the null-energy condition.  We also discuss how the equation of state parameter \teq{w = p / \rho} evolves in a universe where the dark energy is caused by a ghost condensate.  Furthermore, we comment on the anthropic argument for a small cosmological constant and how it is weakened by considering the possibility that the size of the primordial density perturbations created by inflation also varies over the landscape of possible universes.
	
	Finally, we discuss a problem in elementary fluid mechanics that had eluded a definitive treatment for several decades: the reverse sprinkler, commonly associated with Feynman.  We provide an elementary theoretical description compatible with its observed behavior.

	\end{abstract}

\tableofcontents

\listoffigures

\end{frontmatter}

\pagestyle{headings}

\chapter{Introduction}
\label{chap:intro}

\begin{epigraphs}
\qitem{\hskip1in est aliquid, quocumque loco, quocumque recessu \\
\hskip1in unius sese dominum fecisse lacertae.}
{Juvenal, Satire III}
\qitem{\hskip1in He thought he saw a Argument \\
\hskip1in That proved he was the Pope: \\
\hskip1in He looked again, and found it was \\
\hskip1in A Bar of Mottled Soap. \\
\hskip1in ``A fact so dread,'' he faintly said, \\
\hskip1in ``Extinguishes all hope!''}
{Lewis Carroll, {\it Sylvie and Bruno Concluded}}
\end{epigraphs}

This dissertation is essentially a collection of the various theoretical investigations that I pursued as a graduate student and that progressed to a publishable state.  It is difficult, {\it a posteriori}, to come up with a theme that will unify them all.  Even the absurdly broad title that I have given to this document fails to account at all for Chapter \ref{chap:sprinkler}, which concerns a long-standing problem in elementary fluid mechanics.  Therefore I will not attempt any such artificial unification here.

I have made an effort, however, to make this thesis more than collation of previously published papers.  To that end, I have added material that reviews and clarifies the relevant physics for the reader.  Also, as far as possible, I have complemented the previously published research with discussions of recent advances in the literature and in my own understanding.

Chapter \ref{chap:massless} in particular was written from scratch and is intended as a review of the relationship between massless particles, Lorentz invariance (LI), and local gauge invariance.  In writing it I attempted to answer the charge half-seriously given to me as a first-year graduate student by Mark Wise of figuring out why we religiously follow the commandment of promoting the global gauge invariance of the Dirac Lagrangian to a local invariance in order to obtain an interacting theory.  Consideration of the role of local gauge invariance in quantum field theories (QFT's) with massless, interacting particles also helps to motivate the research described in Chapter \ref{chap:goldstones}.

Chapter \ref{chap:goldstones} brings up spontaneous Lorentz violation, which is the idea that perhaps the quantum vacuum of the universe is not a Lorentz singlet (or, to put it otherwise, that empty space is not empty).  The idea that gravity might be mediated by the Goldstone bosons of such a symmetry breaking is attractive because it offers a possible solution to two of the greatest obstructions to a quantum description of gravity: the non-renormalizability of linear gravity, and the cosmological constant problem.

The work described in Chapter \ref{chap:phenoLV} seeks to place experimental limits on how large spontaneous Lorentz violation can be when coupled to ordinary gravity.  This line of research is independent from the ideas of Chapter \ref{chap:goldstones} and applies to a wide variety of models in which cosmological physics takes place in a background that is not a Lorentz singlet.

Chapter \ref{chap:cosmological} begins with a brief overview of the cosmological constant problem, one of the greatest puzzles in modern theoretical physics.  The next three sections of that chapter concern original results that are connected to that problem.  Section \ref{sect:nec} in particular has applications beyond the cosmological constant problem, as it offers a theorem that helps connect the energy conditions of General Relativity (GR) with considerations of stability.

All of this work concerns both QFT and GR, our two most powerful (though mutually incompatible) tools for describing the universe at a fundamental level.  In Chapter \ref{chap:sprinkler} we consider an amusing problem about introductory college physics that, surprisingly, had evaded a completely satisfactory treatment for several decades.

\section{Notation and conventions}

We work throughout in units in which \teq{\hbar = c = 1}.  Electrodynamical quantities are given in the Heaviside-Lorentz system of units in which the Coloumb potential of a point charge $q$ is
\beq
\Phi = \frac{q}{4 \pi r}~. \nonumber
\eeq

We also work in the convention in which the Fourier transform and inverse Fourier transform in $n$ dimensions are
\beq
f(x) = \int \frac{d^n k}{(2\pi)^{n/2}} \tilde f(k) e^{-i k \cdot x}; ~~~ \tilde f(x) = \int \frac{d^n k}{(2\pi)^{n/2}} f(x) e^{i k \cdot x}~. \nonumber
\eeq

Lorentz 4-vectors are written as \teq{x = (x^0, x^1, x^2, x^3)}, where $x^0$ is the time component and $x^1, x^2$, and $x^3$ are the $\vh x, \vh y$, and $\vh z$ space components respectively.  Spatial vectors are denoted by boldface, so that we also write $x = (x^0, \vv x)$.  Unit spatial vectors are denoted by superscript hats.  Greek indices such as $\mu, \nu, \rho$, etc.  are understood to run from 0 to 3, while Roman indices such as $i, j, k$, etc. are understood to run from 1 to 3.  Repeated indices are always summed over, unless otherwise specified.

We take $g^{\mu\nu}$ to represent the full metric in GR, while \teq{\eta^{\mu \nu} = \mbox{diag} (+1, -1, -1, -1)} is the Minkowski metric of flat space-time.  Indices are raised and lowered with the appropriate metric.  The square of a tensor denotes the product of the tensor with itself, with all the indices contracted pairwise with the metric.  Thus, for instance, the d'Alembertian operator in flat spacetime is
\beq
\bert = - \partial^2 = - \partial_\mu \partial^\mu = - \eta^{\mu \nu} \partial_\mu \partial_\nu
= - \partial_0^2 + \vv \nabla^2~. \nonumber
\eeq
We define the Planck mass as \teq{M_{\sl{Pl}} = \sqrt{1 / 8 \pi G}}, where $G$ is Newton's constant.  For linear gravity we expand the metric in the form \teq{g^{\mu\nu} = \eta^{\mu \nu} + M_{\sl{Pl}}^{-1} h^{\mu\nu}} and keep only terms linear in $h$.  In Chapter \ref{chap:massless} we will work in units in which $M_{\sl{Pl}} = 1$.  Elsewhere we will show the factors of $M_{\sl{Pl}}$ explicitly.

We use the chiral basis for the Dirac matrices
\beq
\gamma^\mu = \left( \begin{array} {c c} 0 & \sigma^\mu \\ \bar \sigma^\mu & 0 \end{array} \right)~, ~~ \gamma^5 = \left( \begin{array} {r c} -1 & 0 \\ 0 & 1 \end{array} \right)~, \nonumber
\eeq
where \teq{\sigma^\mu = (1, \vv \sigma)}, \teq{\bar \sigma^\mu = (1, - \vv \sigma)}, and the $\sigma^i$'s are the Pauli matrices
\beq
\sigma^1 = \left( \begin{array} {c c} 0 & 1 \\ 1 & 0 \end{array} \right)~, ~~
\sigma^2 = \left( \begin{array} {c r} 0 & -i \\ i & 0 \end{array} \right)~, ~~
\sigma^3 = \left( \begin{array} {c r} 1 & 0 \\ 0 & -1 \end{array} \right)~. \nonumber
\eeq
All other conventions are the standard ones in the literature.

In writing this thesis, I have used the first person plural (``we'') whenever discussing scientific arguments, regardless of their authorship.  I have used the first person singular only when referring concretely to myself in introductory of parenthetical material.  I feel that this inconsistency is justified by the avoidance of stylistic absurdities.
\chapter{Massless mediators}
\label{chap:massless}

\begin{epigraphs}
\qitem{\hskip.5in Did he suspire, that light and weightless down \\ \hskip.5in perforce must move.}{William Shakespeare, {\it Henry IV, part ii}, Act 4, Scene 3}
\qitem{You lay down metaphysic propositions which infer universal consequences, and then you attempt to limit logic by despotism.}{Edmund Burke, {\it Reflections on the Revolution in France}}
\end{epigraphs}

\section{Introduction}
\label{sect:masslessintro}

I have sometimes been asked by scientifically literate laymen (my father, for instance, who is a civil engineer, and my ophthalmologist) to explain to them how a particle like the photon can be said to have no mass.  How would a particle with zero mass be distinguishable from no particle at all?  My answer to that question has been that in modern physics a particle is not defined as a small lump of stuff (which is the mental image immediately conveyed by the word, as well as the non-technical version of the classical definition of the term) but rather as an excitation of a field, somewhat akin to a wave in an ocean.  In that sense, masslessness means something technical: that the excitation's energy goes to zero when its wavelength is very long.  I have then added that masslessness also means that those excitations must always propagate at the speed of light and can never appear to any observer to be at rest.

Here I will attempt a fuller treatment of this problem.  Much of the professional life of a theoretical physicist consists of ignoring technical difficulties and underlying conceptual confusion, in the hope that something publishable and perhaps even useful might emerge from his labor.  If the theorist had to proceed in strictly logical order, the field would advance very slowly.  But, on the other hand, the only thing that can ultimately protect us from being seriously wrong is sufficient clarity about the basics.  In modern physics, long-range forces (electromagnetism and gravity) are understood to be mediated by massless particles with spin $j \geq 1$.  The description of such massless particles in quantum field theory (QFT) is therefore absolutely central to our current understanding of nature.

Therefore, I have decided to use the opportunity afforded by the writing of this thesis to review the subject.  My goals are to elucidate why a relativistic description of massless particles with spin $j \geq 1$ naturally requires something like local gauge invariance (which is not a physical symmetry at all, but a mathematical redundancy in the description of the physics) and to clarify under what circumstances one might expect to evade this requirement.

I shall conclude with a discussion of how these considerations apply to whether some of the major outstanding problems of quantum gravity could be addressed by considering gravity to be an emergent phenomenon in some theory without fundamental gravitons.  Nothing in this chapter will be original in the least, but it will provide a motivation for some of the original work presented in Chapter \ref{chap:goldstones}.

\subsection{Unbearable lightness}
\label{subsect:unbearablelightness}

In his undergraduate textbook on particle physics, David Griffiths points out that massless particles are meaningless in Newtonian mechanics because they carry no energy or momentum, and cannot sustain any force.  On the other hand, the relativistic expression for energy and momentum:
\beq
p^\mu = \left( E, \vv p \right) = \gamma m \left(1, \vv v \right)
\label{pmu}
\eeq
allows for non-zero energy-momentum for a massless particle if \teq{\gamma \equiv \left(1-\vv v ^2 \right)^{-1/2}  \to \infty}, which requires \teq{\abs{\vv v} \to 1}.   Equation (\ref{pmu}) doesn't tell us what the energy-momentum is, but we assume that the relation \teq{p^2 = m^2} is valid for $m=0$, so that a massless particle's energy $E$ and momentum $\vv p$ are related by
\beq
E = \abs{\vv p}~.
\label{Epmassless}
\eeq
Griffiths adds that
\begin{quote}
Personally I would regard this ``argument'' as a joke, were it not for the fact that [massless particles] are known to exist in nature.  They do indeed travel at the speed of light and their energy and momentum {\it are} related by [Eq. (\ref{Epmassless})] (\cite{griffiths}).
\end{quote}
The problem of what actually determines the energy of the massless particle is solved not by special relativity, but by quantum mechanics, via Planck's formula \teq{E = \omega}, where $\omega$ is an angular frequency (which is an essentially wave-like property).  Thus massless particles are the creatures of QFT {\it par excellence}, because, at least in current understanding, they can only be defined as relativistic, quantum-mechanical entities.  Like other subjects in QFT, describing massless particles requires arguments that would seem absurd were it not for the fact that they yield surprisingly useful results that have given us a handle on observable natural phenomena.

\begin{figure} []
\bigskip
\begin{center}
	\begin{fmffile}{fmfYukawa}
	\begin{fmfgraph*}(30,22)
		\fmfpen{thick}
		\fmfleft{i1,i2}
		\fmfright{o1,o2}
		\fmf{fermion}{i1,g1,i2}
		\fmf{fermion}{o1,g2,o2}
		\fmf{dashes,label=$q$}{g1,g2}
	\end{fmfgraph*}
	\end{fmffile}
\end{center}
\bigskip
\caption[Feynman diagram for scattering mediated by scalar field]{\small Feynman diagram for the scattering of two particles that interact through the exchange of a mediator.}
\label{fig:mediator}
\end{figure}
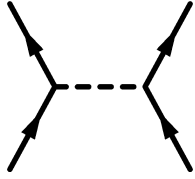

We need massless particles because we regard interaction forces as resulting from the exchange of other particles, called ``mediators.''  Figure \ref{fig:mediator} shows the Feynman diagram that represents the leading perturbative term in the amplitude for the scattering of two particles (represented by the solid lines) that interact via the exchange of a mediator (represented by the dashed line).  We can calculate this Feynman diagram in QFT and match the result to what we would get in non-relativistic quantum mechanics from an interaction potential $V(r)$ (see, e.g., Section 4.7 in \cite{peskin&schroeder}).  The result is
\beq
V(r) = - \frac{g^2}{4 \pi} \frac{e^{-\mu r}}{r}~,
\label{Yukawa}
\eeq
where $g$ is the coupling constant that measures the strength of the interaction and $\mu$ is the mediator's mass.  Therefore, a long-range force requires \teq{\mu = 0}.  In order to accommodate the observed properties of the long-range electromagnetic and gravitational interactions, we also need to give the mediator a on-zero spin.  We will see that this is non-trivial.

\subsection{Overview}
\label{subsect:masslessoverview}

In this chapter we shall first briefly review how one-particle states are defined in QFT and how their polarizations correspond to basis states in irreducible representations of the Lorentz group.  We will emphasize the difference between the case when the mass $m$ of the particle is positive and the case when it is zero.  We shall proceed to use these tools to build a field $\A$ that transforms as a Lorentz 4-vector, first for $m>0$ and then for $m=0$.  We shall conclude that the relativistic description of a massless spin-1 field requires the introduction of local gauge invariance.  Similarly, we will point out how the relativistic description of a massless spin-2 particle that transforms like a two-index Lorentz tensor requires something like diffeomorphism invariance (the fundamental symmetry of GR).  Our discussion of these matters will rely heavily on the treatment given in \cite{weinbergI}.

We will then seek to formulate a solid understanding of the meaning of local gauge invariance and diffeomorphism invariance as redundancies of the mathematical description required to formulate a relativistic QFT with massless mediators.  To this end we will also review the Weinberg-Witten theorem (\cite{weinbergwitten}) and conclude by considering how it might be possible to do without gauge invariance and evade the Weinberg-Witten theorem in an attempt to write a QFT of gravity without UV divergences.

\section{Polarizations and the Lorentz group}
\label{sect:polarizations}

We define one-particle states to be eigenstates of the 4-momentum operator $P^\mu$ and label them by their eigenvalues, plus any other degrees of freedom that may characterize them:
\beq
P^\mu \ket{p, r} = p^\mu \ket{p, r}~.
\label{1particle}
\eeq
Under a Lorentz transformation $\Lambda$ that takes $p$ to $\Lambda p$, the state transforms as
\beq
\ket{p, r} \to U(\Lambda) \ket{p,r}
\label{1particleLorentz}
\eeq
where $U(\Lambda)$ is a unitary operator in some representation of the Lorentz group.  The 4-momentum itself transforms in the fundamental representation, so that
\beq
U^\dagger (\Lambda) P^\mu U(\Lambda) = \Lambda^\mu_{~\nu} P^\nu~.
\label{4momentumLorentz}
\eeq
The 4-momentum of the transformed state is therefore given by
\beq
P^\mu U(\Lambda) \ket{p, r} = U(\Lambda) \left[ U^\dagger (\Lambda) P^\mu U(\Lambda)\right] \ket{p,r} =
U(\Lambda) \Lambda^\mu_{~\nu} p^\nu \ket{p,r} = (\Lambda p)^\mu U(\Lambda) \ket{p,r}~,
\label{eigenvaluetransform}
\eeq
which implies that \teq{U(\Lambda)\ket{p,r}} must be a linear combination of states with 4-momentum $\Lambda p$:
\beq
U(\Lambda) \ket{p, r} = \sum_{r'} c_{r r'}(p, \Lambda) \ket{\Lambda p, r'}~.
\label{cmatrix}
\eeq
If the matrix \teq{c_{r r'}(p, \Lambda)} in Eq. (\ref{cmatrix}), for some fixed $p$, is written in block-diagonal form, then each block gives an irreducible representation of the Lorentz group.  We will call particles in the same irreducible representation ``polarizations.''  The number of polarizations is the dimension of the corresponding irreducible representation.\footnote{Notice that in this choice of language a Dirac fermion has four polarizations: the spin-up and spin-down fermion, plus the spin-up and spin-down antifermion.}

\subsection{The little group}
\label{subsect:littlegroup}

For a particle with mass given by \teq{m = \sqrt{p^2} \geq 0}, let us choose an arbitrary reference 4-momentum $k$ such that \teq{k^2 = m^2}.  Any 4-momentum with the same invariant norm can be written as
\beq
p^\mu = K(p)^\mu_{~\nu} k^\nu
\label{referencek}
\eeq
for some appropriate Lorentz transformation $K(p)$.

Let us then define the ``little group'' as the group of Lorentz transformations $I$ that leaves the reference $k^\mu$ invariant:
\beq
I^\mu_{~\nu} k^\nu = k^\mu~.
\label{littlegpI}
\eeq
Then Eq. (\ref{cmatrix}) can be approached by considering \teq{D_{r r'}(I) =  c_{r r'}(p=k, \Lambda = I)} so that
\beq
U(I) \ket{k, r} = \sum_{r'} D_{r r'}(I) \ket{k, r'}
\eeq
and defining 1-particle states with other 4-momenta by:
\beq
\ket{p, r} = N(p) U(K(p)) \ket{k, r}~,
\eeq
where $N(p)$ is a normalization factor.  If we impose that
\beq
\langle k', r' | k, r \rangle = \delta_{r'r} \delta^3(\vv k' - \vv k)
\label{orthonormalk}
\eeq
for states with 4-momentum $k$, then
\beqa
\langle p', r' | p, r \rangle &=& N^\ast(p') N(p) \bra{k',r'}U^\dagger(K(p')) U(K(p))\ket{k\phantom{'}\!,r} \nn
&=& N^\ast(p') N(p) D_{r'r}\left(K^{-1}(p') K(p)\right) \delta^3(\vv k' - \vv k)~.
\label{bracketp1}
\eeqa
Since the $\delta$-function in the second line vanishes unless \teq{\vv k' = \vv k}, this implies that the overlap is zero unless \teq{\vv p' = \vv p}, and the $D$ matrix in Eq. (\ref{bracketp1}) is therefore trivial:
\beq
\langle p', r' | p, r \rangle = \abs{N(p)}^2 \delta_{r'r} \delta^3(\vv k' - \vv k)~.
\label{bracketp2}
\eeq

We wish to rewrite Eq. (\ref{bracketp2}) in terms of \teq{\delta^3(\vv p' - \vv p)}, to which we have argued it must be proportional.  It is not difficult to show that \teq{d^3 p / p^0} is a Lorentz-invariant measure when integrating on the mass shell \teq{p^0 = \sqrt{\vv p^2 + m^2}}.  This implies that \teq{\delta^3(\vv k' - \vv k) = \delta^3( \vv p' - \vv p) p^0 / k^0} and we therefore have that
\beq
\langle p', r' | p, r \rangle = \abs{N(p)}^2 \delta_{r'r} \delta^3(\vv p' - \vv p) p^0/k^0~.
\label{orthonormalp}
\eeq
Equation (\ref{orthonormalp}) naturally leads to the choice of normalization
\beq
N(p) = \sqrt{k^0/p^0}~.
\label{Np}
\eeq

\subsection{Massive particles}
\label{subsect:massivelittlegp}

A massive particle will always have a rest frame in which its 4-momentum is \teq{k^\mu = (m, 0, 0, 0)}.  This is, therefore, the natural choice of reference 4-momentum.  It is easy to check that the little group is then $SO(3)$, which is the subgroup of the Lorentz group that includes only rotations.

The generators of $SO(3)$ may be written as
\beq
J^i = i \epsilon^{i j k} x^j \partial^k~,
\label{rgenerators}
\eeq
which are the angular momentum operators and which obey the commutation relation
\beq
\left[J^i, J^j \right] = i \epsilon^{i j k} J^k~.
\label{rcommutator}
\eeq
The Lie algebra of $SO(3)$ is the same as that of $SU(2)$, because both groups look identical in the neighborhood of the identity.  In quantum mechanics, the intrinsic angular momentum of a particle (its spin) is a label of the dimensionality of the representation of $SU(2)$ that we assign to it.  A particle of spin $j$ lives in the $2j+1$ dimensional representation of $SU(2)$.

The generators of $SO(1,3)$ may be written as
\beq
J^{\mu \nu} = i \left( x^\mu \partial^\nu - x^\nu \partial^\mu \right)~,
\label{lgenerators}
\eeq
which are clearly anti-symmetric in the indices and which obey the commutation relation
\beq
\left[ J^{\mu \nu}, J^{\rho \sigma} \right] = i \left( \eta^{\nu \rho} J^{\mu \sigma} - \eta^{\mu \rho} J^{\nu \sigma} - \eta^{\nu \sigma} J^{\mu \rho} + \eta^{\mu \sigma} J^{\nu \rho} \right)~.
\label{lcommutator}
\eeq
We may write the six independent components of $J^{\mu \nu}$ as two three-component vectors:
\beq
K^i = J^{0 i}~; ~~~ L^i = \frac 1 2 \epsilon^{i j k} J^{j k}~,
\label{rotationsboosts}
\eeq
where $\vv K$ is the generator of boosts and $\vv L$ is the generator of rotations.  Using Eqs. (\ref{lcommutator}) and (\ref{rotationsboosts}), one can immediately show that these satisfy the commutation relations:
\beq
\left[ L^i, L^j \right] = i \epsilon^{i j k} L^k~; ~~~ \left[ L^i, K^j \right] = i \epsilon^{i j k} K^k~; ~~~ \left[ K^i, K^j \right] = - i \epsilon^{i j k} J^k~.
\label{rotationsboostscommutator}
\eeq

Let us define two new 3-vectors:
\beq
\vv J_\pm = \frac 1 2 \left( \vv L \pm i \vv K \right)~.
\label{Jpm}
\eeq
Using Eq. (\ref{rotationsboostscommutator}) we can write their commutators as
\beq
\left[ J^i_\pm, J^j_\pm \right] = i \epsilon^{i j k} J^k_\pm~; ~~~ \left[J^i_\pm, J^j_\mp \right] = 0~.
\label{Jpmcommutator}
\eeq
That is, both $\vv J_+$ and $\vv J_-$ separately satisfy the commutation relation for angular momentum, and they also commute with each other.  This means that we can identify all finite-dimensional representations of the Lorentz group $SO(1,3)$ by pair of integer or half-integer spins $(j_+, j_-)$ that correspond to two uncoupled representations of $SO(3)$.  The Lorentz-transformation property of a left-handed Weyl fermion $\psi_L$ corresponds to $(1/2,0)$, while $(0,1/2)$ corresponds to the right-handed Weyl fermion $\psi_R$.  A massive Dirac fermion corresponds to the representation \hbox{$(1/2,0) \oplus (0,1/2)$}.

A Lorentz 4-vector (that is, a quantity that transforms under the fundamental representation of $SO(3,1)$), corresponds to \hbox{$(1/2,1/2)$}.  This indicates that it can be decomposed into a spin-1 and a spin-0 part, since \hbox{$1/2 \otimes 1/2 = 1 \oplus 0$}.  Or, to put it otherwise, a general Lorentz vector has four independent components, three of which may be matched to the three polarizations of a $j=1$ particle and one to the single polarization of a $j=0$ particle.

\subsection{Massless particles}
\label{subsect:masslesslittlegp}

Since a massless particle has no rest frame, the simplest reference 4-momentum is \teq{k =(1, 0, 0, 1)}.  The corresponding little group clearly contains as a subgroup rotations about the $z$-axis.  The little group can be parametrized as
\beq
I(\delta, \eta, \phi)^\mu_{~\nu} = \Lambda(\delta, \eta)^\mu_{~\rho} \Lambda(\phi)^\rho_{~\nu}~,
\label{masslessI}
\eeq
where
\beq
\Lambda(\phi)^\mu_{~\nu} = \left( \begin{array}{c c c c}
1 & 0 & 0 & 0 \\
0 & \cos \phi & \sin \phi & 0 \\
0 & -\sin \phi & \cos \phi & 0 \\
0 & 0 & 0 & 1 \end{array} \right)
\label{O2}
\eeq
and
\beq
\Lambda(\delta,\eta)^\mu_{~\nu} = \left( \begin{array}{c c c c}
1 + \zeta & \delta & \eta & -\zeta \\
\delta & 1 & 0 & -\delta \\
\eta & 0 & 1 & -\eta \\
\zeta & \delta & \eta & 1-\zeta \end{array} \right) ~,
\label{Iboostmatrix}
\eeq
with \teq{\zeta = \left(\delta^2 + \eta^2 \right)/2}.

It can be readily checked that
\beq
\Lambda(\delta_1,\eta_1)^\mu_{~\rho} \Lambda(\delta_2,\eta_2)^\rho_{~\nu} =
\Lambda(\delta_1 + \delta_2, \eta_1 + \eta_2)^\mu_{~\nu}~,
\label{iso2translation}
\eeq
which implies that the little group is isomorphic to the group of rotations (by an angle $\phi$) and translations (by a vector $(\delta, \eta)$) in two dimensions.\footnote{The Lorentz transformation in Eq. (\ref{Iboostmatrix}) is, of course, not a physical translation.  It just happens that the group of such matrices is isomorphic to the group of translations on the plane.}  Unlike $SO(3)$, this group, $ISO(2)$, is not semi-simple, i.e., it has invariant abelian subgroups: the rotation subgroup defined by Eq. ({\ref{O2}) and the translation subgroup defined by Eq. (\ref{Iboostmatrix}).

This leads to the important consequence that massless one-particle states \teq{\ket{p, r}} can have only two polarization, called ``helicities,'' given by the component of the angular momentum along its direction of motion.  The physical reason for this is that only the angular momentum component associated with the rotations in Eq. (\ref{O2}) can define discrete polarizations.  Helicities are Lorentz-invariant, unlike the polarizations of a massive particle.

It is clear that massless particles in QFT are different from massive ones.  It is possible to understand some of the properties of massless particles by considering them as massive and then taking the \teq{m \to 0} limit carefully, but this discussion should make it apparent that this limiting procedure is fraught with danger.  We shall explore this issue in the construction of the vector field.

\section{The vector field}
\label{sect:vectorfield}

We seek a causal, free quantum field $A^\mu$ that transforms like a Lorentz 4-vector.  By analogy to the procedure used to obtain free quantum fields with spin 0 and 1/2 (see, e.g., Chapters 2 and 3 in \cite{peskin&schroeder}, or Sections 5.2 to 5.5 in \cite{weinbergI}), we start by writing
\beq
A^\mu(x) = \int \frac{d^3 p}{(2 \pi)^{3/2}} \sum_r 
\left( \epsilon^\mu_r (\vv p) a_r(\vv p) e^{i p \cdot x} + \epsilon^{\mu \ast}_r (\vv p) a^\dagger_r (\vv p) e^{-i p \cdot x} \right) ~,
\label{Amu}
\eeq
where the index $r$ runs over the physical polarizations of the field, while $a$ and $a^\dagger$ are the creation and destruction operators for particles of the corresponding momentum and polarization that obey bosonic commutation relations, and \hbox{$p^\mu = \left(\sqrt{m^2 + \vv p^2}, \vv p \right)$}.

Let $K(p)$ be the Lorentz transformation (boost) that takes a particle of mass $m$ from rest to a 4-momentum $p$.  It can be shown that the measure \teq{d^3 p / p^0} is Lorentz-invariant when integrating on the mass-shell \teq{p^2 = m^2}.  Since both $p^\mu$ and $A^\mu$ are Lorentz 4-vectors, we must have
\beq
\epsilon^\mu_r(\vv p) = \sqrt{\frac{m}{p^0}} K(p)^\mu_{~\nu} \epsilon^\nu_r(0)~.
\label{epsilon0}
\eeq

Now consider the behavior of $\epsilon^\mu_r(0)$ under an infinitesimal rotation.  For our field $A^\mu(x)$ in Eq. (\ref{Amu}) to have a definite spin $j$, we must have that
\beq
\vv L^\mu_{~\nu} \epsilon^\nu_r(0) = \vv S^{(j)}_{r r'} \epsilon^\mu_{r'}(0) ~,
\label{epsilonrot}
\eeq
where the three components of $\vv S^{(j)}$ are the standard spin matrices for spin $j$.  Equation (\ref{epsilonrot}) follows immediately from requiring $\epsilon^\mu_r(0)$ to transform under rotations as both a 4-vector and as a spin-$j$ object.\footnote{It should perhaps also be pointed out that in Eq. (\ref{epsilonrot}) the indices $\mu, \nu$ in the left-hand side indicate components of the three matrices $L^i$ defined in Eq. (\ref{rotationsboosts}).  In Eq. (\ref{lcommutator}) $\mu, \nu$ labeled the matrices themselves.}

For the rotation generators in the fundamental representation of $SO(1,3)$ we have:
\beqa
(L^i)^0_{~0} = (L^i)^0_{~j} = (L^i)^j_{~0} = 0~, \\
(L^i)_{jk} = i \epsilon^i_{~jk}~.
\label{L}
\eeqa
Therefore, for \hbox{$(\vv L^2)^\mu_{~\nu} = \sum_i (L^i)^\mu_{~\rho} (L^i)^\rho_{~\nu}$}, we have
\beqa
(\vv L^2)^0_{~0} = (\vv L^2)^j_{~0} =(\vv L^2)^0_{~j} = 0~; ~~~ (\vv L^2)^j_{~k} = 2 \eta^j_{~k}~.
\label{L2}
\eeqa
Meanwhile, recall that, for the spin matrices,
\beq
(\vv S^{(j)}\,^2)_{r r'} = j(j+1) \delta_{rr'}~.
\label{S2}
\eeq
Using Eqs. (\ref{epsilonrot}), (\ref{L2}), and (\ref{S2}) we therefore obtain that
\beq
\epsilon^i_r(0) = \frac{j(j+1)}{2} \epsilon^i_r(0)~; ~~~ j(j+1) \epsilon^0_r (0) = 0~.
\label{epsilon}
\eeq

Equation (\ref{epsilon}), combined with Eq. (\ref{epsilon0}), leaves us only two posibilities if the field $A^\mu(x)$ in Eq. (\ref{Amu}) is to transform as a 4-vector: 

\bi
\item Either $j=0$ and $\epsilon^0(0)$ is the only non-vanishing component,
\item or $j=1$ and the three $\epsilon^i(0)$'s are the only non-vanishing components
\ei
This agrees with the claim made at the end of the previous section, which we had based on $1/2 \otimes 1/2 = 1 \oplus 0$.  Let us explore both possibilities.

\subsection{Vector field with $j=0$}
\label{subsect:vectorj0}

For the $j=0$ case we can chose the conventionally normalized \hbox{$\epsilon^0(0) = i \sqrt{m/2}$}, which, by Eq. (\ref{epsilon0}) gives
\beq
\epsilon^\mu (\vv p) = i p^\mu \sqrt{\frac 1 {2 p^0}}~.
\label{epsj0}
\eeq
One can then compare the resulting form for $A^\mu(x)$ in Eq. (\ref{Amu}) to the form for a free scalar field and conclude that this vector field has the form
\beq
A^\mu (x) = \partial^\mu \phi (x)
\label{Adphi}
\eeq
for $\phi(x)$ a free, Lorentz scalar field.  Notice that as the field $\phi$ has a single physical polarization, so also does $A^\mu$, and that even though our construction of the vector field assumed an $m > 0$ in Eq. (\ref{epsilon0}), the $m \to 0$ limit in this case is perfectly sensible.\footnote{This kind of massless, spinless vector field will appear again in the discussion of the ``ghost condensate'' mechanism in Chapter \ref{chap:phenoLV}.}

\subsection{Vector field with $j=1$}
\label{subsect:vectorj1}

Now consider the case where the vector field has $j=1$.  Following the popular convention we write
\beq
\epsilon^\mu_{r=\pm1}(0) = \mp \frac 1 {2 \sqrt m} \left(\eta^\mu_{~1} \pm i \eta^\mu_{~2} \right)
\label{transverse}
\eeq
and
\beq
\epsilon^\mu_{r=0}(0) = \sqrt{\frac 1 {2m}} \eta^\mu_{~3}~.
\label{longitudinal}
\eeq
We may check that the raising and lowering operators \hbox{$S^{(1)}_\pm = S^{(1)}_1 \pm i S^{(1)}_2$} act appropriately on these polarization vectors.  For a plane-wave propagating along the $i=3$ spatial direction, $r=\pm1$ correspond to two transverse, circular polarizations of the vector field, while $r=0$ corresponds to the longitudinal polarization.

We may rewrite the field $A^\mu$ in terms of polarization vectors that are mass-independent by introducing
\beq
\tilde \epsilon^\mu_r(0) = \sqrt{2m} \epsilon^\mu_r(0)
\label{epsilontilde}
\eeq
then we have that Eq. (\ref{Amu}) becomes
\beq
A^\mu(x) = \int \frac{d^3 p}{(2\pi)^{3/2}} \frac 1 {\sqrt{2 p^0}} \sum_{r=-1}^1\left(
\tilde \epsilon^\mu_r (\vv p) a_r(\vv p) e^{i p \cdot x} + \tilde \epsilon^\mu_r(\vv p)  a^\dagger_r (\vv p) e^{-i p \cdot x} \right) ~,
\label{Amu2}
\eeq
where \hbox{$\tilde \epsilon^\mu_r(\vv p) = K(p)^\mu_{~\nu} \tilde \epsilon^\nu_r(0)$}.  The field in Eq. (\ref{Amu2}) obeys the equation of motion
\beq
\left( \bert - m^2 \right) A^\mu (x) = 0~.
\label{KG}
\eeq
Notice also that
\beq
p_\mu \tilde \epsilon^\mu_r (\vv p) = p_\mu K^\mu_{~\nu}(p) \tilde \epsilon^\nu_r(0) =
\left(K^{-1}(p) p \right)_\nu \tilde \epsilon^\nu_r(0) = m \tilde \epsilon^0_r(0) = 0
\label{lorenzepsilon}
\eeq
implies that
\beq
\partial_\mu \A =0~.
\label{lorenz}
\eeq

In the limit $m \to 0$ the boost $K(p)$ becomes the identity and \hbox{$\tilde \epsilon^\mu_r (\vv p) = \tilde \epsilon^\mu_r(0)$} for all $\vv p$.  The field then obeys both \teq{\bert \A =0} and \teq{\partial_\mu \A =0}.\footnote{Therefore taking the \teq{m \to 0} limit of the spin-1 vector field automatically gives us the massless field in the Lorenz gauge.}  The fact that there are complications in this limit is revealed by using Eq. (\ref{epsilon0}) and the form of $\tilde \epsilon^\mu_r(0)$'s to obtain
\beq
\Pi^{\mu \nu}(\vv p) \equiv \sum_{r=-1}^1 \tilde \epsilon^\mu_r(\vv p) \tilde \epsilon^\nu_r(\vv p) = \eta^{\mu \nu} + \frac{p^\mu p^\nu}{m^2}~.
\label{Pi}
\eeq
Notice that $\Pi^{\mu\nu}p_\nu = 0$, while $\Pi^{\mu \nu} k_\nu = k^\mu$ for $k \cdot p = 0$, which means $\Pi^{\mu\nu}$ is a projection unto the space orthogonal to $p_\mu$.  Equation (\ref{Pi}) clearly is not finite as $m \to 0$.  This will be a problem if we try to directly couple $A^\mu$ to anything in a Lorentz-invariant way, 
\beq
\La_{\sl{int}} \propto A^\mu j_\mu~,
\label{Aint}
\eeq
because then the rate at which $A^\mu$'s would be emitted by the interaction would be proportional to
\beq
\sum_r \abs{\tilde \epsilon^\mu_r(\vv p) \vev{j_\mu}}^2 = \Pi^{\mu \nu}(\vv p) \vev{j_\mu} \vev{j_\nu}^\ast
= \vev {j^\mu} \vev{j_\mu}^\ast + \frac{1}{m^2} \abs{p \cdot \vev j}^2~,
\label{emission}
\eeq
which clearly diverges as $m \to 0$ unless we impose that $p \cdot \vev j = 0$.  That is, in the presence of an interaction of the form Eq. (\ref{Aint}), we must require that the current to which the field couples be conserved,
\beq
\partial_\mu \vev{j^\mu} = 0~,
\label{jconserved}
\eeq
in order to avoid an infinite rate of emission.

As emphasized earlier in this chapter, the spin of a massless particle must point either parallel or anti-parallel to its direction of propagation.  These possibilities correspond to the longitudinal polarizations \hbox{$\tilde \epsilon^\mu_{\pm 1}$}.  A massless particle cannot have a longitudinal polarization \hbox{$\tilde \epsilon^\mu_0$}.  The requirement of current conservation in Eq. (\ref{jconserved}) ensures that the longitudinal polarization decouples from the current $j^\mu$ in the $m \to 0$ limit, so that it cannot be produced by the interaction in Eq. (\ref{Aint}).

\subsection{Massless $j=1$ particles}
\label{subsect:masslessj1}

Let us now try to construct a genuinely massless vector field with non-zero spin $j$.  To that effect we adopt an arbitrary reference momentum \teq{\vv k = (0, 0, 1)} and a corresponding light-like reference 4-momentum \teq{k = (1, 0, 0, 1)}.  Let $K(p)$ be now defined as the Lorentz transformation that takes a massless particle with reference momentum $k$ to a general momentum $p$.  We can write this transformation as the composition of a rotation (from the direction of $\vv k$ to the direction of $\vv p$) followed by a boost along the direction of $\vv p$ that scales the magnitude.  Then
\beq
\epsilon^\mu_r(\vv p) = K(p)^\mu_{~\nu} \epsilon^\nu_r(\vv k)~.
\label{epsilonk} 
\eeq

We now require that $\epsilon^\mu_r(\vv k)$ transform as both a massless particle with helicity \teq{r=\pm j} and as a 4-vector.  For rotations by an angle $\phi$ around the axis of $\vv k$, we must have
\beq
e^{ir\phi} \epsilon^\mu_r(\vv k) = \Lambda(\phi)^\mu_{~\nu} \epsilon^\nu_r(\vv k)~,
\label{epsilonkrot}
\eeq
where $\Lambda(\phi)^\mu_{~\nu}$ is the Lorentz transformation matrix corresponding to the rotation, given in Eq. (\ref{O2}).  For Eq. (\ref{epsilonkrot}) to be true of a general $\phi$ in the $j=1$ case, we must have
\beq
\epsilon^\mu_{\pm 1}(\vv k) \propto (0,1, \pm i, 0)~
\eeq
and we might as well normalize this solution to match the $\tilde \epsilon^\mu_r$'s in Eq. (\ref{epsilontilde}), giving
\beq
\epsilon^\mu_{\pm 1}(\vv k) = \frac{1}{\sqrt{2}} (0,1, \pm i, 0)~.
\label{epsilonmassless}
\eeq
These are the same polarization vectors that we obtained previously in the $m \to 0$ limit of the massive vector field.

But the little group for massless particles is larger than the $O(2)=U(1)$ group represented by Eq. (\ref{O2}), as was seen in Subsection \ref{subsect:masslesslittlegp}.  For our field to transform as a 4-vector we would also require that
\beq
\epsilon^\mu_r(\vv k) = \Lambda(\delta,\eta)^\mu_{~\nu} \epsilon^\nu_r(\vv k)~,
\label{epsilonboost}
\eeq
where \teq{\Lambda(\delta,\eta)^\mu_{~\nu}} was given in Eq. (\ref{Iboostmatrix}).  Plugging in the polarization 4-vectors in Eq. (\ref{epsilonmassless}) we can see immediately that this is impossible because, under the transformation \teq{\Lambda(\delta,\eta}),
\beq
\epsilon^\mu_{\pm 1}(\vv k) \to \epsilon^\mu_{\pm 1}(\vv k) + \frac{k^\mu}{\abs{\vv k}} \frac{\delta \pm i \eta}{\sqrt 2} ~.
\label{boostedepsilon}
\eeq
Thus we are forced to accept that the one-particle states of a massless spin-1 vector field are not Lorentz-covariant under the action of their little group, but only covariant up to a term proportional to the reference $k^\mu$.  If we then construct the general states using Eq. (\ref{epsilonk}) and
\beq
A^\mu(x) = \int \frac{d^3 p}{(2\pi)^{3/2}} \frac 1 {\sqrt{2 p^0}} \sum_{r=\pm1}\left(
\epsilon^\mu_r (\vv p) a_r(\vv p) e^{i p \cdot x} + \epsilon^\mu_r(\vv p)  a^\dagger_r (\vv p) e^{-i p \cdot x} \right)
\label{Amumassless}
\eeq
we see that we are forced to accept that $A^\mu(x)$ transforms under a general Lorentz transformation $\Lambda$ as:
\beq
A^\mu(x) \to \Lambda^\mu_{~\nu} A^\nu(\Lambda x) + \partial^\mu \Omega(x, \Lambda)
\label{Omega}
\eeq
where $\Omega$ is some function of the coordinates $x$ and the parameters of the Lorentz transformation $\Lambda$.

Equation (\ref{Omega}) should, in my opinion, be regarded as a disaster.  Massless spin-1 quantum fields, which we need in order to explain the observed properties of the electromagnetic interaction, are incompatible with one of the most sacred principles of modern physics: Lorentz covariance.\footnote{This statement may seem peculiar in light of the fact that the Lorentz group was first discovered as the symmetry of the Maxwell equations of classical electrodynamics.  But those equations are written in terms of the fields $\vv E$ and $\vv B$.  The scalar and vector potentials ($A^0$ and $\vv A$ respectively) enter classical electrodynamics only as computational aids.  It is quantum mechanics which {\it requires} a formulation in terms of $\A$.}  It is not, however, an irretrievable disaster, and in fact there will be a rich silver lining to it.

We can ``save'' Lorentz covariance by announcing that two fields related by the transformation
\beq
\A \to \A + \partial^\mu \Omega
\label{gaugetransform}
\eeq
describe the same physics, so that the second term in Eq. (\ref{Omega}) becomes irrelevant.\footnote{This irresistibly brings to my mind a scene from the Woody Allen movie comedy {\it Bananas} in which victorious rebel commander Esposito announces from the Presidential Palace that ``from this day on, the official language of San Marcos will be Swedish...  Furthermore, all children under 16 years old are now 16 years old.''}  We can couple such an $\A$ if the interaction is of the form \teq{\La_{int} \propto \A j_\mu} for a conserved current $j_\mu$, because in that case the coupling is invariant under transformations of the form in Eq. (\ref{gaugetransform}).  Notice that this requirement on the coupling of $\A$ agrees with what we imposed earlier, by Eq. (\ref{emission}), in order to avoid an infinite rate of emission for the vector field in the \teq{m \to 0} limit.

It is easy to construct a genuinely Lorentz-covariant two-index field strength tensor that is invariant under Eq. (\ref{gaugetransform}):
\beq
F_{\mu\nu} = \partial_\mu A_\nu - \partial_\nu A_\mu~.
\label{F}
\eeq
Lorentz-invariant couplings to this field strength would be gauge-invariant, but the presence of derivatives in Eq. (\ref{F}) means that the resulting forces must fall off faster with distance than an inverse-square law (i.e., they cannot be long-range forces).

\section{Why local gauge invariance?}
\label{sect:gaugeinvariance}

The Dirac Lagrangian for a free fermion, \teq{\La = \bar\psi (i \slash\partial - m) \psi} is invariant under the global $U(1)$ gauge transformation $\psi \to e^{i \alpha}\psi$.  This global symmetry, by Noether's theorem, implies conservation of the current:
\beq
j^\mu = \PgP~.
\label{U1j}
\eeq
In the established model of quantum electrodynamics, this Lagrangian is transformed into an interacting theory by making the gauge invariance local: The phase $\alpha$ is allowed to be a function of the space-time point $x$.  This requires the introduction of a gauge field $\A$ with the the transformation property
\beq
A_\mu \to A_\mu + \partial_\mu \alpha
\label{gaugetransformU1}
\eeq
and the use of a covariant derivative \teq{D_\mu = \partial_\mu - i A_\mu} instead of the usual derivative $\partial_\mu$.  This procedure automatically couples $\A$ to the conserved current in Eq. (\ref{U1j}) so that the coupling is invariant under transformations of the form Eq (\ref{gaugetransformU1}).  We than add a Lorentz-invariant kinetic term \teq{-F_{\mu\nu}^2/4} for the field $\A$.  The generalization to non-abelian gauge groups is well known, as is the Higgs mechanism to break the gauge invariance spontaneously and give the field $\A$ a mass.

This is what we are taught in elementary courses on QFT, but the question remains: Why do we promote a global symmetry of the free fermion Lagrangian to a local symmetry?  Equation (\ref{Omega}) provides a deeper insight into the physical meaning of local gauge invariance: a massless particle, having no rest frame, cannot have its spin point along any axis other than that of its motion.  Therefore, it can have only two polarizations.  By describing it as a 4-vector, spin-1 field $\A$ (which has three polarizations) a mathematical redundancy is introduced.

This redundancy is local gauge invariance.  A field with local gauge symmetry is coupled to the conserved current of the corresponding global gauge symmetry in order to make the coupling locally gauge-invariant.  The procedure described of promoting the global gauge symmetry to a local gauge invariance is therefore required in order to couple fermions in a Lorentz-invariant way via a long-range, spin-1 force.

\subsection{Expecting the Higgs}

Remarkably, local gauge invariance also comes to our aid in writing sensible QFT's for the short-range weak nuclear interaction.  At low energies, this interaction is naturally described as being mediated by massive, spin-1 vector fields.  The Lagrangian for such a mediator must look like
\beq
\La = - \frac 1 4 F^2_{\mu\nu} + \frac 1 2 m^2 A^2 - A_\mu J^\mu~,
\label{LmassiveA}
\eeq
where $J^\mu$ is the current to which it couples.  But in the case of the weak nuclear interaction this current is not conserved.  At energy scales much higher than the $m$ in Eq. (\ref{LmassiveA}), we therefore expect the same problem we found in Subsection \ref{subsect:vectorj1} of a divergent emission rate for the longitudinal polarization, unless other higher-derivative operators, which were not relevant at low energies, have come to our rescue.

In the standard model of particle physics, the resolution of this problem is to make the mediators of the weak nuclear interaction gauge bosons, and then to break that gauge invariance spontaneously by introducing a scalar Higgs field with a non-zero VEV, thus giving the bosons the mass that accounts for the short range of the force they mediate.  At high energies the gauge invariance is restored.  The problematic longitudinal polarization disappears and is transmuted into the Goldstone boson of the spontaneously broken symmetry.  Since the Goldstone boson has no spin, it does not have the problem of a divergent rate of emission.  This is the reason why many billions of dollars have been spent in the search for that yet-unseen Higgs boson, a search soon to come to a head with the turning on of the Large Hadron Collider (LHC) at CERN next year.

\subsection{Further successes of gauge theories}

Gauge theories as descriptions of the fundamental particle interactions have other very attractive attributes.  It was shown by 't Hooft that these theories are always renormalizable, i.e., that the infinities that plague QFT's can all be absorbed into a redefinition of the bare parameters of the theory, namely the masses and the coupling constants (\cite{thooft}).  Politzer (\cite{politzer}) and, independently, Gross and Wilczek (\cite{gross&wilczek}), showed that the renormalization flow of the coupling constants in non-abelian gauge theories provides a natural explanation of the observed phenomenon of asymptotic freedom, whereby the nuclear interactions become more feeble at higher energies.

It is also widely believed, though not strictly demonstrated, that QCD, the theory in which the strong nuclear force is mediated by the bosons of an $SU(3)$ gauge theory, accounts for confinement, i.e., for the fact that the strongly interacting fermions (quarks) never occur alone and can appear only in bound states that are singlets of $SU(3)$.  These successes illustrate what we meant when we said in Subsection \ref{subsect:masslessj1} that having to accept local gauge symmetry was a disaster with a rich silver lining.  For interesting accounts of the history of local gauge invariance in classical and quantum physics, see \cite{ORaifStraumann,JacksonOkun}.

\section{Massless $j=2$ particles and diffeomorphism invariance}

We could repeat the sort of procedure used in Subsection \ref{subsect:masslessj1} in order to try to construct Lorentz-covariant $h^{\mu\nu}$ out of the two helicities of a $j=2$ massless field.  This procedure would similarly fail, requiring us to accept the transformation rule:
\beq
h^{\mu\nu} (x) \to \Lambda^\mu_{~\rho} \Lambda^\nu_{~\sigma} h^{\rho \sigma} (\Lambda x)
+ \partial^\mu \xi^\nu(x, \Lambda) + \partial^\nu \xi^\mu(x, \Lambda)~.
\label{xi}
\eeq
Saving Lorentz covariance would then require announcing that states related by a transformation of the form
\beq
h^{\mu\nu} \to h^{\mu\nu} + \partial^\mu \xi^\nu + \partial^\nu \xi^\mu
\label{diffeomorphism}
\eeq
are physically equivalent.  We can construct a four-index field strength tensor $R_{\mu\nu\rho\sigma}$ invariant under Eq. (\ref{diffeomorphism}) that is anti-symmetric in $\mu,\nu$, anti-symmetric in $\rho,\sigma$, and symmetric under exchange of the two pairs.  But to accommodate a long-range force we would need to couple $h^{\mu\nu}$ to a quantity $\Theta^{\mu \nu}$ such that 
\beq
\partial_\mu \vev{\Theta^{\mu \nu}} = 0~.
\label{Thetaconserved}
\eeq
This $\Theta^{\mu \nu}$ is the stress-energy tensor obtained from translational invariance
\beq
x^\mu \to x^\mu - \xi^\mu~,
\label{translation}
\eeq
through Noether's theorem.\footnote{If there were another conserved $\Theta^{\prime\mu\nu}$, there would have to be another conserved 4-vector besides $p^\mu$, namely \teq{p^{\prime\mu} = \int d^3 x\, \Theta^{\prime0\mu}}.  Kinematics would then allow only forward collisions.}  Invariance under Eq. (\ref{diffeomorphism}) corresponds to promoting the translational symmetry in Eq. (\ref{translation}) to a local invariance by letting $\xi^\mu$ be a function of $x$.  It turns out that the theory constructed in this way matches linearized GR around a flat background with $h_{\mu\nu}$ being the graviton field.

It is well known that one can reconstruct the full GR uniquely from linear gravity by a self-consistency procedure (\cite{lineartoGR,deserGR,feynmanGR}).  Therefore a relativistic QFT in flat spacetime with a massless spin-2 particle mediating a long-range force essentially implies GR.  In full GR the invariance under Eq. (\ref{diffeomorphism}) is a consequence of the invariance of the theory under diffeomorphisms:
\beq
x^\mu \to x'^\mu (x)~.
\label{fulldiffeomorphism}
\eeq
Remarkably, we may therefore think of diffeomorphism invariance as a redundancy required by the relativistic description of a massless spin-2 particle.
  
\section{The Weinberg-Witten theorem}

The Weinberg-Witten theorem\footnote{According to the authors, a less general version of their theorem was formulated earlier by Sidney Coleman, but was not published.} rules out the existence of massless particles with higher spin in a very wide class of QFT's (\cite{weinbergwitten}).  In their original paper, the authors present their elegant proof very succinctly.  This review is longer than the paper itself, which may be justified by the importance of this result in further clarifying the need for local gauge invariance in relativistic theories that accommodate long-range forces such as are observed in nature.

Let $\ket{p,\pm j}$ and $\ket{p', \pm j}$ be two one-particle, massless states of spin $j$, labelled by their light-like 4-momenta $p$ and $p'$, and by their helicity (which we take to be the same for the two particles).  We will be considering the matrix elements
\beq
\bra{p', \pm j} j^\mu \ket{p, \pm j}~; ~~~ \bra{p', \pm j} T^{\mu \nu} \ket{p, \pm j}~,
\label{JTmatrix}
\eeq
where $j^\mu$ is a conserved current (i.e., \hbox{$\partial_\mu \vev{j^\mu} =0$}) and $T^{\mu \nu}$ is a conserved stress-energy tensor (i.e. \hbox{$\partial_\mu \vev{T^{\mu \nu}} = 0$}).

\subsection{The $j > 1/2$ case}
\label{subsect:wwj1}

If we assume that the massless particles in question carry a non-zero conserved charge \hbox{$Q = \int d^3 x \, J^0$}, so that (suppressing the helicity label for now)
\beq
Q \ket{p} = q \ket{p}~,
\label{chargeq}
\eeq
where $q \neq 0$, then evidently
\beq
\bra{p'} Q \ket{p\phantom{'}\!} = q \delta^3 (\vv p' - \vv p)~.
\label{qdelta}
\eeq
Meanwhile, we also have that
\beqa
\bra{p'} Q \ket{p\phantom{'}\!} &=& \int d^3 x \, \bra{p'} j^0(t, \vv x) \ket{p\phantom{'}\!} = 
\int d^3 x \,  \bra{p'} e^{i \vv P \cdot \vv x} \, j^0(t, 0) \, e^{-i \vv P \cdot \vv x} \ket{p\phantom{'}\!} \nn
&=& \int d^3 x \, e^{i (\vv p' - \vv p) \cdot \vv x} \bra{p'} j^0(t,0) \ket{p\phantom{'}\!} = (2 \pi)^3 \delta^3 (\vv p' - \vv p) \bra{p'} j^0(t,0) \ket{p\phantom{'}\!}~,
\label{jdelta} 
\eeqa
so that combining Eqs. (\ref{qdelta}) and (\ref{jdelta}) gives
\beq
\lim_{p' \to p} \bra{p'} j^0 (t,0) \ket{p\phantom{'}\!} = \frac{q}{(2 \pi)^3}~,
\label{jlim0}
\eeq
which, by Lorentz covariance, implies that
\beq
\lim_{p' \to p} \bra{p'} j^\mu (t,0) \ket{p\phantom{'}\!} = \frac{q p^\mu}{E(2 \pi)^3} \neq 0~.
\label{jlimmu}
\eeq
Notice that Eq. (\ref{jlimmu}) implies current conservation, because \hbox{$p^2=0$}.

For any light-like $p$ and $p'$
\beq
(p' + p)^2 = 2 (p' \cdot p) = 2(|\vv p'| \abs{\vv p} - \vv p' \cdot \vv p) = 2 |\vv p'| \abs{\vv p} (1 - \cos \theta) \geq 0 ~,
\eeq 
where $\theta$ is the angle between the momenta.  If $\theta \neq 0$, then \hbox{$(p' + p)$} is time-like and we can therefore choose a frame in which it has no space component, so that
\beq
p = (\abs{\vv p}, \vv p)~; ~~~ p' = (\abs{\vv p}, - \vv p)
\eeq
(i.e., the two particles propagate in opposite directions with the same energy).  In this frame, consider rotating the particles by an angle $\phi$ around the axis of $\vv p$:
\beq
\ket{p, \pm j} \to e^{\pm i \phi j} \ket{p, \pm j}~; ~~~ \ket{p', \pm j} \to e^{\mp i \phi j} \ket{p, \pm j}~.
\label{spinrotation}
\eeq
The Lorentz covariance of the matrix element of $j^\mu$ then implies that
\beq
e^{\pm 2 i \phi j} \bra{p', \pm j} j^\mu(t,0) \ket{p\phantom{'}, \pm j} = \Lambda(\phi)^\mu_{~\nu} \bra{p', \pm j} j^\nu(t,0) \ket{p\phantom{'}, \pm j}~,
\label{jspinlorentz}
\eeq
where $\Lambda(\phi)$ is the Lorentz transformation corresponding to a rotation by an angle $\phi$ around the direction of $\vv p$.  But $\Lambda(\phi)$ contains no Fourier components other than $e^{\pm i \phi}$ and 1, so Eq. (\ref{jspinlorentz}) implies that the matrix elements vanish for $j > 1/2$.  In the limit $p' \to p$, we then arrive at a contradiction with Eq. (\ref{jlimmu}).  Therefore no relativistic QFT with a conserved current can have massless spin-1 particles (either fundamental or composite) that have Lorentz-covariant spectra and are charged under the conserved current.

\subsection{The $j > 1$ case}
\label{subsect:wwj2}

If the massless particles in question carry no conserved charge, we may still consider the matrix elements of the stress-energy tensor $T^{\mu \nu}$.  By the same kind of argument as in Subsection \ref{subsect:wwj1}
\beq
\lim_{p' \to p} \bra{p'} T^{\mu \nu} (t,0) \ket{p\phantom{'}\!} = \frac{p^\mu p^\nu}{E(2 \pi)^3} \neq 0~.
\label{Tlimmu}
\eeq
Notice again that this stress-energy is conserved because $p^2 = 0$.

Then combining Eq. (\ref{spinrotation}) with relativistic covariance implies that
\beq
e^{\pm 2 i \phi j} \bra{p', \pm j} T^{\mu \nu}(t,0) \ket{p\phantom{'}, \pm j} = 
\Lambda(\phi)^\mu_{~\rho} \Lambda(\phi)^\nu_{~\sigma}
\bra{p', \pm j}T^{\rho \sigma}(t,0) \ket{p\phantom{'}, \pm j}~.
\label{Tspinlorentz}
\eeq
The fact that $\Lambda(\phi)$ contains only the Fourier components $e^{\pm i \phi}$ and 1 then implies that the matrix elements must vanish for \teq{j > 1}, contradicting Eq. (\ref{Tlimmu}) in the limit $p' \to p$.  Therefore no relativistic QFT with a conserved stress-energy tensor can have massless spin-2 particles (either fundamental or composite) that have Lorentz-covariant spectra.

\subsection{Why are gluons and gravitons allowed?}
\label{subsect:gluonsgravitons}

Evidently, the Weinberg-Witten theorem does not forbid photons, because they carry no conserved charge.  It also does not forbid the $W^\pm$ and $Z$ bosons because they are massive.  But the Standard Model contains charged, massless spin-1 particles (the gluons) as well as massless spin-2 particles (the gravitons).  How is this possible?  The resolution of this question helps to clarify the necessity for local gauge invariance.

In a Yang-Mills theory,
\beq
\La_{\sl{YM}} = - \frac 1 4 F^a_{\mu \nu} F^{a \mu \nu} + \La_{\sl{matter}} (\psi, D_\mu \psi)
\label{YM}
\eeq
the gauge-invariant current
\beq
j^\mu_a = \frac{\delta S_{\sl{matter}}}{\delta A^a_\mu}
\label{YMj}
\eeq
is not conserved, because it obeys the equation \hbox{$D_\mu \vev {j^\mu_a}=0$}, rather than  \hbox{$\partial_\mu \vev {j^\mu_a}=0$}.  Furthermore, \teq{\vev{j^\mu_a}} vanishes for one-particle gauge field states.  Therefore considering the matrix elements of this $j^\mu_a$ between gauge boson states in Yang-Mills theory would avail us nothing because the limit in Eq. (\ref{jlimmu}) would be zero.

What we actually want is a current that measures the flow of charge in the absence of matter (i.e., for the Yang-Mills bosons alone) and that is conserved in the sense \hbox{$\partial_\mu \vev {{\cal J}^\mu_a}=0$}:
\beq
{\cal J}^\mu_a = -F_c^{\mu\nu} f_{cab} A_{b\nu}~,
\label{gaugecurrent}
\eeq 
where the $f$'s are the structure constants of the gauge group.  Conservation follows immediately from the equation of motion for Eq. (\ref{YM}).  This is, in fact, the conserved current obtained through Noether's theorem from the global gauge invariance of Eq. (\ref{YM}) without matter.  But the current in Eq. (\ref{gaugecurrent}) is obviously not gauge-invariant.  Therefore, under the action of a Lorentz transformation $\Lambda$,
\beq
{\cal J}^\mu_a \to \Lambda^\mu_{~\nu} {\cal J}^\nu_a + \partial^\mu \Omega_a
\eeq
and it is not, consequently, Lorentz-covariant.  If we tried making it Lorentz-covariant by introducing an unphysical extra polarization of the gauge boson, then the theorem would fail because the helicities would not be Lorentz-invariant, invalidating the choice-of-frame procedure used to arrive at Eq. (\ref{spinrotation}).

To put this in another way, in a gauge theory the physical $\ket{p, \pm j}$ states are actually equivalence classes, because two states related by a gauge transformation represent the same physics.  A technical way of thinking about this is that the physical states are elements of the BRST cohomology (\cite{BRST}).  Therefore, matrix elements such as those in Eq. (\ref{JTmatrix}) are only well-defined if the operator $j^\mu$ is BRST-closed, which requires the operator to be gauge-invariant.  It is well known that Yang-Mills theories do not allow the construction of gauge-invariant conserved currents.

The case of the graviton is very closely analogous to that of the Yang-Mills bosons.  In Einstein-Hilbert gravity
\beq
S = \int d^4 x \, \sqrt{-g} \left[ R + \La_{\sl{matter}}(\phi, \nabla_\mu \phi, g_{\mu\nu})  \right]~,
\eeq
where the field $\phi$ stands for all possible matter fields of any spin.  The covariant stress-energy tensor
\beq
T^{\mu\nu} = \frac 1 {\sqrt{-g}} \frac{\delta S_{\sl{matter}}}{\delta g_{\mu \nu}}
\label{Tmatter}
\eeq
obeys \hbox{$\nabla_\mu \vev{T^{\mu\nu}} = 0$} rather than \hbox{$\partial_\mu \vev{T^{\mu\nu}} = 0$}, and $ \vev{T^{\mu\nu}} = 0$ for any state with only gravitational fields.  What we want is therefore not $T$, but rather
\beq
\Theta^\mu_{~\nu} =
\frac{\partial R} {\partial (\partial_\mu g_{\alpha \beta} )} (\partial_\nu g_{\alpha \beta})
- g^\mu_{\phantom a \nu} R~.
\label{gravityT}
\eeq
But recall that the Ricci scalar $R$ contains not only the metric and its first derivatives, but also terms linear in its second derivatives.  In order to define $\Theta$ we therefore need to do the usual trick of integrating by parts and setting the boundary terms to zero in order to get rid of the second derivatives in $R$.  This means that $R$ is no longer a covariant scalar and therefore $\Theta$ is not a covariant tensor, but rather a pseudotensor.

It is well known that gravitational energy cannot be defined in a covariant way.  For instance, the energy of gravity waves on a flat background is localizable only for waves traveling in a single direction, which is not a coordinate-invariant condition (see, for instance, Chapter 33 in \cite{diracGR}).  A general Lorentz transformation of the graviton field $h^{\mu\nu}$ will destroy this condition.  This means that the stress-energy pseudotensor $\Theta^{\mu\nu}$ for gravitons involves a field $h^{\mu\nu}$ that does not transform like a Lorentz tensor.  Its matrix elements are therefore not Lorentz-covariant.  Once again, if we attempt to remedy this by introducing unphysical extra polarizations of the gravitons, the Lorentz invariance of the helicity is lost.

Otherwise stated, in a theory with diffeomorphism invariance like GR, the physical states are equivalence classes, because two states related by a coordinate transformation represent the same physics.  The matrix elements in Eq. (\ref{JTmatrix}) are only well defined if the operator $T^{\mu \nu}$ is BRST-closed, but GR admits no local BRST-closed operators, and thus evades the Weinberg-Witten theorem.

Notice that even in theories with a local symmetry, such as QCD or GR, the Weinberg-Witten theorem does rule massless particles of higher spin that carry a conserved charge associated with a symmetry that commutes with the local symmetry.  For instance, the authors of \cite{weinbergwitten} point out that their result forbids QCD from having flavor non-singlet massless bound states with \teq{j \geq 1}, since flavor symmetries commute with the $SU(3)$ local gauge symmetry.  Similarly, a $j=1$ gauge theory cannot produce composite gravitons with Lorentz-covariant spectra, because translations in flat Minkowski space-time commute with the gauge symmetry.  Gauge theories admit the conserved, Lorentz-covariant Belinfante-Rosenfeld stress-energy tensor (\cite{belinfanterosenfeld}).

\subsection{Gravitons in string theory}
\label{subsect:strings}

String theories have a massless spin-2 particle in their spectrum.  This discovery killed the original versions of string theory as possible descriptions of the strong nuclear interaction (which was the context in which they had been proposed) and made modern string theory a candidate for a quantum theory of gravity (see, for instance, Chapter 1 in \cite{GSW}).  The reason why this result does not violate the Weinberg-Witten theorem is that it is not possible to define a conserved stress-energy tensor in string theory.

Consider a string propagating in a $D$-dimensional background space-time with metric \teq{g_{a b}}, where \teq{a, b = 0, 1, \ldots D-1}.  If $S$ is the action in the background, then
\beq
T^{a b} = \frac{1}{\sqrt{-g}} \frac{\delta S}{\delta g_{a b}}
\label{stringT}
\eeq
is not well defined because a consistent string theory requires imposing superconformal symmetry on the background, which in turn automatically requires \teq{g_{a b}} to obey an equation of motion (at low energies this equation of motion corresponds to the Einstein field equation of GR).  The functional derivative in Eq.(\ref{stringT}) cannot be defined because there is no consistent off-shell definition of the background action $S$: The exact equation of motion for \teq{g_{a b}} in string theory does not come from extremizing the action with respect to the background metric, but rather from a constraint required for consistency.\footnote{I thank John Schwarz for clarifying this point for me.}

In general, we expect that a theory with {\it emergent} diffeomorphism invariance would not have a stress-energy tensor.  The reason is that in the low-energy effective action (i.e., in GR) the graviton couples to a stress-energy tensor which is not observable because it is not diffeomorphism-covariant.  If the fundamental theory itself has no diffeomorphism invariance, then it should not have a stress-energy tensor at all (see \cite{seiberg}).

\section{Emergent gravity}
\label{sect:emergentgravity}

The Weinberg-Witten theorem can be read as the proof that massless particles of higher spin cannot carry conserved Lorentz-covariant quantities.  Local gauge invariance and diffeomorphism invariance are natural ways of making those quantities mathematically non-Lorentz-covariant without spoiling physical Lorentz covariance.  It is possible and interesting nonetheless, to consider other ways of accommodating massless mediators with higher spin.  Despite the successes of gauge theories, the fact remains that there is no clearly compelling {\it a priori} reason to impose local gauge invariance as an axiom, and that such an axiom has the unattractive consequence that it makes our mathematical description of physical reality inherently redundant  (see, for instance, Chapter III.4 in \cite{zee}).

Also, while local gauge invariance guarantees renormalizability for spin 1, it is well known that quantizing $h_{\mu\nu}$ in linear gravity does not produce a perturbatively renormalizable field theory.  One attractive solution to this problem would be to make the graviton a composite, low-energy degree of freedom, with a natural cutoff scale $\Lambda_{\sl{UV}}$.  The Weinberg-Witten theorem represents a significant obstruction to this approach, because the result applies equally to fundamental and to composite particles.  Indeed, ruling out emergent gravitons was the authors' purpose for establishing that theorem.

In a recent public lecture (\cite{sidneyfest}), Witten has made the strong claim that ``whatever we do, we are not going to start with a conventional theory of non-gravitational fields in Minkowski spacetime and generate Einstein gravity as an emergent phenomenon.''  His reasoning is that identifying emergent phenomena requires first defining a box in 3-space and then integrating out modes with wavelengths shorter than the length of the edges of the box (see Fig. \ref{fig:emergent}).  But Einstein gravity implies diffeomorphism invariance, and a general coordinate transformation spoils the definition of our box.  Witten's conclusion is that gravity can be emergent only if the notion on the space-time on which diffeomorphism invariance operates is simultaneously emergent.  This is a plausible claim, but it goes beyond what the Weinberg-Witten theorem actually establishes.

\begin{figure} []
\bigskip
\begin{center}
\includegraphics[scale=0.32]{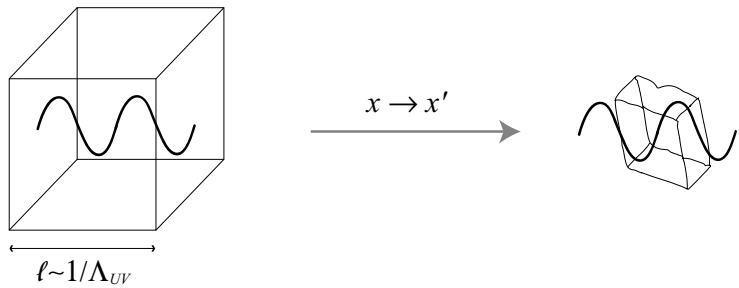}
\end{center}
\caption[Schematic of Witten's argument against an emergent theory of gravity]{\small Schematic representation of Witten's argument that a general coordinate transformation spoils the box used to define the modes that are integrated out in order to identify the emergent low-energy physics for energy scales well below $\Lambda_{\sl{UV}}$.}
\label{fig:emergent}
\end{figure}

In 1983, Laughlin explained the observed fractional quantum Hall effect in two-dimensional electronic systems by showing how such a system could form an incompressible quantum fluid whose excitations have charge $e/3$ (\cite{laughlin}).  That is, the low-energy theory of the interacting electrons in two spatial dimensions has composite degrees of freedom whose charge is a fraction of that of the electrons themselves.  In 2001, Zhang and Hu used techniques similar to Laughlin's to study the composite excitations of a higher-dimensional system (\cite{zhang&hu}).  They imagined a four-dimensional sphere in space, filled with fermions that interact via an $SU(2)$ gauge field.  In the limit where the dimensionality of the representation of $SU(2)$ is taken to be very large, such a theory exhibits composite massless excitations of integer spin 1, 2 and higher.

Like other theories from solid state physics, Zhang and Hu's proposal falls outside the scope of the Weinberg-Witten theorem because the proposed theory is not Lorentz-invariant:  The vacuum of the theory is not empty and has a preferred rest-frame (the rest frame of the fermions).  However, the authors argued that in the three-dimensional boundary of the four-dimensional sphere, a relativistic dispersion relation would hold.  One might then imagine that the relativistic, three-dimensional world we inhabit might be the edge of a four-dimensional sphere filled with fermions.  Photons and gravitons would be composite low-energy degrees of freedom, and the problems currently associated with gravity in the UV would be avoided.  The authors also argue that massless bosons with spin 3 and higher might naturally decouple from other matter, thus explaining why they are not observed in nature.

In Chapter \ref{chap:goldstones} we will discuss another proposal, dating back to the work of Dirac (\cite{diracLV}) and Bjorken (\cite{bjorken1}) for obtaining massless mediators as the Goldstone bosons of the spontaneous breaking of Lorentz violation.  Such an arrangement evades the Weinberg-Witten theorem because the Lorentz invariance of the theory is realized non-linearly in the Goldstone bosons.  Therefore the matrix elements in Eq. (\ref{JTmatrix}) will not be Lorentz-covariant.
\chapter{Goldstone photons and gravitons}
\label{chap:goldstones}

In this chapter we will address some issues connected with the construction of models in which massless mediators are obtained as Goldstone bosons of the spontaneous breaking of Lorentz invariance (LI).  This presentation is based largely on previously published work \cite{LV1,CPT04}.

\section{Emergent mediators}

In 1963, Bjorken proposed a mechanism for what he called the ``dynamical generation of quantum electrodynamics'' (QED) (\cite{bjorken1}).  His idea was to formulate a theory that would reproduce the phenomenology of standard QED, without invoking local $U(1)$ gauge invariance as an axiom.  Instead, Bjorken proposed working with a self-interacting fermion field theory of the form
\begin{equation}
\La = \bar\psi(i\partial\!\!\!/-m)\psi - \lambda(\PgP)^2 .
\label{bjorken}
\end{equation}
Bjorken then argued that in a theory such as that described by Eq. (\ref{bjorken}), composite ``photons'' could emerge as Goldstone bosons resulting from the presence of a condensate that spontaneously
broke LI.

Conceptually, a useful way of understanding Bjorken's proposal is to think of it as as a resurrection of the ``lumineferous \ae ther'' (\cite{Nambu1,Nambu2}):  ``empty'' space is no longer really empty.  Instead, the theory has a non-vanishing vacuum expectation value (VEV) for the current $j^\mu=\PgP$.  This VEV, in turn, leads to a massive background gauge field $\A \propto j^\mu$, as in the well-known London equations for the theory of superconductors (\cite{London}).  Such a background spontaneously breaks Lorentz invariance and produces three massless excitations of $\A$ (the Goldstone bosons) proportional to the changes $\delta j^\mu$ associated with the three broken Lorentz transformations.\footnote{In Bjorken's work, $\A$ is just an auxiliary or interpolating field.  Dirac had discussed somewhat similar ideas in \cite{diracLV}, but, amusingly, he was trying to write a theory of electromagnetism with only a gauge field and no fundamental electrons.  In both the work of Bjorken and the work of Dirac, the proportionality between $\A$ and $j^\mu$ is crucial.} 

Two of these Goldstone bosons can be interpreted as the usual transverse photons.  The meaning of the third photon remains problematic.  Bjorken originally interpreted it as the longitudinal photon in the temporal-gauge QED, which becomes identified with the Coulomb force (see also \cite{Nambu1}).  More recently, Kraus and Tomboulis have argued that the extra photon has an exotic dispersion relation and that its coupling to matter should be suppressed (\cite{KrausTomboulis}).

Bjorken's idea might not seem attractive today, since a theory such as Eq. (\ref{bjorken}) is not renormalizable, while the work of 't Hooft and others has demonstrated that a locally gauge-invariant theory can always be renormalized (\cite{thooft}).  Furthermore, as detailed in Section \ref{sect:gaugeinvariance}, the gauge theories have had other very significant successes.  Unless we take seriously the line of thought pursued in Chapter \ref{chap:massless} that local gauge invariance is suspect because it is a redundancy of the mathematical description rather than a genuine physical symmetry, there would not appear to be, at this stage in our understanding of fundamental physics, any compelling reason to abandon local gauge invariance as an axiom for writing down interacting QFT's.\footnote{According to Mark Wise, though, in the 1980's Feynman considered Bjorken's proposal as an alternative to postulating local gauge invariance.}  Furthermore, the arguments for the existence of a LI-breaking condensate in theories such as Eq. (\ref{bjorken}) have never been solid.\footnote{For Bjorken's most recent revisiting of his proposal, in the light of the theoretical developments since 1963, see \cite{Bjorken2}.}

In 2002 Kraus and Tomboulis resurrected Bjorken's idea for a different purpose of greater interest to contemporary theoretical physics: making a composite graviton (\cite{KrausTomboulis}).  They proposed what Bjorken might call ``dynamical generation of gravity.''  In this scenario a composite graviton would emerge as a Goldstone boson from the spontaneous breaking of Lorentz invariance in a theory of self-interacting fermions.  Being a Goldstone boson, such a graviton would be forbidden from developing a potential, thus providing a solution to the ``large cosmological constant problem:'' the \teq{\Lambda h^\mu_{~\mu}} tadpole term for the graviton would vanish without fine-tuning (see Section \ref{sect:cosmointro}).  This scheme would also seem to offer an unorthodox avenue to a renormalizable quantum theory of gravity, because the fermion self-interactions could be interpreted as coming from the integrating out, at low energies, of gauge bosons that have acquired large masses via the Higgs mechanism, so that Einstein gravity would be the low energy behavior of a renormalizable theory.  This proposal would, of course, radically alter the nature of gravitational physics at very high energies.  Related ideas had been previously considered in, for instance, \cite{ohanian}.

In \cite{KrausTomboulis}, the authors consider fermions coupled to gauge bosons that have acquired masses beyond the energy scale of interest.  Then an effective low-energy theory can be obtained by integrating out those gauge bosons.  We expect to obtain an effective Lagrangian of the form
\begin{eqnarray}
\La &=&
\bar\psi(i\partial\!\!\!/-m)\psi+\sum_{n=1}^\infty\lambda_{n}(\PgP)^{2n}
\nonumber \\ && +\sum_{n=1}^\infty\mu_{n} \left[ \tensorbilinear \right]^{2n} +
\,\ldots ~,
\label{bothbilinears}
\end{eqnarray}
where we have explicitly written out only two of the power series in fermion bilinears that we would in general expect to get from integrating out the gauge bosons.

One may then introduce an auxiliary field for each of these fermion
bilinears.  In this example we shall assign the label $A^\mu$ to the
auxiliary field corresponding to $\PgP$, and the label $h^{\mu\nu}$ to
the field corresponding to $\tensorbilinear$.  It is possible to write
a Lagrangian that involves the auxiliary fields but not their
derivatives, so that the corresponding algebraic equations of motion
relating each auxiliary field to its corresponding fermion bilinear
make that Lagrangian classically equivalent to
Eq. (\ref{bothbilinears}).  In this case the new Lagrangian would be
of the form
\begin{eqnarray}
\La ' &=& (\eta^{\mu\nu} + h^{\mu\nu})\tensorbilinear -
\bar\psi(A\!\!\!/ + m)\psi + \ldots \nonumber \\ && -V_A(A^2)
-V_h(h^2) + \ldots ~,
\label{auxiliary}
\end{eqnarray}
where $A^2 \equiv A_\mu A^\mu$ and $h^2 \equiv h_{\mu \nu}
h^{\mu \nu}$.  The ellipses in Eq. (\ref{auxiliary}) correspond to
terms with other auxiliary fields associated with more complicated
fermion bilinears that were also omitted in Eq. (\ref{bothbilinears}).

We may then imagine that instead of having a single fermion species we
have one very heavy fermion, $\psi_1$, and one lighter one, $\psi_2$.
Since Eq. (\ref{auxiliary}) has terms that couple both fermion species
to the auxiliary fields, integrating out $\psi_1$ will then produce
kinetic terms for $A^\mu$ and $h^{\mu\nu}$.

In the case of $A^\mu$ we can readily see that since it is minimally
coupled to $\psi_1$, the kinetic terms obtained from integrating out
the latter must be gauge-invariant (provided a gauge-invariant regulator
is used).  To lowest order in derivatives of $A^\mu$, we must then get
the standard photon Lagrangian $-F_{\mu\nu}^2/4$.  Since
$A^\mu$ was also minimally coupled to $\psi_2$, we then have, at low
energies, something that has begun to look like QED.

If $A^\mu$ has a non-zero VEV, LI is spontaneously broken, producing
three massless Goldstone bosons, two of which may be interpreted as
photons (see \cite{KrausTomboulis} for a discussion of how the exotic
physics of the other extraneous ``photon'' can be suppressed).  The
integrating out of $\psi_1$ and the assumption that $h^{\mu\nu}$ has a
VEV, by similar arguments, yield a low-energy approximation to
linearized gravity.

Fermion bilinears other than those we have written out explicitly in
Eq. (\ref{bothbilinears}) have their own auxiliary fields with their
own potentials.  If those potentials do not themselves produce VEV's
for the auxiliary fields, then there would be no further Goldstone
bosons, and one would expect, on general grounds, that those extra
auxiliary fields would acquire masses of the order of the
energy-momentum cutoff scale for our effective field theory, making
them irrelevant at low energies.

The breaking of LI would be crucial for this kind of mechanism, not
only because we know experimentally that photons and gravitons are
massless or very nearly massless, but also because it allows us to evade the Weinberg-Witten theorem (\cite{weinbergwitten}), as we discussed in Section \ref{sect:emergentgravity}.

Let us concentrate on the simpler case of the auxiliary field $A^\mu$.
For the theory described by Eq. (\ref{auxiliary}), the equation of
motion for $A^\mu$ is
\begin{equation}
\frac{\partial \La'}{\partial A_{\mu}} = -\PgP - V'(A^2) \cdot 2
A^{\mu} = 0.
\label{algebraic}
\end{equation}

Solving for $\PgP$ in Eq. (\ref{algebraic}) and substituting into both
Eq. (\ref{bothbilinears}) and Eq. (\ref{auxiliary}) we see that the
condition for the Lagrangians $\La$ and $\La'$ to be classically
equivalent is a differential equation for $V(A^2)$ in terms of the
coefficients $\lambda_{n}$:
\begin{equation}
V(A^2) = 2 A^2 [V'(A^2)] -  \sum_{n=1}^{\infty}\lambda_{n} 2^{2n}
A^{2n} [V'(A^2)]^{2n}.
\label{differen}
\end{equation}

It is suggested in \cite{KrausTomboulis} that for some values of
$\lambda_n$ the resulting potential $V(A^2)$ might have a minimum away
from $A^2=0$, and that this would give the LI-breaking VEV needed.  It
seems to us, however, that a minimum of $V(A^2)$ away from the origin
is not the correct thing to look for in order to obtain LI breaking.
The Lagrangian in Eq. (\ref{auxiliary}) contains $A^\mu$'s not just in
the potential but also in the ``interaction'' term $A_\mu \PgP$, which
is not in any sense a small perturbation as it might be, say, in QED.
In other words, the classical quantity $V(A^2)$ is not a useful
approximation to the quantum effective potential for the auxiliary
field.

In fact, regardless of the values of the $\lambda_{n}$,
Eq. (\ref{differen}) implies that $V(A^2=0)=0$, and also that at any
point where $V'(A^2)=0$ the potential must be zero.  Therefore, the
existence of a classical extremum at $A^2=C \neq 0$ would imply that
$V(C)=V(0)$, and unless the potential is discontinuous somewhere, this
would require that $V'$ (and therefore also $V$) vanish somewhere
between $0$ and $C$, and so on \textit{ad infinitum}. Thus the
potential $V$ cannot have a classical minimum away from $A^2=0$,
unless the potential has poles or some other discontinuity.

A similar observation applies to any fermion bilinear for which we
might attempt this kind of procedure and therefore the issue arises as
well when dealing with the proposal in \cite{KrausTomboulis} for
generating the graviton.  It is not possible to sidestep this
difficulty by including other auxiliary fields or other fermion
bilinears, or even by imagining that we could start, instead of from
Eq. (\ref{bothbilinears}), from a theory with interactions given by an
arbitrary, possibly non-analytic function of the fermion bilinear
$F(\mbox{bilinear})$.  The problem can be traced to the fact
that the equation of motion of any auxiliary field of this kind will
always be of the form
\begin{equation}
0 = -(\mbox{bilinear}) - V'(\mbox{field}^2)\cdot 2 \,\mbox{field} .
\label{general}
\end{equation}

The point is that the vanishing of the first derivative of the
potential or the vanishing of the auxiliary field itself will always,
classically, imply that the fermion bilinear is zero.  Classically at
least, it would seem that the extrema of the potential would
correspond to the same physical state as the zeroes of the auxiliary
field.

\section{Nambu and Jona-Lasinio model (review)}

The complications we have discussed that emerge when one tries to
implement LI breaking as proposed in \cite{KrausTomboulis} do not, in
retrospect, seem entirely surprising. A VEV for the auxiliary field
would classically imply a VEV for the corresponding fermion bilinear,
and therefore a trick such as rewriting a theory in a form like
Eq. (\ref{auxiliary}) should not, perhaps, be expected to uncover a
physically significant phenomenon such as the spontaneous breaking of
LI for a theory where it was not otherwise apparent that the fermion
bilinear in question had a VEV.  Let us therefore turn our attention
to considering what would be required so that one might reasonably
expect a fermion field theory to exhibit the kind of condensation that
would give a VEV to a certain fermion bilinear.

If we allowed ourselves to be guided by purely classical intuition, it
would seem likely that a VEV for a bilinear with derivatives (such as
$\tensorbilinear$) might require nonstandard kinetic terms in the
action.  Whether or not this intuition is correct, we abandon consideration of such bilinears here
as too complicated.

The simplest fermion bilinear is, of course, $\bar\psi\psi$.  Being a
Lorentz scalar, $\langle \bar\psi\psi \rangle \neq 0$ will not break LI.  This kind
of VEV was treated back in 1961 by Nambu and Jona-Lasinio, who
used it to spontaneously break chiral symmetry in one of the early
efforts to develop a theory of the strong nuclear interactions, before
the advent of quantum chromodynamics (QCD) (\cite{NJL}).  It might be
useful to review the original work of Nambu and Jona-Lasinio, as it
may shed some light on the study of the possibility of giving VEV's to
other fermion bilinears that are not Lorentz scalars.

In their original paper, Nambu and Jona-Lasinio start from a
self-interacting massless fermion field theory and propose that the
strong interactions be mediated by pions, which appear as Goldstone
bosons produced by the spontaneous breaking of the chiral symmetry
associated with the transformation $\psi \mapsto
\exp{(i\alpha\gamma^5)} \psi$.  This symmetry breaking is produced by
a VEV for the fermion bilinear $\bar{\psi}\psi$.  In other words,
Nambu and Jona-Lasinio originally proposed what, by close analogy to
Bjorken's idea, would be the ``dynamical generation of the strong
interactions.''\footnote{Historically, though, Bjorken was motivated
by the earlier work of Nambu and Jona-Lasinio.}

Nambu and Jona-Lasinio start from a non-renormalizable quantum field
theory with a four-fermion interaction that respects chiral symmetry:
\begin{equation}
\La = i \bar{\psi} \partial\!\!\!/ \psi - \frac{g}{2}[(\PgP)^2 -
(\bar{\psi}\gamma^\mu \gamma^5 \psi)^2] .
\label{NJL}
\end{equation}

\begin{figure}[]
\bigskip
\begin{center}



\includegraphics[scale=1.2]{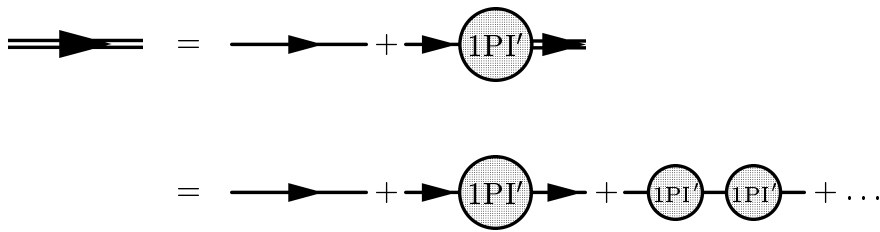}

\end{center}
\caption[Diagrammatic Schwinger-Dyson equation]{\small Diagrammatic Schwinger-Dyson equation.  The double line represents the primed propagator, which incorporates the
self-energy term.  The single line represents the unprimed propagator.
1PI$'$ stands for the sum of one-particle irreducible graphs with the
primed propagator.}
\label{sde}
\end{figure}

In order to argue for the presence of a chiral symmetry-breaking
condensate in the theory described by Eq. (\ref{NJL}), Nambu and
Jona-Lasinio borrowed the technique of self-consistent field theory
from solid state physics (see, for instance, \cite{selfconsistent}).
If one writes down a Lagrangian with a free and an interaction part,
$\La = \La_0 + \La_i$, ordinarily one would then proceed to
diagonalize $\La_0$ and treat $\La_i$ as a perturbation.  In
self-consistent field theory one instead rewrites the Lagrangian as
$\La = (\La_0 + \La_s) + (\La_i - \La_s) = \La_0' + \La_i'$, where
$\La_s$ is a self-interaction term, either bilinear or quadratic in
the fields, such that $\La_0'$ yields a linear equation of motion.
Now $\La_0'$ is diagonalized and $\La_i'$ is treated as a perturbation.

In order to determine what the form of $\La_s$ is, one requires that
the perturbation $\La_i'$ not produce any additional self-energy
effects.  The name ``self-consistent field theory'' reflects the fact
that in this technique $\La_i$ is found by computing a self-energy via
a perturbative expansion in fields that already are subject to that
self-energy, and then requiring that such a perturbative expansion not
yield any additional self-energy effects.

Nambu and Jona-Lasinio proceed to make the ansatz that for
Eq. (\ref{NJL}) the self-interaction term will be of the form $\La_s =
-m \bar{\psi}\psi$.  Then, to first order in the coupling constant
$g$, they proceed to compute the fermion self-energy $\Sigma'(p)$,
using the propagator $S'(p) = i(p\!\!\!/ - m)^{-1}$, which corresponds
to the Lagrangian $\La_0' = \bar{\psi}(i \partial\!\!\!/ - m)\psi$
that incorporates the proposed self-energy term.

The next step is to apply the self-consistency condition using the
Schwinger-Dyson equation for the propagator
\begin{equation}
S'(x-y) = S(x-y) + \int d^4 z \, S(x-z) \Sigma'(0) S'(z-y)~,
\label{Schwinger-Dyson}
\end{equation}
which is represented diagrammatically in Fig. \ref{sde}.
The primes indicate quantities that correspond to a free Lagrangian
$\La_0'$ that incorporates the self-energy term, whereas the unprimed
quantities correspond to the ordinary free Lagrangian $\La_0$.  For
$\Sigma'$ we will use the approximation shown in Fig. \ref{ladder},
valid to first order in the coupling constant $g$.

After Fourier transforming Eq. (\ref{Schwinger-Dyson}) and summing the
left side as a geometric series, we find that the self-consistency
condition may be written, in our approximation, as
\begin{equation}
m = \Sigma'(0) = \frac{g m i}{2\pi^4} \int \frac{d^4 p}{p^2 - m^2 + i
\epsilon} .
\label{NJLselfconsistent}
\end{equation}

If we evaluate the momentum integral by Wick rotation and regularize
its divergence by introducing a Lorentz-invariant energy-momentum
cutoff $ p^2 < \Lambda^2 $ we find
\begin{equation}
\frac{2\pi^2 m}{g \Lambda^2} = m \left[ 1 -
\frac{m^2}{\Lambda^2}\log{\left( \frac{\Lambda^2}{m^2} + 1 \right)}
\right] .
\label{NJLselfconsistent2}
\end{equation}

\begin{figure}[]
\bigskip
\begin{center}


\includegraphics[scale=1.2]{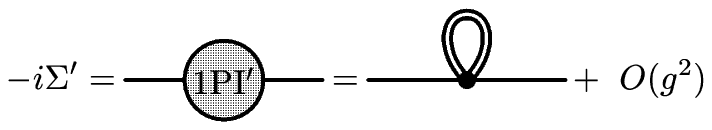}

\end{center}
\caption[Primed self-energy in a theory with a four-fermion interaction]{\small Diagrammatic equation for the primed self-energy.  We
will work to first order in the fermion self-coupling constant $g$.}
\label{ladder}
\end{figure}

This equation will always have the trivial solution $m=0$, which
corresponds to the vanishing of the proposed self-interaction term
$\La_i$.  But if
\begin{equation}
0 < \frac{2\pi^2}{g\Lambda^2} < 1
\label{NJLcondition}
\end{equation}
then there may also be a non-trivial solution to
Eq. (\ref{NJLselfconsistent2}), i.e., a non-zero $m$ for which the
condition of self-consistency is met.  For a rigorous treatment of the
relation between non-trivial solutions of this self-consistent
equation and local extrema in the Wilsonian effective potential for
the corresponding fermion bilinears, see \cite{Aoki} and the
references therein.

In this model (which from now on we shall refer to as NJL), we see
that if the interaction between fermions and antifermions is
attractive ($g > 0$) and strong enough ($\frac{2\pi^2}{g \Lambda^2} <
1$) it might be energetically favorable to form a fermion-antifermion
condensate.  This is reasonable to expect in this case because the
particles have no bare mass and thus the energy cost of producing them
is small.  The resulting condensate would have zero net charge, as
well as zero total momentum and spin.  Therefore it must pair a
left-handed fermion $\psi_L = \frac{1}{2}(1-\gamma^5)\psi$ with the
antiparticle of a right-handed fermion $\psi_R = \frac{1}{2}
(1+\gamma^5)\psi$, and vice versa.  This is the mass-term
self-interaction $\La_i = -m\bar{\psi}\psi = -m (\bar{\psi}_L \psi_R +
\bar{\psi}_R \psi_L)$ that NJL studies.

After QCD became the accepted theory of the strong interactions, the
ideas behind the NJL mechanism remained useful.  The $u$ and $d$
quarks are not massless (nor is $u$-$d$ flavor isospin an exact
symmetry) but their bare masses are believed to be quite small
compared to their effective masses in baryons and mesons, so that the
formation of $\bar{u}u$ and $\bar{d}d$ condensates represents the
spontaneous breaking of an approximate chiral symmetry.  Interpreting
the pions (which are fairly light) as the pseudo-Goldstone bosons
generated by the spontaneous breaking of the approximate $SU(2)_R
\times SU(2)_L$ chiral isospin symmetry down to just $SU(2)$, proved a
fruitful line of thought from the point of view of the phenomenology
of the strong interaction.\footnote{For a treatment of this subject,
including a historical note on the influence of the NJL model in the
development of QCD, see Chap. 19, Sec. 4 in \cite{weinbergII}.}

Condition Eq. (\ref{NJLcondition}) has a natural interpretation if we
think of the interaction in Eq. (\ref{NJL}) as mediated by massive
gauge bosons with zero momentum and coupling $e$.  For it to be
reasonable to neglect boson momentum in the effective theory, the mass
$\mu$ of the bosons should be $\mu > \Lambda$.  If $e^2 < 2\pi^2$ then
$ g = e^2/\mu^2 < 2\pi^2/\Lambda^2$, which violates
Eq. (\ref{NJLcondition}).  Therefore for chiral symmetry breaking to
happen, the coupling $e$ should be quite large, making the
renormalizable theory nonperturbative.  This is acceptable because the
factor of $1/\mu^2$ allows the perturbative calculations we have
carried out in the effective theory Eq. (\ref{NJL}).  This is why the
NJL mechanism is modernly thought of as a model for a phenomenon
of non-perturbative QCD.

\section{An NJL-style argument for breaking LI}

We have reviewed how NJL formulated a model that exhibited a non-zero
VEV for the fermion bilinear $\bar{\psi}\psi$.  The next simplest
fermion bilinear that we might consider is $\PgP$, which was the one
that Bjorken, Kraus, and Tomboulis considered when they discussed the
``dynamical generation of QED.''  This particular fermion bilinear is
especially interesting because it corresponds to the $U(1)$ conserved
current, and also because it is the simplest bilinear with an odd
number of Lorentz tensor indices, so that a non-zero VEV for it would
break not only LI but also charge (C), charge-parity (CP), and
charge-parity-time (CPT) reversal invariance.  C and CP may not be
symmetries of the Lagrangian, as indeed they are not in the standard
model, but by a celebrated result CPT must be an invariance of any
reasonable theory (see \cite{CPT} and references therein).  This
invariance, however, may well be spontaneously broken, as it would be
by any VEV with an odd number of Lorentz indices.

Before proceeding, however, it may be advisable to try to develop some
physical intuition about what would be required for a fermion bilinear
like $\PgP$ to exhibit a VEV.  If we choose a representation of the
gamma matrix algebra and use it to write out $(\PgP)^2$ for an
arbitrary Dirac bispinor $\psi$, we may check that $(\PgP)^2 \geq 0$ for
the choice of mostly negative metric \teq{g^{\mu\nu} =
\mbox{diag}(1,-1,-1,-1)}.  That is, $\PgP$ is time-like.  This has an
intuitive explanation, based on the observation that $\PgP$ is a
conserved fermion-number current density.  Classically a charge
density $\rho$ moving with a velocity $\vv{v}$ will produce a current
$j^\mu = (\rho, \rho \vv{v})$ (in units of $c = 1$).  Therefore the
relativistic requirement that the charge density not move faster than
the speed of light in any frame of reference implies that $j^2 \geq
0$.  Considerations of causality make it natural to expect that
something similar would be true of $\PgP$.

For any time-like Lorentz vector $n^\mu$ it is possible to find a
Lorentz transformation that maps it to a vector $n'^\mu$ with only one
non-vanishing component: $n'^0$.  For a constant current density
$j^\mu$, this means that for $j^\mu$ to be non-zero there must be a
charge density $j^0$, which has a rest frame.  Therefore we only
expect to see a VEV for $\PgP$ if our theory somehow has a vacuum with
a non-zero fermion number density.  The consequent spontaneous breaking
of LI may be seen as the introduction of a preferred reference frame:
the rest frame of the vacuum charge.

In the literature of finite density quantum field theory and of color
superconductivity (see, for instance, \cite{colorsuperconduct1} and
\cite{colorsuperconduct2}), the Lagrangians discussed are explicitly
non-Lorentz-invariant because they contain chemical potential terms of
the form $f\cdot\bar{\psi}\gamma^0\psi$ .  This term appears in
theories whose ground state has a non-zero fermion number because, by
the Pauli exclusion principle, new fermions must be added just above
the Fermi surface, i.e., at energies higher than those already
occupied by the pre-existing fermions, while holes (which can be
thought of as antifermions) should be made by removing fermions at
that Fermi surface.  The result is an energy shift that depends on the
number of fermions already present and which has opposite signs for
fermions and antifermions, as illustrated in Fig. \ref{fig:findensity}.

\begin{figure} []
\bigskip
\begin{center}
\includegraphics[scale=.7]{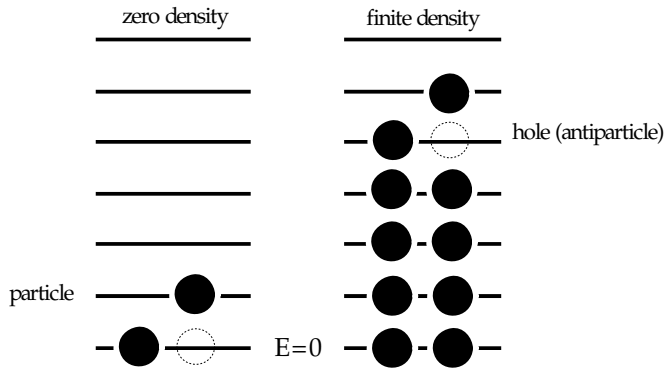}
\end{center}
\caption[Fermion and antifermion energies at finite densities]{\small Fermion and antifermion energies in QFT, at zero density (left) and at finite density (right).  Finite density introduces a chemical potential term \teq{-f \cdot \bar \psi \gamma^0 \psi} into the fermion Lagrangian.}
\label{fig:findensity}
\end{figure}

The physical picture that emerges is now, hopefully, clearer:  A
theory with a VEV for $\PgP$ is one with a condensate that has non-zero
fermion number.  This means that only theories with some form of
attractive interaction between particles with the same sign in fermion
number may be expected to produce such a VEV.  The situation is
closely analogous to BCS superconductivity (\cite{BCS}), in which a
phonon-mediated attractive interaction between electrons allows the
presence of a condensate with non-zero electric charge.  Note that in
the NJL model, the condensate was composed of fermion-antifermion
pairs, and therefore clearly $\langle \bar{\psi}\gamma^0\psi \rangle  = 0$, which
implies $\langle \PgP \rangle = 0$.  It should now be clear why a VEV for
$\PgP$ would break not only LI but also C, CP, and CPT.  This picture also helps to clarify the nature of the Goldstone bosons that we will be invoking as mediators of the electromagnetic interaction:  They are density waves in the background ``Dirac sea,'' whose energy at infinite wavelengths vanishes because they are then proportional to the broken boosts.

There is an easy way to write a theory that will have a VEV for a
$U(1)$ conserved current: to couple a massive photon to such a current
via a purely imaginary charge.  To see this, let us write a Proca
Lagrangian for a massive photon field with an external source:
\begin{equation}
\La = -\frac{1}{4}F_{\mu\nu}^2  + \frac{\mu^2}{2}A^2 - j_\mu A^\mu .
\label{Proca}
\end{equation}

The equation of motion for the photon field is
\begin{equation}
\partial_\mu F^{\mu\nu} = j^\nu - \mu^2 A^\nu .
\label{EOMProca}
\end{equation}

At energy scales well below the photon mass $\mu$, the kinetic term
$-F_{\mu\nu}^2 /4$ may be neglected with respect to the mass term
$\mu^2 A^2 /2$.  We may then integrate out the photon at zero momentum
by solving the equation of motion Eq. (\ref{EOMProca}) for the photon
field $A^\mu$ with its conjugate momenta $F^{\mu\nu}$ set to zero, and
substituting the result back into the Lagrangian in Eq. (\ref{Proca}).
The resulting low-energy effective field theory has the Hamiltonian
\begin{equation}
\mathcal H_{\mathrm{effective}} = \frac{j^2}{2 \mu^2} .
\label{Procaeff}
\end{equation}

Nothing interesting happens if the source is a timelike current
density, since in that case Eq. (\ref{Procaeff}) has its minimum at
$j^\mu=0$.  But if we were to make the charge coupling to the photon
imaginary (e.g., $j^\mu = i e \PgP$ for $e$ real), then $j^2$ is
actually always negative (recall that $(\PgP)^2$ is always positive)
and we get a ``potential'' with the wrong sign, so that the energy can
be made arbitrarily low by decreasing $j^2$.  If we make $j^\mu$
dynamical by adding to the Lagrangian terms corresponding to the field
that sets up the current, we might expect, for certain parameters
in the theory, that the energy be minimized for a finite value of
$j^\mu$.

By making the charge purely imaginary, our effective theory at energy
scales much lower than the photon mass $\mu$ will look similar to
Eq. (\ref{NJL}), except that the four-fermion interaction in the
effective Lagrangian will be $e^2(\PgP)^2/2\mu^2$ (with an overall
positive, rather than a negative, sign).  What this means is that
fermions are attracting fermions and antifermions are attracting
antifermions, rather than what we had in NJL (and in QED):
attraction between a fermion and an antifermion.  Condensation, if it
occurs, will here produce a net fermion number, spontaneously breaking
C, CP, and CPT.\footnote{Dyson argued that a
theory with a long-range attraction between particles of the same fermion number
would be unstable and used this to suggest that perturbative series in
QED would diverge after renormalization of the charge and mass
\cite{Dyson}.  As we will see at the end of this section, the ``photon'' mass $\mu$ will prevent the instability in our case.}

Let us analyze this situation again more rigorously using
self-consistent field theory methods, following Nambu and
Jona-Lasinio.  For this we consider a fermion field with the usual
free Lagrangian $\La_0 = \bar{\psi}(i\partial\!\!\!/ - m_0)\psi$ and
pose as our self-consistent ansatz:
\begin{equation}
\La_s = -(m - m_0)\bar{\psi}\psi - f \bar{\psi}\gamma^0\psi .
\label{Ansatz}
\end{equation}

The corresponding momentum-space propagator for $\La_0' = \La_0 +
\La_s$ is, therefore,
\begin{equation}
S'(k) = i(k\!\!\!/ - f\gamma^0 - m)^{-1} .
\label{primedpropagator}
\end{equation}

Now let us suppose that the interaction term looks like
\begin{equation}
\La_i = \frac{g}{2}(\PgP)^2 .
\label{ajvinteraction}
\end{equation}

To obtain the Feynman rules corresponding to
Eq. (\ref{ajvinteraction}) we note that this is what we would obtain
in massive QED if we replaced the charge $e$ by $ie$ and the usual
photon propagator by $i g^{\mu\nu}/\mu^2$, with $g = e^2/\mu^2$.  Therefore
to compute the self-energy we will rely on the identity represented in
Fig. \ref{contractions}.  (In QED the second diagram on the right-hand
side of Fig. \ref{contractions} would vanish by Furry's theorem, but
in our case the propagator in the loop will have a chemical potential
term that breaks the C invariance on which Furry's theorem depends.)

\begin{figure}[]
\bigskip
\begin{center}
\begin{fmffile}{fmfcontractions}

\begin{equation}
\parbox{25mm}{\begin{fmfgraph}(25,20)
  \fmfleft{i} \fmfright{o}
  \fmf{plain,tension=2/3}{i,z,z,o}
  \fmfdot{z}
\end{fmfgraph}}
= \parbox{25mm}{\begin{fmfgraph}(25,20)
  \fmfleft{i} \fmfright{o}
  \fmf{plain}{i,z1}
  \fmf{photon,tension=1/3}{z1,z2}
  \fmf{plain,right,tension=0}{z2,z1}
  \fmf{plain}{z2,o}
\end{fmfgraph}}
+ \parbox{25mm}{\begin{fmfgraph*}(25,20)
  \fmfipair{i,o,z,zl}
  \fmfiequ{i}{(0,.5h)}
  \fmfiequ{o}{(w,.5h)}
  \fmfiequ{z}{(.5w,.5h)}
  \fmfiequ{zl}{(.5w,.8h)}
  \fmfiv{l=$$}{zl}
  \fmfi{plain}{i--z}
  \fmfi{plain}{z--o}
  \fmfi{photon}{z--zl}
  \fmfi{plain}{fullcircle scaled .4h shifted (.5w,h)}
\end{fmfgraph*}} \nonumber
\end{equation}
\end{fmffile}
\end{center}
\caption[Four-fermion vertex as two massive, zero momentum photon exchanges]{\small The four-fermion vertex in the self-interacting theory
may be seen as the sum of two photon-mediated interactions with a
massive photon that carries zero momentum and is coupled to the
fermion via a purely imaginary charge.}
\label{contractions}
\end{figure}
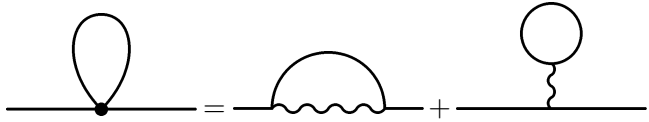

To leading order in $g$, the self-energy is
\begin{equation}
\Sigma(0) =  2ig \int \frac{d^4 k}{(2\pi)^4}\frac{3 (k_0 - f) \gamma^0
+ 3 k_i\gamma^i - 2m}{k_0^2-\vv{k}^2-m^2+f^2-2fk_0 + i\epsilon\sigma}
\label{sigmaunsimplified}
\end{equation}
where $\sigma$ (a function of $|\vv{k}|$, $f$, and $m$)
takes values $\pm 1$ so as to enforce the standard Feynman
prescription for shifting the $k^0$ poles:  positive $k^0$ poles are
shifted down from the real line, while negative poles are shifted up.

At first sight it might appear as if the self-energy in
Eq. (\ref{sigmaunsimplified}) could not be used to argue for the
breaking of LI, because the shift in the integration variable $k
\mapsto k' = (k^0 - f,\vv{k})$ would wipe out $f$ dependence.  This,
however, is not the case, as we will see.  We may carry out the $dk^0$
integration, for which we must find the corresponding poles.  These
are located at
\begin{equation}
k_0 = f \pm \sqrt{\vv{k}^2+m^2} .
\label{k0poles}
\end{equation}

From now on, without loss of generality, we will take $f$ to be
positive.  The contour integral that results from closing the $d^0k$
integral of Eq. (\ref{sigmaunsimplified}) in the complex plane will vanish unless
$f < \sqrt{\vv{k}^2+m^2}$, because otherwise both poles in
Eq. (\ref{k0poles}) will lie on the same side of the imaginary axis.
In light of the Feynman prescription used for the shifting of the
poles away from the real axis, it would then be possible to close the
contour at infinity so that there would be no poles in the interior.
The pole-shifting prescription, through its effect on the $dk^0$
integral, is what introduces an actual $f$ dependence into the
expression for the self-energy.

By the Cauchy integral formula, we have
\begin{eqnarray}
\Sigma(0) &=& \frac{-g}{4\pi^3} \int d^3 k \left[
\frac{3\sqrt{\vv{k}^2+m^2}\gamma^0  +
2m}{2\sqrt{\vv{k}^2+m^2}}\right. \nonumber \\ &&
\left. \phantom{\frac{1}{1}}
\times \theta (\sqrt{\vv{k}^2+m^2} - f) -\frac{3}{2}\gamma^0 \right]~,
\label{sigmasimplified}
\end{eqnarray}
where the second term in the right-hand side subtracts the
contribution from closing the contour out at infinity in the complex
plane (note the branch cut in the logarithm that results from
computing that part of the contour integral explicitly).  We will
introduce the cutoff $\vv{k}^2 < \Lambda^2$ to make the integral in
Eq. (\ref{sigmasimplified}) finite.\footnote{Carrying out the $dk^0$
integration separately from the spatial integral is legitimate and
useful in light of the form of Eq. (\ref{sigmaunsimplified}), which
does not lend itself naturally to Wick rotation.  But the use of a
non-Lorentz-invariant regulator may cause concern that any breaking of
LI we might arrive at could be an artifact of our choice of regulator.
An alternative is to regulate Eq. (\ref{sigmasimplified}) dimensionally by replacing $d^3 k$ with $d^{d-1} k$.
The resulting equations are more complicated and the dependence on the
range of energies where our non-renormalizable theory is valid is
obscured, but the overall argument does not change.  It is also
possible to multiply the integrand in Eq. (\ref{sigmaunsimplified}) by
a cutoff in Minkowski space $\theta(\Lambda^2 + k^2) = \theta
(\Lambda^2 + k_0^2 - \vv{k}^2)$.  For $\vv{k}^2 < \Lambda^2$ we get
the same result as in Eq. (\ref{sigmasimplified}).  For $\vv{k}^2 >
\Lambda^2$ we must impose the condition that $k_0^2 > \vv{k}^2 -
\Lambda^2$.  It should be pointed out that previous work on LI breaking has used 3-momentum
cutoffs in computing self-energies \cite{Soldati}, although in that
case there seems to be a physical interpretation for such a cutoff
which does not apply to the present discussion.  The original work of
Nambu and Jona-Lasinio \cite{NJL} considers cutoffs in Euclidean
4-momentum and in 3-momentum, arriving in both cases at similar
conclusions.}

Note that the Heaviside step function $\theta (\sqrt{\vv{k}^2+m^2} -
f) $ in Eq. (\ref{sigmasimplified}) is always unity if $m > f$, so
that there will be no $f$ dependence at all in
Eq. (\ref{sigmasimplified}) unless $m \leq f$.  Assuming that $m \leq
f$ we have
\begin{eqnarray}
\Sigma(0)&=& \frac{-g}{2\pi^2} \left[-(f^2-m^2)^{3/2}\, \gamma^0
+ m^3\log{(f+\sqrt{f^2-m^2})} \right. \nonumber \\  &&
- m^3 \log{(\Lambda + \sqrt{\Lambda^2+m^2})} \nonumber \\  &&
\left. \vphantom{\frac{0}{1}} + m\Lambda \sqrt{\Lambda^2+m^2} - m
f\sqrt{f^2-m^2} \right] .
\label{sigmaregularized}
\end{eqnarray}

\begin{figure*}[]
\bigskip
\centering \subfigure[]{\includegraphics[scale=0.45]{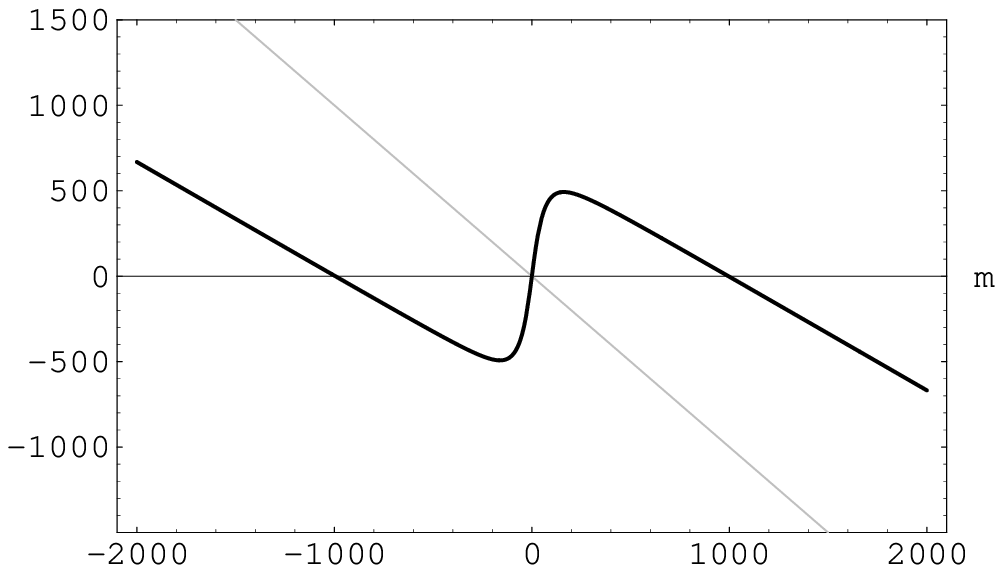}}
\subfigure[]{\includegraphics[scale=0.45]{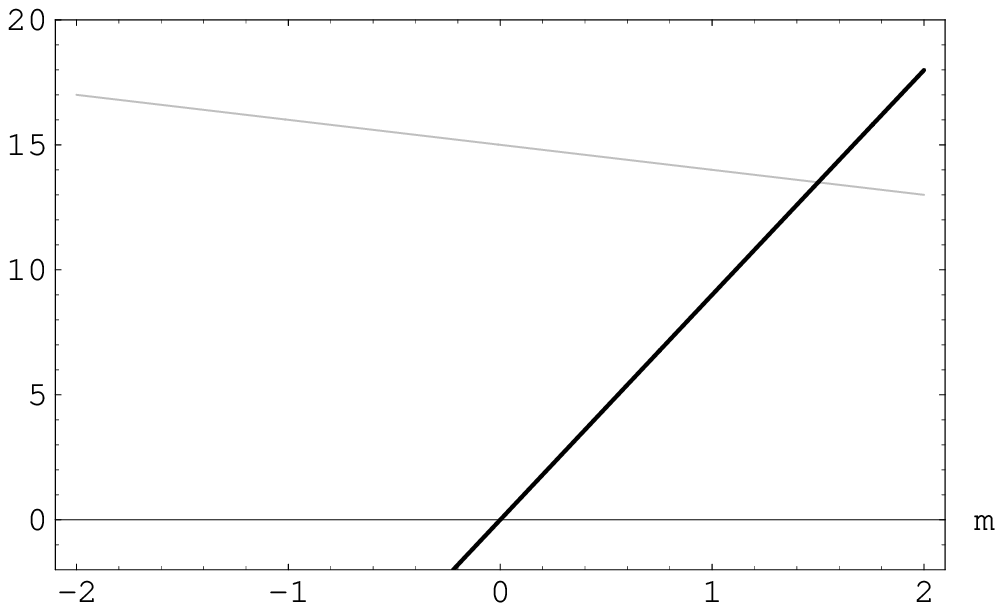}}
\subfigure[]{\includegraphics[scale=0.45]{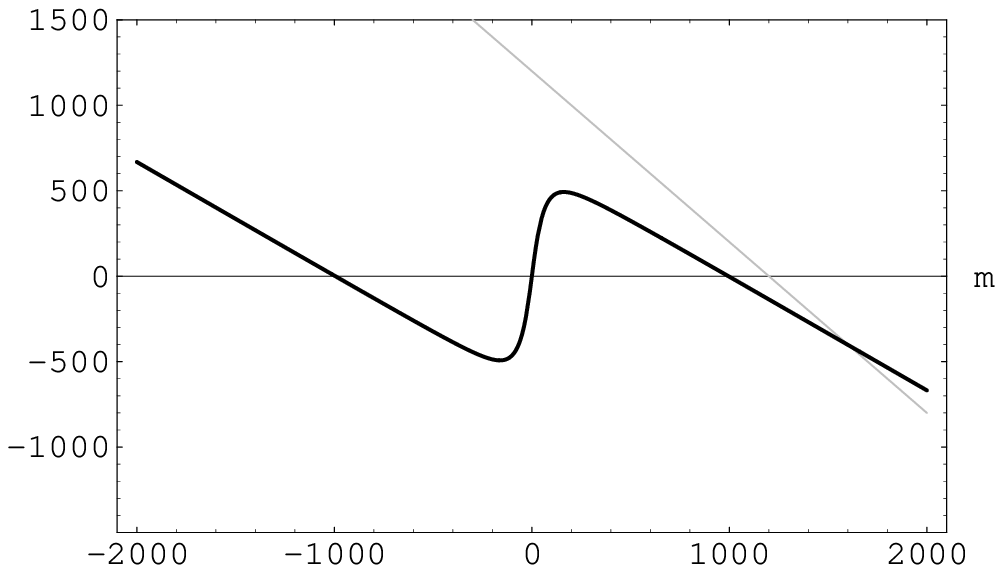}} \\
\subfigure[]{\includegraphics[scale=0.45]{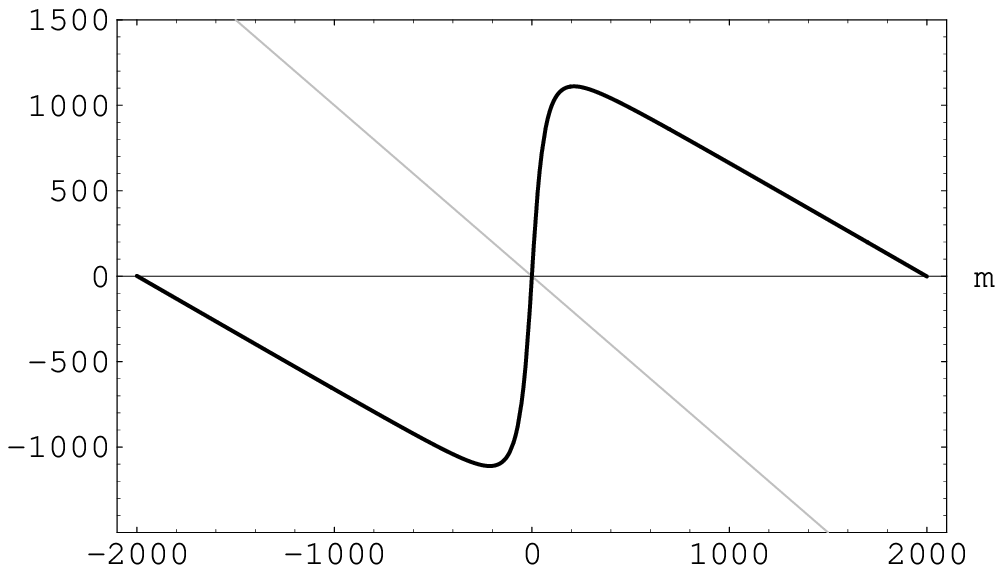}}
\subfigure[]{\includegraphics[scale=0.45]{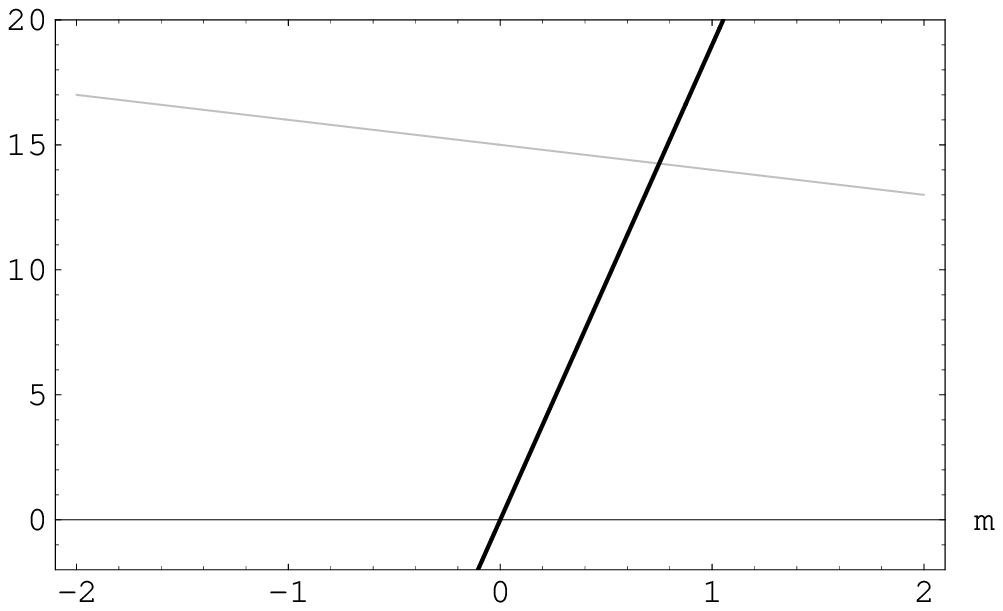}}
\subfigure[]{\includegraphics[scale=0.45]{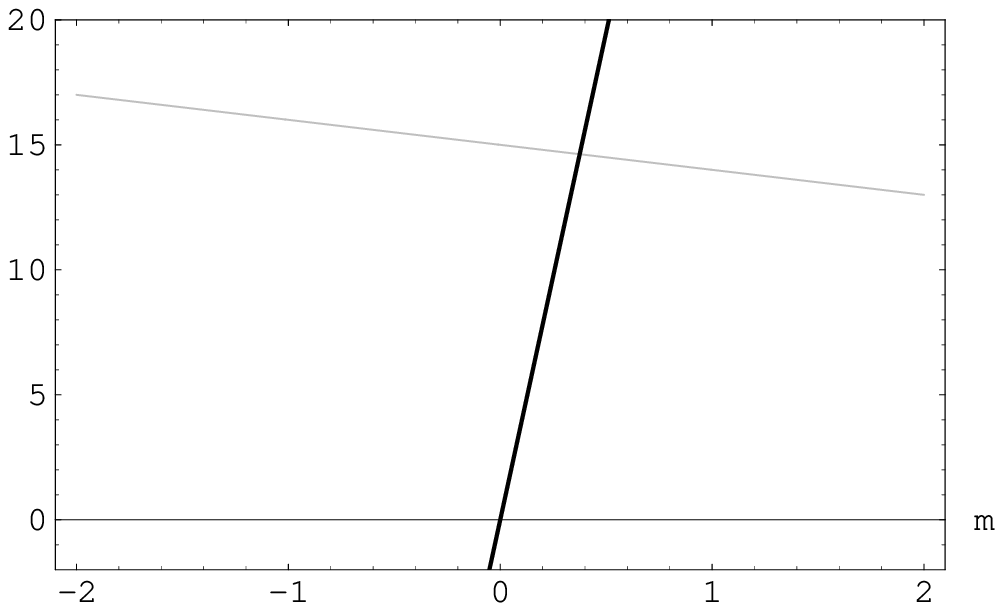}}

\caption[Plots self-consistent equations for the fermion mass $m$]
{\small Plots of the left-hand side (in gray) and right-hand
side (in black) of equation Eq. (\ref{forgraphs}).  Define $\alpha
\equiv \frac{g}{2\pi^2}$.  For each plot the parameters are: (a)
$\Lambda = 100$, $m_0 = 0$, $\alpha = 0.001$.  (b) $\Lambda = 100$,
$m_0 = 15$, $\alpha = 0.001$.  (c) $\Lambda = 100$, $m_0 = 1200$,
$\alpha = 0.001$.  (d) $\Lambda = 100$, $m_0 = 0$, $\alpha = 0.002$.
(e) $\Lambda = 100$, $m_0 = 15$, $\alpha = 0.002$.  (f)  $\Lambda =
200$, $m_0 = 15$, $\alpha = 0.001$.}
\label{sdesoln1}
\end{figure*}

As before, we use the Schwinger-Dyson equation
Eq. (\ref{Schwinger-Dyson}), and after summing up the right-hand side
as a geometric series, we arrive at the self-consistency condition for
our ansatz Eq. (\ref{Ansatz}):
\begin{eqnarray}
m_0 - m -f\gamma^0 &=& -\Sigma(0) \nonumber \\ &=&
\frac{g}{2\pi^2}\left[-(f^2-m^2)^{3/2} \, \gamma^0 \right. \nonumber \\ && 
+ m^3\log{\left( \frac{f+\sqrt{f^2-m^2}}{\Lambda +
\sqrt{\Lambda^2+m^2}}\right)} \nonumber \\ &&
+ m\Lambda \sqrt{\Lambda^2 + m^2} \nonumber \\ &&
- m f\sqrt{f^2-m^2} \left. \phantom{\frac{0}{1}} \right] .
\label{ajvselfconsistent}
\end{eqnarray}

Clearly Eq. (\ref{ajvselfconsistent}) will not admit a non-trivial
solution $f \neq 0$ unless $g$ is positive, which agrees with our
intuition that the theory must exhibit attraction between particles of
the same fermion number.  The self-consistent condition
Eq. (\ref{ajvselfconsistent}) may be separated into two simultaneous
equations:
\begin{equation}
f  =  \frac{g}{2 \pi^2} (f^2 - m^2)^{3/2}
\label{ajvselfconsistentsep1}
\end{equation}
and
\begin{eqnarray}
m_0 - m & = & \frac{g m}{2 \pi^2} \left[m^2
\log{\left(\frac{f+\sqrt{f^2-m^2}}{\Lambda + \sqrt{\Lambda^2 +
m^2}}\right)} \right. \nonumber \\   && +
\left. \vphantom{\frac{0}{1}} \Lambda \sqrt{\Lambda^2+ m^2} - f
\sqrt{f^2-m^2} \right] .
\label{ajvselfconsistentsep2}
\end{eqnarray}
It is important to bear in mind that
Eqs. (\ref{ajvselfconsistentsep1}) and (\ref{ajvselfconsistentsep2})
were written under the assumption that $f \geq m$.  For $f < m$ the
$f$ dependence of the self-energy in Eq. (\ref{sigmaunsimplified})
disappears.  The trivial, Lorentz-invariant solution $f = 0$ to the
self-consistent equations will always be present for any $m$, as
should be the case when spontaneous breaking of a symmetry is observed.

Equation (\ref{ajvselfconsistentsep1}) can be readily solved for $f$
as a function of $m$ (imposing the condition that $f$ be real and
positive), and the resulting $f(m)$ can be substituted into
Eq. (\ref{ajvselfconsistentsep2}) to yield
\begin{eqnarray}
m_0 - m & = & \frac{g m}{2 \pi^2} \left[m^2
\log{\left(\frac{f(m)+\sqrt{f^2(m)-m^2}}{\Lambda + \sqrt{\Lambda^2 +
m^2}}\right)} \right. \nonumber \\  && + \left. \vphantom{\frac{0}{1}}
\Lambda \sqrt{\Lambda^2+ m^2} - f(m) \sqrt{f^2(m)-m^2} \right] .
\label{forgraphs}
\end{eqnarray}

Equation (\ref{forgraphs}) cannot be solved algebraically, but we may
study some of its properties graphically.  In Fig. \ref{sdesoln1} we
have plotted the left-hand side and the right-hand side of
Eq. (\ref{forgraphs}) for various values of the parameters $g$, $m_0$,
and $\Lambda$.  As plot (a) illustrates, $m_0 = 0$ implies $m = 0$,
i.e., we cannot dynamically generate both a chemical potential and a
mass term.  For $m=m_0=0$ we have
\begin{equation}
f = \pi\sqrt{2/g} .
\label{fforzerom}
\end{equation}

Plot (b) in Fig. \ref{sdesoln1} shows a $0 < m_0 \ll \Lambda$ for
which the corresponding $m$ will be significantly less than $m_0$.
Plot (c) in the same figure illustrates that a very large $m_0$ is
needed before $m > m_0$, but such solutions are not physically
meaningful because $m_0$ itself is already well beyond the energy
scale for which our effective theory is supposed to hold.  By
comparing plot (b) to plot (e) we may see the effect of increasing $g$
for a given $m_0$ and $\Lambda$.  A comparison of plots (b) and (f)
should illustrate the effect of increasing $\Lambda$ with the other
parameters fixed.

The plots in Fig. \ref{sdesoln2} illustrate the progression, as the
parameter $\Lambda$ is increased for fixed $\alpha$, from an unstable
theory in which bare masses $m_0$ on the order of $\Lambda$ are mapped
to $m > \Lambda$, to a theory that maps such bare masses to $m <
\Lambda$.  Such an analysis of Eq. (\ref{forgraphs}) reveals that the
condition for this mass stability is
\begin{equation}
0 < \frac{2 \pi^2}{g \Lambda^2} < 1 ~,
\label{physicalcondition}
\end{equation}
which is reminiscent of the condition
Eq. (\ref{NJLcondition}) for chiral symmetry breaking in the NJL model
(except that now the interaction has the opposite sign).  Combining
Eq. (\ref{physicalcondition}) with Eq. (\ref{fforzerom}) (which was
exact for $m_0$ but may serve approximately for $m_0$ small) we arrive
at the requirement
\begin{equation}
0 < f^2 < \Lambda^2 ~,
\label{stability}
\end{equation}
which would surely have to hold if our theory were stable.
Indeed, we may interpret Eq. (\ref{stability}) as saying that if we
pick physically good parameters $g$, $m_0$, and $\Lambda$ we will have
a stable theory with finite chemical potential $f$.  The parameters
for plots (a), (b), (d), (e), and (f) in Fig. \ref{sdesoln1} all give
examples of such stable theories.  As in NJL, the good parameters
involve $g^{-1/2}$ large with respect to $\Lambda$, suggesting that
Eq. (\ref{ajvinteraction}) should be a low-energy approximation to a
non-perturbative interaction of a full renormalizable theory that
allows attraction between particles of the same fermion number sign.

\begin{figure*}[]
\bigskip
\centering \subfigure[]{\includegraphics[scale=0.5]{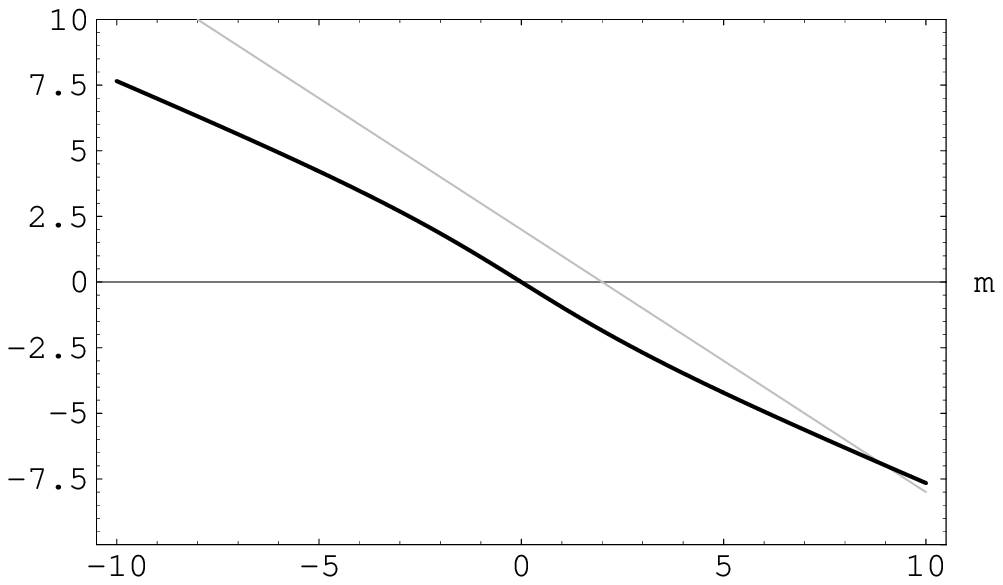}}
\subfigure[]{\includegraphics[scale=0.5]{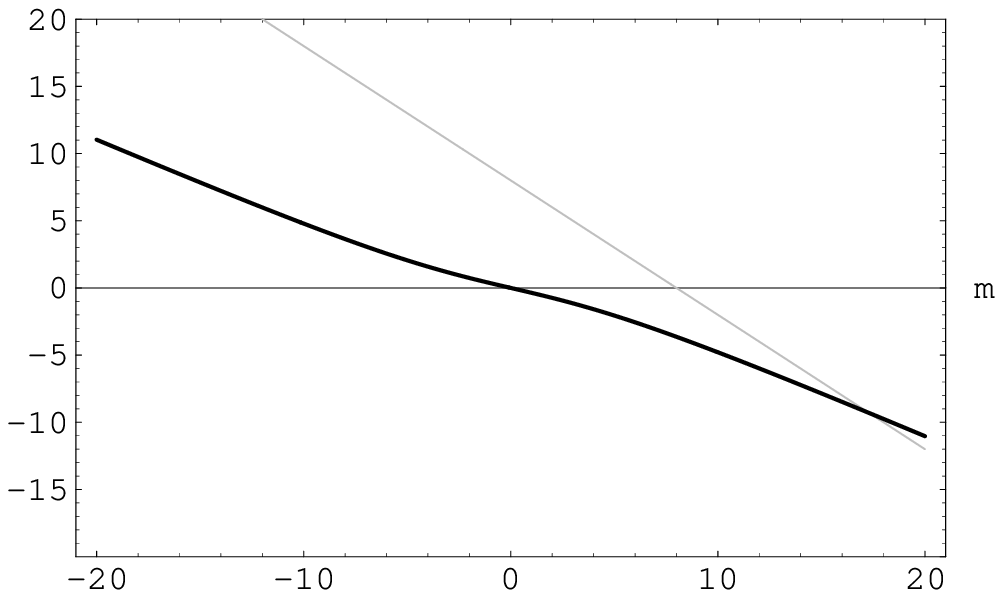}}
\subfigure[]{\includegraphics[scale=0.5]{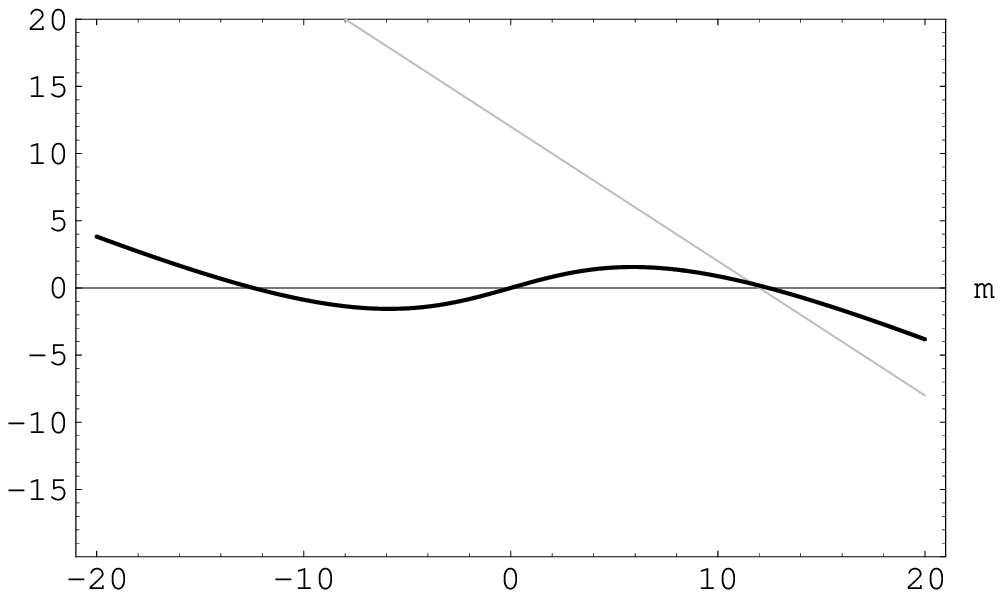}}
\subfigure[]{\includegraphics[scale=0.5]{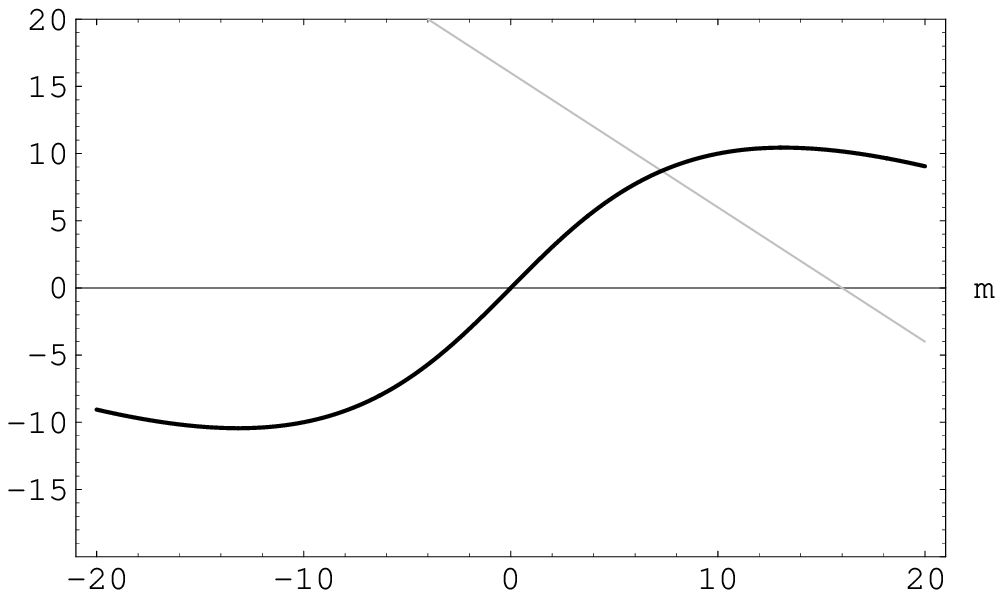}}

\caption[Further plots of self-consistent equations for the fermion mass $m$]{\small Plots of the left-hand side (in gray) and right-hand
side (in black) of equation Eq. (\ref{forgraphs}).  For all of them
$\alpha \equiv \frac{g}{2\pi^2} = 0.01$.  (a) $\Lambda = m_0 = 2$.
(b) $\Lambda = m_0 = 8$.  (c)  $\Lambda = m_0 = 12$.  (d) $\Lambda =
m_0 = 16$.}
\label{sdesoln2}
\end{figure*}

The issue of how the form of the self-consistent equations will depend
on the choice of regulator for the integral in
Eq. (\ref{sigmaunsimplified}) is not an entirely straightforward
matter.  But it seems to be a solid conclusion that, for positive
fermion self-coupling $g$, the solutions to such self-consistent
equations show the presence of LI-breaking vacua.  In the next section
of this paper we offer an alternative approach that strengthens this
conclusion and that sheds further light on the issue of stability.

\section{Consequences for emergent photons}

The theory
\begin{equation}
\La = \bar{\psi}(i\partial\!\!\!/ - m_0)\psi + \frac{g}{2}(\PgP)^2
\label{ajv}
\end{equation}
is equivalent to
\begin{equation}
\La' = \bar{\psi}(i\partial\!\!\!/ -A\!\!\!/- m_0)\psi - \frac{A^2}{2g} .
\label{ajv'}
\end{equation}

Since we argued that Eq. (\ref{ajv}) may spontaneously break LI by
giving a finite $\langle \PgP \rangle$, we conclude that $A^\mu$ in Eq. (\ref{ajv'})
would also have a finite VEV, since, by the algebraic equation of
motion,
\begin{equation}
A^\mu = - g \PgP .
\label{bilineartoA}
\end{equation}

This interpretation agrees with the observation that Eq. (\ref{ajv'})
has a vector boson field whose mass term carries the wrong sign if
$g>0$, indicating that the zero-field state is not a good vacuum.  To
find the correct vacuum for the theory we must carry out the path
integral over the fermion field to obtain the effective action
$\Gamma[A]$, and then minimize that quantity.  Figure \ref{fig:effective} shows
the radiative corrections to $\Gamma[A]$ as a perturbative series, in terms of Feynman diagrams.
The field $A^\mu$ is minimally coupled to $\psi$, so that the computation should proceed as
in QED.  By the Ward identity we do not expect a correction to the
mass term for $A^\mu$, as long as an adequate regulator is used.  But
we do expect to get terms in the effective action that go as $A^4$ and
higher even powers of the auxiliary field.

\begin{figure} []
\bigskip
\begin{center}
\begin{fmffile}{fmfeffective}

\begin{equation}
\Gamma[A] = V(A) +
\parbox{20mm}{\begin{fmfgraph}(20,15)
  \fmfleft{i} \fmfright{o}
  \fmf{photon}{i,z1}
  \fmf{plain,right,tension=1/3}{z1,z2}
  \fmf{plain,right,tension=1/3}{z2,z1}
  \fmf{photon}{z2,o}
\end{fmfgraph}}
+ \parbox{20mm}{\begin{fmfgraph*}(20,15)
  \fmfipair{i,o,z,ii,oo,zz}
  \fmfiequ{i}{(0,0)}
  \fmfiequ{o}{(w,0)}
  \fmfiequ{z}{(.5w,h)}
  \fmfiequ{ii}{(.25w,.25h)}
  \fmfiequ{oo}{(.75w,.25h)}
  \fmfiequ{zz}{(.5w,.7h)}
  \fmfi{photon}{i--ii}
  \fmfi{photon}{o--oo}
  \fmfi{photon}{z--zz}
  \fmfi{plain}{ii--oo--zz--ii}
\end{fmfgraph*}}
+  \parbox{20mm}{\begin{fmfgraph*}(20,15)
  \fmfipair{i,o,x,z,ii,oo,xx,zz}
  \fmfiequ{i}{(0,0)}
  \fmfiequ{o}{(w,0)}
  \fmfiequ{x}{(0,h)}
  \fmfiequ{z}{(w,h)}
  \fmfiequ{ii}{(.25w,.25h)}
  \fmfiequ{oo}{(.75w,.25h)}
  \fmfiequ{xx}{(.25w,.75h)}
  \fmfiequ{zz}{(.75w,.75h)}
  \fmfi{photon}{i--ii}
  \fmfi{photon}{o--oo}
  \fmfi{photon}{x--xx}
  \fmfi{photon}{z--zz}
  \fmfi{plain}{ii--oo--zz--xx--ii}
\end{fmfgraph*}}
+ \, \ldots \nonumber
\end{equation}
\end{fmffile}
\end{center}

\caption[Radiative corrections for the effective potential of the auxiliary field $\A$]{\small Correction of the effective potential of the auxiliary field $\A$ from integrating out the fermion.  The first graph does not contribute by the Ward identity, while the second vanishes by Furry's theorem.}
\label{fig:effective}
\end{figure}

Since we have reason to believe that QED is stable for any value of
the charge $e$, it therefore seems logical to expect that the
effective action for $A^\mu$ in Eq. (\ref{ajv'}) gives it a finite
time-like VEV, which would imply a finite VEV for $\PgP$ in the theory
of Eq. (\ref{ajv}).  We argued in the previous section that $g$ must
be large for the theory described by Eq. (\ref{ajv}) to be stable.  This too seems natural in
light of Eq. (\ref{ajv'}), because a large $g$ makes the $A^2$ term
small, so that the instability created by it may be easily controlled
by the interaction with the fermions, yielding a VEV for $A^\mu$ that
lies within the energy range of the effective theory.  Figure \ref{fig:effectivesketch} schematically represents how the radiative corrections to the effective action give a finite VEV for $\A$.

\begin{figure*}[]
\bigskip
\begin{center}
\includegraphics[scale=.25]{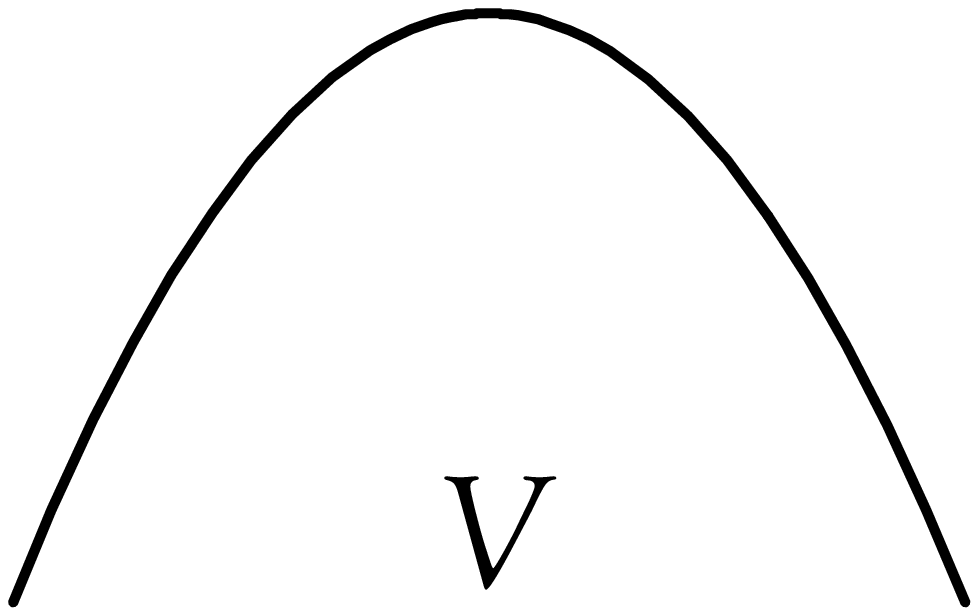} \raise20pt\hbox{\LARGE $~~\rightarrow~~$} \includegraphics[scale=.25]{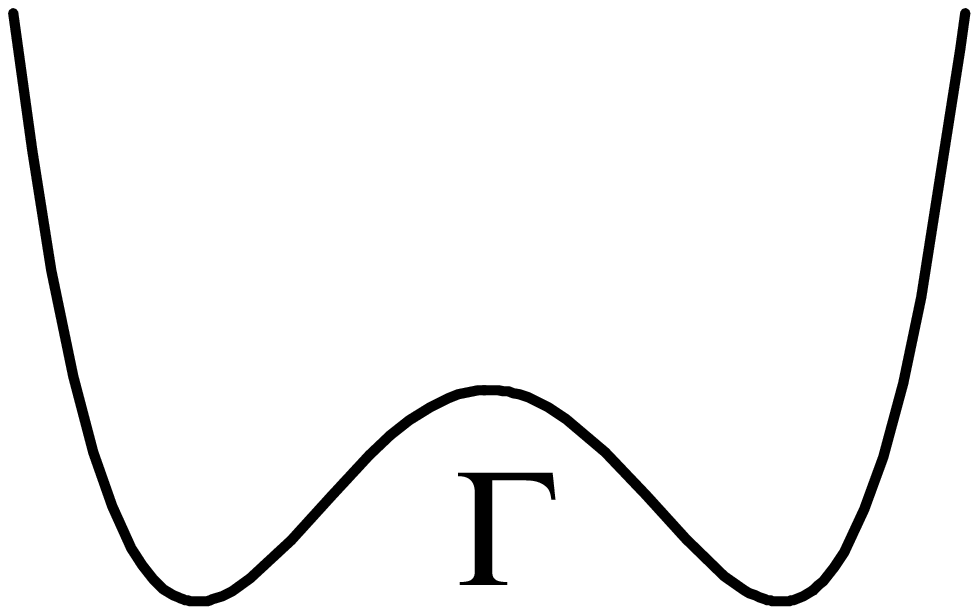}
\end{center}
\caption[Graphic representation of how radiative corrections give a finite $\vev{\A}$]{\small Radiative corrections make the effective potential $\Gamma[A]$ stable and give $\A$ a non-zero VEV.}
\label{fig:effectivesketch}
\end{figure*}

Armed with Eq. (\ref{ajv'}) it would seem possible to carry out the
program proposed by Bjorken, and by Kraus and Tomboulis, in order to
arrive at an approximation of QED in which the photons are composite
Goldstone bosons.  It is conceivable that a complicated theory of
self-interacting fermions, perhaps one with non-standard kinetic
terms, might similarly yield a VEV for $\tensorbilinear$, allowing the
project of dynamically generating linearized gravity to go forward.

It would have been more encouraging if we had been able to obtain a non-zero \teq{\vev \PgP} through a more natural mechanism than invoking an imaginary charge.  Non-abelian gauge theories (such as QCD) exhibit attraction between particles of the same fermion number (and, like abelian theories with imaginary charge, they exhibit anti-screening).  So far, however, attempts to find a non-abelian gauge theory with non-zero \teq{\vev \PgP} have failed, possibly because in such theories the attraction between fermion and antifermion is stronger than the attraction between fermions (see, for instance, \cite{bergesrajagopal}).
 \chapter{Phenomenology of spontaneous Lorentz violation}
 \label{chap:phenoLV}

\begin{epigraphs}
\qitem{What lies behind the Principle of Relativity?  This is a philosophical question, not a scientific one.  You will have your own opinion; here is ours.  We think the Principle of Relativity as used in special relativity rests on one word: emptiness.  Space is empty.}
{Edwin F. Taylor and John A. Wheeler, {\it Spacetime Physics}, Chap. 3}
\end{epigraphs}

\section{Introduction}

Lorentz invariance (LI), the fundamental symmetry of Einstein's
special relativity, states that physical results should not change
after an experiment has been boosted or rotated.  In recent years, and
particularly since the publication of work on the possibility of
spontaneously breaking LI in bosonic string field theory
(\cite{strings}), there has been considerable interest in the prospect
of violating LI.  More recent motivations for work on Lorentz
non-invariance have ranged from the explicit breaking of LI in the
non-commutative geometries that some have proposed as descriptions of
physical space-time (see \cite{Madore} and references therein), and in
certain supersymmetric theories considered by the string community
(\cite{OoguriVafa,frey}), to the possibility of explaining puzzling cosmic
ray measurements by invoking small departures from LI
(\cite{ColemanGlashow}) or modifications to special relativity itself
(\cite{pavlopoulos,MagueijoSmolin,Amelino}).  It has also been suggested that
anomalies in certain chiral gauge theories may be traded for
violations of LI and CPT (\cite{Klinkhamer}).  Extensions of the standard model have been proposed that are meant to capture the low-energy effects of whatever new
high-energy physics (string theory, non-commutative geometry, loop
quantum gravity, etc.) might be introducing violations of LI
(\cite{SME}).

Our own investigation of composite massless mediators in Chapters \ref{chap:massless} and \ref{chap:goldstones} led us to consider the question of how a reasonable QFT might spontaneously break LI through a timelike Lorentz vector VEV \teq{\langle \PgP \rangle \neq 0}.  This breaking of LI can be thought of conceptually as the introduction of a preferred frame: the rest frame of the fermion number density.  If some kind of gauge coupling were added to the theory without
destroying this LI breaking, the fermion number density would also be
a charge density, and the preferred frame would be the rest frame of a
charged background in which all processes are taking place.  This
allows us to make some very general remarks in Section \ref{sect:chargedbackground} on the resulting
LI-violating phenomenology for electrodynamics and on experimental
limits to our non-Lorentz-invariant VEV.  This discussion will be based on work previously published in \cite{LV1}.

Experimental data put very tight constraints on Lorentz violating
operators that involve Standard Model particles \cite{kostelecky}, but
the bounds are more model-independent on Lorentz violation that
appears only in couplings to gravity \cite{nelson,burgess}. One broad
class of Lorentz-breaking gravitational theories are the so-called
vector-tensor theories in which the space-time metric $g^{\mu\nu}$ is
coupled to a vector field $S^\mu$ that does not vanish in the
vacuum. Consideration of such theories dates back to \cite{nordtvedt}
and their potentially observable consequences are extensively
discussed in \cite{Will}. These theories have an unconstrained
vector field coupled to gravity. Theories with a unit constraint on
the vector field were proposed as a means of alleviating the
difficulties that plagued the original unconstrained theories
(\cite{jacobson0}).

The phenomenology of these theories with the unit constraint has
been recently explored. It has been proposed as a toy model for
modifying dispersion relations at high energy (\cite{modifieddisp}). The
spectrum of long-wavelength excitations is discussed in
\cite{jacobson}, where it was found that all polarizations have a
relativistic dispersion relation, but travel with different
velocities. Applications of these theories to cosmology have been
considered in \cite{carroll,lim}. Constraints on these theories are
weak, as for instance, there are no corrections to the Post-Newtonian
parameters $\gamma$ and $\beta$ (\cite{jacobson2}). The status of this
class of theories, also known as ``\ae ther-theories,'' is reviewed in
\cite{unitvtreview}.

In Section \ref{sect:long-range} we will show that the general low-energy effective action at the two-derivative level of the Goldstones of spontaneous Lorentz violation by a timelike vector VEV minimally coupled to gravity corresponds to the vector-tensor theory of gravity with the unit constraint.  This will allow us to place observational constraints of very general validity on this kind of Lorentz violation, from solar system tests of gravity.  This discussion will be based on work previously published in \cite{LV2}.  Finally, in Section \ref{sect:cosmicsolid} we shall discuss the physical meaning of this kind of Lorentz violation and its relation to some other models that have appeared recently in the literature.

\section{Phenomenology of Lorentz violation by a background source}
\label{sect:chargedbackground}

Following up on the idea presented in Chapter \ref{chap:goldstones}, imagine that the fermions of
the universe have some interaction that plays the role of
Eq. (\ref{ajvinteraction}) in giving a VEV to $\PgP$, and that in
addition they have a $U(1)$ gauge coupling (at this stage we have
abandoned the project of producing composite photons).  Then the
$U(1)$ gauge field may interact with a charged background and we would
be breaking LI in electrodynamics by introducing a preferred frame:
the rest frame of the background source.

The possibility of a vacuum that breaks LI and has non-trivial optical
properties has already been investigated in \cite{AndrianovSoldati,
Soldati}. This work, however, deals with significantly more
complicated models, both in terms of the interactions that
spontaneously break LI and of the optical properties of the resulting
vacuum. To obtain a phenomenology for our own simpler proposal, we
consider a free photon Lagrangian of the form
\begin{equation}
\La_0^{\mathrm{photon}} = -\frac{1}{4}F_{\mu\nu}^2 -j_\mu A^\mu ~,
\label{photonlagrangian}
\end{equation}
where $j^\mu = e \langle \PgP \rangle$, thought of as an external source.
The corresponding propagator for the free photon is
\begin{equation}
\langle T\{A^\mu(x)A^\nu(y)\}\rangle = D^{\mu\nu}_F (x-y) + \langle A^\mu(x) \rangle_j
\, \langle A^\nu(y) \rangle_j ~,
\label{photonpropagator}
\end{equation}
where $D^{\mu\nu}(x-y)$ is the connected photon propagator
and $\langle A^\mu(x) \rangle_j$ is the expectation value of $A^\mu$ in the presence
of the external source.

If we take $j^\mu$ constant and naively attempt to calculate the
classical expectation value of $A^\mu$ in the presence of a constant
source by integrating the Green function for electrodynamics, we will
get a volume divergence.  We may attempt to regulate this volume
divergence by introducing a photon mass $\mu$, which gives the result
\begin{equation}
\langle A^\mu(x)\rangle_j = \frac{j^\mu}{\mu^2} .
\label{withphotonmass}
\end{equation}
(It is trivial to check that this is a solution to
$\bert A^\mu - \mu^2 A^\mu = - j^\mu$, the wave equation for the
massive photon field with a source.)  This is not satisfactory because
the disconnected term in Eq. (\ref{photonpropagator}) will be
proportional to $\mu^{-4}$ and Feynman diagrams computed with our
modified photon propagator would produce results that depend strongly
on what we took for a regulator.  In fact the mass is physical and
analogous to the effective photon mass first described by the London
brothers in their theory of the electromagnetic behavior of
superconductors \cite{London}.  (Using the language of particle
physics we may say that, in the presence of a $U(1)$ gauge field, the
VEV $\langle \PgP \rangle$ spontaneously breaks the gauge invariance 
and gives a mass to the boson, as in the Higgs mechanism.)

Photons in a superconductor propagate through a constant
electromagnetic source.  In a simplified picture, we may think of it
as a current density set up by the motion of charge carriers of mass
$m$ and charge $e$, moving with a velocity $\vv u$.  The proper
charge density is $\rho_0$.  The proper velocity of the charge
carriers is $\eta^\mu = (1, \vv u)/\sqrt{1-u^2}$.  The source is
then $j^\mu = \rho_0 \eta^\mu = \rho_0 p^\mu / m$, where $p^\mu$ is
the classical energy momentum of the charge carriers.  We may think of
$m$ and $\rho_0$ as deriving from the solutions to the parameters in a
self-consistent equation such as we had in Eq. (\ref{forgraphs}).

The canonical energy momentum $P^\mu$ of the system is $ P^\mu =
m\eta^\mu + eA^\mu = mj^\mu/\rho_0 + eA^\mu$.  As is discussed in the
superconductivity literature (see, for instance, Chap. 8 in
\cite{Kittel}), the superconducting state has zero canonical
energy momentum, which leads to the London equation
\begin{equation}
j^\mu = -\frac{e\rho_0}{m}A^\mu .
\label{Londonright}
\end{equation}
With this $j^\mu$ inserted into the right-hand side of
$\bert A^\mu = - j^\mu$ (the wave equation for the photon field in
the Lorenz gauge), we find that we have a solution to the wave
equation of a massive $A^\mu$ with no source and a mass $\mu^2 =
e\rho_0/m$:
\begin{equation}
\bert A^\mu - \frac{e \rho_0}{m} A^\mu = 0 .
\label{Londonwaveequation}
\end{equation}

If we solve for $A^\mu$ in Eq. (\ref{Londonright}) and substitute this
back into Eq. (\ref{photonpropagator}), we get that
\begin{equation}
\langle T\{A^\mu(x)A^\nu(y)\} \rangle = D^{\mu\nu}_F(x-y) + \frac{m^2}{e^2 j^2}j^\mu
j^\nu .
\label{photonpropagatorfinal}
\end{equation}

Notice that if $j^\mu (x)$ is not constant, then Fourier
transformation of the second term in Eq. (\ref{photonpropagatorfinal})
will not yield, in Feynman diagram vertices, the usual energy-momentum
conserving delta function.  Therefore, presumed small violations of
energy or momentum conservation in electromagnetic processes could
conceivably be parametrized by the space-time variation of the
background source.\footnote{This line of thought could connect to work
on LI violation from variable couplings as discussed in
\cite{varyingcoupling}.}

With Eq. (\ref{photonpropagatorfinal}) and a rule for external massive
photon legs, one may then go ahead and calculate the amplitude for
various electromagnetic processes with this modified photon
propagator, and parametrize supposed observed violations of LI (see
\cite{phenomenology1,phenomenology2,phenomenology3}) by $j^\mu$.  If
we can make an estimate of the size of the mass $m$ of the
background charges, experimental limits on the photon mass ($< 2
\times 10^{-16}$ eV according to \cite{ReviewPP}) will provide a limit
on the VEV of $\PgP$, in light of Eq. (\ref{Londonright}).

There are other consequences of a VEV $\langle \PgP \rangle \neq 0$ on
which we may speculate.  Such a background may have cosmological effects, a line of
thought that might connect, for instance, with \cite{Fried}.  Also,
it is conceivable that such a VEV might have some relation to the
problem of baryogenesis, since it gives the background finite fermion
number and spontaneously breaks CPT, a violation that can ease the
Sakharov condition of thermodynamical non-equilibrium
\cite{baryogenesis,carrollbar}.

\section{Effective action for the Goldstone bosons of spontaneous Lorentz violation}
\label{sect:lowenergyLV}

Here we begin by considering the general low-energy effective action
for a theory in which Lorentz invariance is spontaneously broken by
the VEV of a Lorentz four-vector $S^\mu$. With an appropriate rescaling, the VEV satisfies \beq \langle
S_\mu S^\mu \rangle = 1 ~, \label{vev} \eeq since we assume the VEV of
$S^{\mu}$ is time-like. The existence of this VEV implies that there
exists a universal rest frame (which we sometimes refer to as the
preferred frame) in which $S^{\mu} =\delta^{\mu}_0$. When the
resulting low-energy effective action is minimally coupled to gravity,
we shall see that it simply becomes the vector-tensor theory with the
unit constraint.

Objects of mass $M_1$ and $M_2$ in a system moving relative to the
preferred-frame can experience a modification to Newton's law of
gravity of the form (\cite{Will,Will2})
\beq U_{\rm Newton} = -G_{\rm N}
\frac{M_1 M_2}{r} \left(1- \frac{\alpha_2}{2} \frac{(\vv w \cdot
\vv r)^2}{r^2}\right)~,
\label{newtonlaw}
\eeq
where $\vv w$ is the
velocity of the system under consideration, such as the solar-system
or Milky Way galaxy, relative to the universal rest frame. The main
purpose of this note is to compute $\alpha_2$ in theories where
Lorentz invariance is spontaneously broken by the VEV of a
four-vector.

The VEV of $S^{\mu}$ spontaneously breaks Lorentz invariance. But as
rotational invariance is preserved in the preferred frame, only the
three boost generators of the Lorentz symmetry are spontaneously
broken. The low-energy fluctuations $ S^{\mu}(x)$ which preserve
Eq. (\ref{vev}) are the Goldstone bosons of this breaking, i.e., those
that satisfy \beq S_{\mu}(x) S^{\mu}(x) = 1 ~. \label{constraint} \eeq
In the preferred-frame the fluctuations can be parameterized as a
local Lorentz transformation \beq S^{\mu}(x) = \Lambda^{\mu}_0(x)=
\frac{1}{\sqrt{1-\vv \phi^2}} \left( \begin{array}{c} 1 \\ \vv \phi \end{array} \right)~, \eeq where $\vv \phi$ is as vector with components $\phi^1$, $\phi^2$, and $\phi^3$.

Under Lorentz transformations $S^{\mu}(x) \rightarrow
\Lambda^{\mu}_{\nu} S^{\nu}(x)$ and the symmetry is realized
non-linearly on the fields $\phi^{i}$. Using this field $S^{\mu}(x)$
we may then couple the Goldstone bosons to Standard Model
fields. Since however, the constraints on Lorentz-violating operators
\footnote{More correctly, operators that appear to be Lorentz
violating when the Goldstone bosons $\phi^{i}$ are set to zero.}
involving Standard Model fields are considerable \cite{kostelecky}, we
instead focus on their couplings to gravity, which are more model-independent because they are always present once the Goldstone bosons
are made dynamical.

The Goldstone bosons are made dynamical by adding in kinetic terms for
them. Since Lorentz invariance is only broken spontaneously, the
action for the kinetic terms should still be invariant under Lorentz
transformations. The only interactions relevant at the two-derivative level and not eliminated by the constraint Eq.
(\ref{constraint}) are\footnote{The other possible term, \teq{\epsilon^{\mu\nu\rho\sigma}\partial_\mu S_\nu \partial_\rho S_\sigma}, is a total derivative.}
\beq {\cal L} = c_1 \partial_{\alpha}S^{\beta}
\partial^{\alpha} S_{\beta} +(c_2+c_3) \partial_{\mu} S^{\mu} \partial
_{\nu} S^{\nu} + c_4 S^{\mu} \partial_{\mu} S^{\alpha} S^{\nu}
\partial_{\nu} S_{\alpha} ~. \eeq
Expanding this action to quadratic
order in $\phi^i$, one finds that the four parameters $c_i$ can be
chosen to avoid the appearance of any ghosts. In particular, we
require $c_1+c_4<0$.\footnote{Notice that in our convention $S^\mu$ is dimensionless and the $c_i$'s have mass dimension two.}

To leading order, the effective action for the Goldstone bosons is:
\beq
\La = \frac{1}{2} \sum_{i=1,2,3} \left[ \left(\partial_\mu \phi^i \right)^2 - \alpha \left( \partial_i \phi^i \right)^2 \right]
\label{eftgoldstones}
\eeq
where \teq{\alpha \equiv (c_2 + c_3)/ c_1}.  By inserting a plane wave ansatz, \teq{\phi^i(x^\mu) \propto \exp \left(i \omega x^0 - i k x^3 \right)}, we see that we have 2 transverse waves, $\phi^1$ and $\phi^2$, with speed \teq{v = \omega/k = 1}, and one longitudinal wave, $\phi^3$, with \teq{v = \sqrt{1+\alpha}}.  Since we've broken LI, massless particles no longer need to travel at light speed.  For $\alpha > 0$, the longitudinal Goldstone boson is superluminal.  We shall return to the issue of superluminality in Section \ref{sect:cosmicsolid}.

This agrees with the result, discussed in \cite{KrausTomboulis} and in Chapter \ref{chap:goldstones}, that spontaneous Lorentz violation gives us not only two transverse Goldstone bosons (which we could identify as emergent photons) but also an extra polarization with an unusual dispersion relation.  In \cite{KrausTomboulis}, where the Lorentz-breaking VEV was imagined to be spacelike, that extra polarization was timelike.  In our case it is a longitudinal polarization because the VEV in Eq. (\ref{vev}) was chosen to be timelike.

\section{The long-range gravitational preferred-frame effect}
\label{sect:long-range}

With gravity present the situation is more subtle. One expects the
gravitons to ``eat'' the Goldstone bosons, producing a more complicated
spectrum \cite{kosteleckygravity,gripaios}. The covariant
generalization of the constraint equation becomes \beq g_{\mu \nu} (x)
S^{\mu}(x) S^{\nu}(x) =1 \label{constraint2} \eeq and in the action
for $S^{\mu}$ we replace $\partial_{\mu} \rightarrow
\nabla_{\mu}$.

Note that there is no Higgs mechanism to give the
graviton a mass.  For a gauge theory we have the covariant derivative \teq{D_\mu = \partial_\mu - i e A_\mu}, so that \teq{(D_\mu \phi)^2} gives a term proportional \teq{\phi^2 A^2}, i.e., a gauge boson mass, when \teq{\vev{\phi} \neq 0}.  For in the case of gravity coupled to a vector field we have
\beq
\nabla_\mu S^\nu = \partial_\mu S^\nu + \Gamma^\nu_{\rho \mu} S^\rho~,
\label{nablaS}
\eeq
with
\beq \Gamma^\nu_{\rho \mu} = \frac{1}{2} \left( \partial_\rho h^\nu_\mu + \partial_\mu h^\nu_\rho - \partial^\nu h_{\rho \mu} \right)
\label{hnabla}
\eeq 
so that there is no way to get a term proportional to \teq{S^2 h^2}.

Compare this the ghost condensate mechanism described in \cite{nima}, where \teq{\La = P(X)} for \teq{X \equiv  g_{\mu\nu} \partial^\mu \phi \partial^\nu \phi}.  If we assume that
\beq
P'(X=c^2_\ast \neq 0) = 0~,
\label{nimapoint}
\eeq
then, in the preferred frame, this implies that
\beq
\langle X \rangle = c^2_\ast = \vev{\dot \phi^2} \neq 0
\label{nimavev}
\eeq
and the $X^2$ term in $P(X)$ gives a graviton mass \teq{\dot \phi^4 h_{00}^2}.  This is different from our case, where we get five massless graviton polarizations with different propagation velocities. 

Going back to our model, we see that local diffeomorphisms can be used to gauge away the three Goldstone
bosons. For under a local diffeomorphism (which preserves the
constraint Eq. (\ref{constraint2})), \beq S^{\prime
\mu}(x^{\prime})=\frac{\partial x^{\prime \mu}}{\partial x^{\nu}}
S^{\nu}(x) \eeq and with $x^{\prime \mu}=x^{\mu } + \epsilon^{\mu}$,
$S^{\mu} \equiv v^{\mu} + \phi^{\mu}$, \beq \phi^{\prime
\mu}(x^{\prime})= \phi^{\mu}(x) +v^{\rho} \partial_{\rho} \epsilon
^{\mu} \eeq from which we can determine $\epsilon^\mu$ to completely
remove $\phi^{\mu}$. Note that in the preferred frame, $\epsilon^{i}$
can be used to remove $\phi^{i}$. In this gauge, the constraint
Eq. (\ref{constraint2}) reduces to \beq
S^0(x)=\left(1-h_{00}(x)/2\right)~.\eeq The residual gauge invariance
left in $\epsilon^0$ can be used to remove $h_{00}$. This is an
inconvenient choice when the sources are static. In a more general
frame with $\langle S^{\mu} \rangle =v^{\mu}$, obtained by a uniform
Lorentz boost from the preferred frame, the constraint
Eq. (\ref{constraint2}) is solved by \beq S^{\mu}(x) = v^{\mu}
\left(1- v^{\rho} v^{\sigma} h_{\rho \sigma}(x) /2 \right)
~. \label{generalconstraint} \eeq

Next we discuss a toy model that provides an example of a more
complete theory, that at low energies reduces to the theory described
above with the vector field satisfying a unit covariant constraint
(\ref{constraint2}).\footnote{For a related example, see
\cite{gripaios}.} Consider the following non-gauge-invariant theory
for a vector boson $A^{\mu}$, \beq {\cal L} =-\frac{1}{2} g_{\mu \nu}
g^{\rho \sigma} \nabla_{\rho} A^{\mu} \nabla_{\sigma} A^{\nu} +
\lambda \left( g_{\mu \nu} A^{\mu} A^{\nu} - v^2 \right)^2 ~. \eeq
Fluctuations about the minimum are given by \beq g_{\mu \nu} =
\eta_{\mu \nu} + h_{\mu \nu} ~~,~~ A^{\mu} = v^{\mu}
+\psi^{\mu}~. \eeq This theory has one massive state $\Phi$ with mass
$M_{\Phi} \propto \lambda^{1/2} v$, which is \beq \Phi= v^{\mu}
\psi_{\mu} + h_{\mu \nu}v^{\mu } v^{\nu}/2~. \eeq In the limit that
$\lambda \rightarrow \infty$ this state decouples from the remaining
massless states. In the preferred frame the only massless states are
$h_{\mu \nu}$, and $\psi^i$. Since we have decoupled the heavy state,
we should expand \beq A^0 = v+\left[ \psi^0 + v h_{00}/2 \right] - v
h_{00}/2 \rightarrow v - v h_{00}/2 ~, \eeq where in the last limit we
have decoupled the heavy state. Note that this parameterization of
$A^0$ is precisely the same parameterization that we had above for
$S^0$. In other words, in the limit that we decouple the only heavy
state in this model, the field $A^{\mu}$ satisfies $g_{\mu \nu} A^{\mu
} A^{\nu}=v^2$, which is the same as the constraint
(\ref{constraint2}) with $A^{\mu} \rightarrow v S^{\mu}$.

In the unitary gauge with $\phi^i=0$, the only massless degrees of
freedom are the gravitons. There are the two helicity modes, which in
the Lorentz-invariant limit correspond to the two spin-2 gravitons,
along with three more helicities that are the Goldstone bosons, for a
total of five. The sixth would-be helicity mode is gauged away by the
remaining residual gauge invariance.

But the model that we started from does have a ghost, since we wrote a
kinetic term for $A^{\mu}$ that does not correspond to the
conventional Maxwell kinetic action. The ghost in the theory is $A^0$,
which in our case is massive. The presence of this ghost means that
this field theory model is not a good high-energy completion for the
low-energy theory involving only $S^{\mu}$ and gravity that we are
considering in this section. We assume that a sensible high energy
completion exists for generic values of the $c_i$'s.

Now we proceed to compute the preferred-frame coefficient $\alpha_2$
appearing in the modification to Newton's law.

The action we consider is \beq S= \int d^4 x \, \sqrt{g} \left( {\cal
 L}_{\rm EH} + {\cal L}_{\rm V} + {\cal L}_{\rm gf} \right)~,
 \label{aether} \eeq with\footnote{The coefficients $c_i$ appearing
 here are related to those appearing in, for example \cite{jacobson},
 by $c_i^{\rm here}=-c^{\rm there}_i/16 \pi G$.} \beq {\cal L}_{\rm
 EH} = -\frac{1}{16 \pi G} R \eeq and \beqa {\cal L}_{V}&=& c_1
 \nabla_{\alpha}S^{\beta} \nabla^{\alpha} S_{\beta} +c_2 \nabla_{\mu}
 S^{\mu} \nabla _{\nu} S^{\nu} + c_3 \nabla _{\mu }S^{\nu} \nabla
 _{\nu} S^{\mu} + c_4 S^{\mu} \nabla_{\mu} S^{\alpha} S^{\nu}
 \nabla_{\nu} S_{\alpha} ~. \eeqa
 This is the most general action involving two derivatives
 acting on $S^{\mu}$ that contributes to the two-point function. Note
 that a coefficient $c_3$ appears, since in curved space-time covariant
 derivatives do not commute. Other terms involving two derivatives
 acting on $S^{\mu}$ may be added to the action, but they are either
 equivalent to a combination of the operators already present (such as
 adding $R_{\mu \nu}S^{\mu}S^{ \nu})$, or they vanish because of the
 constraint Eq. (\ref{constraint2}). We assume generic values for the
 coefficients $c_i$ that in the low energy effective theory give no
 ghosts or gradient instabilities.

As previously discussed, $S^{\mu}$ satisfies the constraint
(\ref{constraint2}). We also assume that it does not directly couple
to Standard Model fields. In the literature, Eq. (\ref{constraint2})
is enforced by introducing a Lagrange multiplier into the action. Here
we enforce the constraint by directly solving for $S^{\mu}$, as given
by Eq. (\ref{generalconstraint}), and then insert that solution back
into the action to obtain an effective action for the metric.

In our approach there is a residual gauge invariance that in the
preferred-frame corresponds to reparameterizations involving
$\epsilon^0$ only. To completely fix the gauge we add the gauge-fixing
term \beq {\cal L}_{\rm gf}= -\frac{\alpha }{2} \left(S^{\rho}
S^{\sigma} S^{\mu} \partial _{\mu} h_{\rho \sigma} \right)^2 ~. \eeq
Neglecting interaction terms, in the preferred frame the gauge-fixing
term reduces to \beq {\cal L}_{\rm gf}= -\frac{\alpha }{2}\left(
\partial _0 h_{00}\right)^2 ~. \eeq Physically, this corresponds in
the $\alpha \rightarrow \infty$ limit to removing all time dependence
in $h_{00}$ without removing the static part, which is the
gravitational potential. This is a convenient gauge in which to
compute when the sources are static.

At the two-derivative level, the only effect in this gauge of the new
operators is to modify the kinetic terms for the graviton. The
dispersion relation for the five helicities will be of the form $E = \beta |\vv k|$, where the velocities
$\beta$ are not the same for all helicities and depend on the
parameters $c_i$ (\cite{jacobson}). This spectrum is different than that
which is found in the ``ghost condensate'' theory, where in addition to
the two massless graviton helicities, there exists a massless scalar
degree of freedom with a non-relativistic dispersion relation $E \propto |\vv k|^2$
(\cite{nima}).

There exists a range for the $c_i$'s in which the theory has no ghosts
and no gradient instabilities (\cite{jacobson}). In particular, for
small $c_i$'s, no gradient instabilities appear if \beq
\frac{c_1+c_2+c_3}{c_1+c_4} > 0 ~~~~\mbox{and}~~~~ \frac{c_1}{c_1+c_4}
> 0~. \eeq The condition for having no ghosts is simply $c_1+c_4 < 0$.

The correction to Newton's law in Eq. (\ref{newtonlaw}) is linear
order in the source. Thus to determine its size we only need to find
the graviton propagator, since the non-linearity of gravity
contributes at higher order in the source. In order to compute that
term we have to specify a coordinate system, of which there are two
natural choices. In the universal rest frame, the sources, such as the
solar system or Milky Way galaxy, will be moving and the computation
is difficult. We instead choose to compute in the rest frame of the
source, which is moving at a speed $|\vv w| \ll 1$ relative to the
universal rest frame. Observers in that frame will observe the Lorentz
breaking VEV $v^{\mu} \simeq (1, -\vv w)$. In the rest frame of the
source, a modified gravitational potential will be
generated. Technically this is because terms in the graviton
propagator $v \cdot k \simeq \vv w \cdot \vv k$ are
non-vanishing. It is natural to assume that dynamical effects align
the universal rest frame where $v^{\mu} = \delta^{\mu}_0$ with the
rest frame of the cosmic microwave background.

In a general coordinate system moving at a constant speed with respect
to the universal frame the Lorentz-breaking VEV will be a general
time-like vector $v^{\mu}$. Thus we need to determine the graviton
propagator for a general time-like constant $v^{\mu}$. Since Lorentz
invariance is spontaneously broken, the numerator of the graviton
propagator is the most general tensor constructed out of the vectors
$v^{\mu}$, $k^{\nu}$ and the tensor $\eta^{\rho \sigma}$. There are 14
such tensors. Writing the action for the gravitons as \beq S=
\frac{1}{2} \int d^4 k \, \tilde{h}^{\alpha \beta}(-k) K_{\alpha \beta
| \sigma \rho}(k) \tilde{h}^{\sigma \rho}(k) \eeq it is a
straightforward exercise to determine the graviton propagator ${\cal
P}$ by solving \beq K_{\alpha \beta | \mu \nu}(k) {\cal P}^{\mu \nu |
\rho \sigma}(k) =\frac{1}{2} \left( \eta_\alpha^\rho \eta_\beta^\sigma
+ \eta_\alpha^\sigma \eta_\beta^\rho \right) ~.\eeq The above set of
conditions leads to 21 linear equations that determine the 14
coefficients of the graviton propagator in terms of the coefficients
$c_i$ and the VEV $v^{\mu}$. Seven equations are redundant and provide
a non-trivial consistency check on our calculation.

Although it is necessary to compute all 14 coefficients in order to
invert the propagator, here we present only those that modify
Newton's law as described previously (assuming stress-tensors are
conserved for sources). These are \beqa {\cal P}^{\alpha \beta | \rho
\sigma} _{\rm Newton} &=& \left\{{\rm A} \eta^{\alpha \beta}
\eta^{\rho \sigma} + {\rm B }(\eta^{\alpha \rho} \eta^{\beta \sigma} +
\eta^{\alpha \sigma} \eta ^{\beta \rho} ) + {\rm C}( v^{\alpha
}v^{\beta} \eta^{\rho \sigma} + v^{\rho } v^{\sigma} \eta^{\alpha
\beta}) \right. \nonumber \\ & & \left. + {\rm D} v^{\alpha }
v^{\beta} v^{\rho} v^{\sigma} + {\rm E} ( v^{\alpha} v^{\rho}
\eta^{\beta \sigma} +v^{\alpha} v^{\sigma} \eta^{\beta \rho} +
v^{\beta} v^{\rho} \eta^{\alpha \sigma} + v^{\beta} v^{\sigma}
\eta^{\alpha \rho}) \right\} ~. \eeqa We find that each of these
coefficients is independent of the gauge parameter $\alpha$. We also
numerically checked that without the presence of the gauge-fixing term
the propagator could not be inverted.

To compute the preferred-frame effect coefficient $\alpha_2$, we only
need to focus on terms in the momentum-space propagator proportional
to $(v \cdot k)^2$. To leading non-trivial order in $G (v\cdot k)^2$
and in the $c_i$'s we obtain, from the linear combination  $A + 2B +
2C + D + 4E$, \beqa g_{00} &=& 1+ 8 \pi G_{\rm N} \int \frac{d^4 k}{(2
\pi)^4} \frac{1}{k^2} \left\{1 -8\pi G_{\rm N} \frac{(v \cdot
k)^2}{k^2} \frac{1}{c_1(c_1+c_2+c_3)}\left[2 c_1^3 + 4c^2_3(c_2 +c_3)
+ \right. \right. \nonumber \\ & & \left. \left. + c^2_1(3 c_2+5 c_3
+3 c_4) + c_1((6 c_3- c_4)(c_3+c_4) +c_2(6 c_3 +c_4)) \right]
\phantom{\frac{1^2}{1}} \!\!\!\!\!\!\!\! \right\} \tilde{T}^{00}(k)~,
\eeqa where in the first line $k$ is a four-vector. Next we use
$v^{\mu}=(1,-\vv w)$, place the source at the origin, substitute
$T^{00} = M \delta ^{(3)}(\vv x)$ or $\tilde{T}^{00}(k) = 2 \pi M
\delta(k^0)$ and use \beq \int \frac{d^3 k}{(2 \pi)^3} \frac{k_i
k_j}{\vv k^4} e^{i \vv k \cdot \vv x} =\frac{1}{8 \pi r} \left[
\delta_{ij} - \frac{x_i x_j}{r^2} \right] \eeq to obtain \beqa
g_{00}&=& 1 - 2 G_{\rm N} \frac{M}{r} \left(1 - \frac{(\vv w \cdot
\vv r)^2}{r^2} \frac{8 \pi G_{\rm N}}{2c_1(c_1+c_2+c_3)}\left[2
c_1^3 + 4c^2_3(c_2 +c_3) +\right. \right. \nonumber \\ & &
\left. \left. + c^2_1(3 c_2+5 c_3 +3 c_4) +c_1((6 c_3 - c_4)(c_3+c_4)
+ c_2(6 c_3 +c_4)) \right] \phantom{\frac{1^2}{1}} \!\!\!\!\!\!\!\!
\right)~, \eeqa where we have only written those terms that give a
correction to Newton's law proportional to $[\vv w\cdot
\vv r/r]^2$. We have also assumed that $|\vv w| \ll 1$ so that
higher powers in $\vv w \cdot \vv r/r$ can be neglected. The
factor of $1/c_1$ in the preferred-frame correction to the metric
arises because when $c_1 \rightarrow 0$ the ``transverse'' components
of $\phi^i$ have no spatial gradient kinetic term. Similarly, the
factor of $1/(c_1+c_2+c_3)$ arises because when $c_1+c_2+c_3
\rightarrow 0$ the ``longitudinal'' component of $\phi^i$ has no
spatial gradient kinetic term. Either of these cases causes a
divergence in the static limit.\footnote{This divergence can of course
be avoided by considering higher-derivative terms in the action for
the Goldstone bosons. This would then give non-relativistic dispersion
relations for these modes, \teq{E \propto |\vv k|^n} for \teq{n>1}, as was the case in \cite{nima}.}

The coefficients $c_i$ redefine Newton's constant measured in solar
system experiments and we find that \beq G_{\rm N } = G \left[1 - 8
\pi G(c_1 + c_4) \right] \simeq \frac{G}{1+ 8 \pi G_{\rm } (c_1 +c_4)}~,
\eeq which agrees with previous computations to linear order in the
$c_i$'s after correcting for the differences in notation
\cite{carroll,unitvtreview}.

The experimental bounds on deviations from Einstein gravity in the
presence of a source are usually expressed as constraints on the
metric perturbation. Since the metric is not gauge-invariant, these
bounds are meaningful only once a gauge is specified. In the
literature, the bounds are typically quoted in harmonic gauge. Here,
the preferred-frame effect is a particular term appearing in the
solution for $h_{00}$. For static sources, the gauge transformation
needed to translate the solution in our gauge to the harmonic gauge is
itself static. But since a static gauge transformation cannot change
$h_{00}$, we may read off the coefficient of the preferred-frame
effect in the gauge that we used.

By inspection \beqa \alpha_2 &=& \frac{8 \pi G_{\rm
N}}{c_1(c_1+c_2+c_3)}\left[2 c_1^3 + 4c^2_3(c_2 +c_3) +c^2_1(3 c_2+5
c_3 +3 c_4) \right. \nonumber \\ & & \left. +c_1( (6c_3 -c_4)(c_3 +
c_4) + c_2(6 c_3 +c_4)) \phantom{1^2} \!\!\!\!\!\! \right]~, \label{alpha2} \eeqa
which can be compared with the experimental bound $|\alpha_2| < 4
\times 10^{-7}$ given in \cite{Will2}.  After \cite{LV2} was published, Foster and Jacobson (\cite{fosterjacobson}) carried out the full computation of $\alpha_2$ in terms of the $c_i$ parameters in the vector-tensor theory with the unit constraint and confirmed that Eq. (\ref{alpha2}) is correct to leading non-trivial order.

The experimental bound on $\alpha_2$ is obtained by considering the torque that the effect in Eq. (\ref{newtonlaw}) would exert on the plane of the orbit of a planet.  For simplicity, let us consider a circular planetary orbit of radius $r$, moving around the sun, whose velocity $\vv w$ with respect to the preferred frame we take to be aligned with the $z$-axis, as shown in Fig. \ref{fig:planet}.  The average torque over one orbit is
\beq
\vv \tau = - \vh x \frac{\alpha_2 G_N M_1 M_2 \vv w^2}{4r} \sin 2 \theta_0~,
\label{torque}
\eeq
where $\theta_0$ is the inclination between the plane of planet's orbit and the axis of $\vv w$.

\begin{figure} []
\bigskip
\begin{center}
\includegraphics[scale=.4]{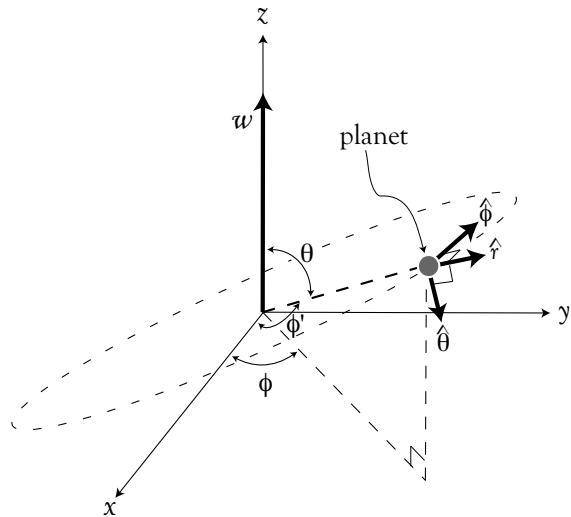}
\end{center}
\caption[Test mass orbiting a source moving with respect to the preferred frame]{\small Diagram of a planet moving in a circular orbit of radius $r$ around the sun (located at the origin), whose velocity with respect to the preferred frame is $\vv w$.  The inclination between the plane of the orbit and the axis of $\vv w$ is $\theta_0$ (the minimum value of the polar angle $\theta$ during the planet's orbit).}
\label{fig:planet}
\end{figure}

This torque would cause the planes or the planets in the solar system to precess at different rates, unless all the orbital planes were perfectly aligned or anti-aligned with the axis of $\vv w$.  If we consider, for instance, the orbits of Earth and Mercury, whose planes are aligned to within a few degrees, and then consider Eq. (\ref{torque}) with 
\bi
	\item $M_1=$ solar mass
	\item $|\vv w| \simeq 10^{-3}$ (the sun's speed with respect to the CMB rest frame)
	\item $\sin 2 \theta_0 \sim O(1)$
\ei
then the fact that Mercury and the Earth have maintained their approximate alignment over the age of the solar system (\teq{\sim 4.5 \times 10^9} years) gives us, roughly, the bound in the literature of \teq{|\alpha_2| \lsim 10^{-7}}.

A considerably stronger constraint on the size of the $c_i$'s can be derived from the fact that a particle moving faster than one of the graviton polarizations would lose energy through gravitational \v{C}erenkov radiation.  In particular, this gravitational \v{C}erenkov radiation would limit the flux of the highest-energy cosmic rays (which are protons moving at nearly the speed of light).  Depending on the exact assumptions regarding the abundance and distribution of cosmic ray sources, the resulting bound can range from \teq{G |c_i| \lsim 10^{-15}} to \teq{G |c_i| \lsim 10^{-31}} (\cite{nelson}).  These limits, however, apply only if the extra graviton polarizations propagate subluminally.  We will have more to say on this issue in Section \ref{sect:cosmicsolid}.

\section{A cosmic solid}
\label{sect:cosmicsolid}

We know that the effect considered in Section \ref{sect:long-range}, the modification of gravity by the presence of a background $S^\mu$ with a rest frame, is present in nature, because the electromagnetic radiation in the CMB has a conserved Poynting 4-vector:
\beq
P^\mu = \frac 1 {8\pi} \left( E^2+B^2, 2 \vv E \times \vv B \right)~.
\label{poynting}
\eeq

This background $P^\mu$ modifies gravity because gravitons can couple to acoustic perturbations in it, as shown in Fig. \ref{fig:CMB}.  This effect is, however, completely negligible, since the characteristic energy scale of the CMB is \teq{T_{\sl{CMB}} \sim 2.7} K, which means that this effect is suppressed by a factor of
\beq
\left( \frac{T_{\sl{CMB}}}{M_{\sl{Pl}}} \right)^2 \sim 10^{-64}~.
\label{suppress}
\eeq
The question remains, however, whether there might be some other background that, unlike the CMB, couples strongly to gravity (and only to gravity, so as to explain why it has not been otherwise detected).  The Goldstone bosons of spontaneous Lorentz violation would correspond to the sound waves in this background, and the modification to gravity comes, as it did in Fig. \ref{fig:CMB} from the mixing of the gravitons with these acoustic modes.

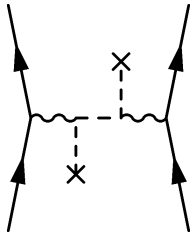
\begin{figure} []
\bigskip
\begin{center}
\begin{fmffile}{fmfCMB}

\begin{fmfgraph*}(30,30)
	\fmfleft{s1,s2}
	\fmfright{t1,t2}
	\fmfforce{(0.4w,0.25h)}{v1}
	\fmfforce{(0.6w,0.75h)}{v2}
	\fmfforce{(0.4w,0.5h)}{i1}
	\fmfforce{(0.6w,0.5h)}{i2}
	\fmfv{decor.shape=cross,decor.size=0.1w}{v1}
	\fmfv{decor.shape=cross,decor.size=0.1w}{v2}
	\fmf{fermion}{s1,g1,s2}
	\fmf{fermion}{t1,g2,t2}
	\fmf{photon}{g1,i1}
	\fmf{photon}{i2,g2}
	\fmf{dashes}{v1,i1,i2,v2}
\end{fmfgraph*}

\end{fmffile}
\end{center}
\caption[Modification to gravity by perturbations in the CMB]{\small Feynman diagram for the (negligible) modification to gravity by the coupling of the graviton to acoustic perturbations in the CMB.}
\label{fig:CMB}
\end{figure}

In \cite{jacobson}, the authors find the propagation velocities of the five graviton polarizations in vector-tensor theories with the unit constraint.  In our language, these are the velocities of the two usual gravitons plus the three acoustic modes in the Lorentz-violating background:

\beq
\begin{array}{l l}
\mbox{\footnotesize 2 transverse traceless metric} & v_{\sl{tt}} = 1/ (1 - c_{13}) \rightarrow 1~, \\
~ & ~ \\
\mbox{\footnotesize 2 transverse Goldstones} & v_{\sl{trv}} = (c_1 - c_1^2/2 +c_3^2/2)/(c_{14})(1-c_{13}) \rightarrow c_1 / (c_{14}) ~, \\
~ & ~ \\
\mbox{\footnotesize 1 longitudinal Goldstone} & v_{\sl{lgt}} = c_{123}(2 - c_{14})/c_{14}(1-c_{13})(2+c_{13}+3c_2)
\\ ~& ~~~~~~~~~~\rightarrow c_{123}/c_{14}~,
\end{array}
\label{soundspeeds}
\eeq 
where \teq{c_{i\ldots k} \equiv G (c_i + \ldots + c_k)} and where the limits correspond to vanishing $c_i$'s.  Since, for general $c_i$'s, there are two distinct sound speeds, one for the longitudinal and one for the transverse modes, our Lorentz-violating background fulfills the canonical definition of a solid.\footnote{This was brought to my attention by Juan Maldacena and Ian Low.}  The transverse sound speed is associated with a shear mode (a deformation which alters the shape but not the volume of a body).  Linear shear waves are absent in a fluid (see, for instance, Chapters III and VI in \cite{LandauElasticity}).

In Section \ref{sect:long-range} we emphasized the difference between our model, which we may now refer to as the ``cosmic solid'' model, and the ``ghost condensate'' of \cite{nima}.  In \cite{nima}, Lorentz invariance is broken by a VEV for a spin-0 vector field \teq{A^\mu = \partial^\mu \phi} with a single degree of freedom, whereas in the cosmic solid model the Lorentz invariance is broken by a spin-1 vector field $A^\mu$ with three degrees of freedom.  Therefore the ghost condensate has a single Goldstone boson, with non-relativistic dispersion relations \teq{E \propto |\vv k|^2}, and it gives the graviton a mass when minimally coupled to it, whereas the cosmic solid has three Goldstone bosons, with dispersion relations \teq{E \propto |\vv k|}, and it does not give the graviton a mass.  It turns out that if the ghost condensate is gauged (i.e., if the ghost condensate field $\phi$ is minimally coupled to a $U(1)$ gauge field $A^\mu$), then the two polarizations of the gauge field provide the two extra degrees of freedom, and the resulting model is equivalent to the cosmic solid (\cite{luty}).  Whether the ghost condensate itself admits a high-energy completion is unresolved (see \cite{markgc, donal}).

It can be seen from Eq. (\ref{soundspeeds}) that the speeds of the Goldstone bosons can be made superluminal without introducing ghosts or other obvious problems in the low-energy effective action.  As pointed out in Section \ref{sect:long-range}, if the Goldstones are required to be subluminal, then $\alpha_2$ no longer gives the strongest constraint on the size of the $c_i$'s because a far more stringent limit applies, from the gravitational \v{C}erenkov radiation of the highest-energy cosmic rays.  Superluminal Goldstone bosons would evade that constraint.  Whether superluminality could result from a reasonable high-energy completion, and whether the initial value problem in the low-energy effective action is well-posed in the presence of superluminal modes, remain open questions.
\chapter{Some considerations on the cosmological constant problem}
\label{chap:cosmological}

\section{Introduction}
\label{sect:cosmointro}

Consider Einstein-Hilbert gravity as an effective theory, containing all the terms compatible with its symmetries:
\beq
S = \int d^4 x \sqrt{-g} \left[{\cal L}_{\sl{matter}} \left(\phi, g_{\mu\nu} \right) - 2 \Lambda + M^2_{\sl{Pl}} R + \cdots \right]~,
\label{einsteinhilbert}
\eeq 
where \teq{M_{\sl{Pl}} \equiv \sqrt{1/8 \pi G}} and
\beq
T_{\mu\nu} \equiv \frac{1}{\sqrt{-g}} \frac{\delta S_{\sl{matter}}}{\delta g^{\sl{$\mu\nu$}}}~.
\label{matterT}
\eeq
The equation of motion for the metric \teq{g^{\mu\nu}} is
\beq
R_{\mu\nu} - \frac{1}{2} g_{\mu\nu} R - \Lambda g_{\mu\nu} = 8 \pi G T_{\mu\nu}~.
\label{fieldeqn}
\eeq 
We would naturally expect that
\beq
\Lambda \sim M^4_{\sl{Pl}} \sim \left(10^{28}~\mbox{eV} \right)^4~.
\label{Lambdalarge}
\eeq

If we let
\beq
g^{\mu\nu} = \eta^{\mu\nu} + \frac 1 {M_{\sl{Pl}}} h^{\mu \nu}~,
\label{ghmetric}
\eeq
where \teq{h^{\mu\nu}} is the graviton field, then the $\Lambda$ term in Eq. (\ref{einsteinhilbert}) gives
\beq
- \sqrt{-g} (2 \Lambda) = -2 \Lambda - \frac{\Lambda}{M_{\sl{Pl}}} h^\mu_{~\mu} - {\cal O}(h^2)~.
\label{tadpole}
\eeq
The first term in the right-hand side of Eq. (\ref{tadpole}) is an irrelevant constant, but the second gives a tadpole diagram for the graviton, as shown in Fig. \ref{fig:tadpole}.  By Eq. (\ref{Lambdalarge}) we would therefore expect this tadpole interaction to be of order \teq{M_{\sl{Pl}}^3}.

Alternatively, one can think of this tadpole diagram, shown in Fig. \ref{fig:tadpole}, as the coupling of a single graviton to the quantum-mechanical vacuum energy.  This corresponds to moving the \teq{\Lambda g_{\mu \nu}} in Eq. (\ref{fieldeqn}) from the left-hand to the right-hand side and thinking of it as the contribution to the matter \teq{T_{\mu \nu}} from the quantum-mechanical vacuum energy.  In quantum field theory, each frequency mode of the free field is a simple harmonic oscillator.   Therefore, each mode has a zero-point energy \teq{E = \omega/2}.  We clearly have to cut off the sum at some scale, but the successes of quantum field theory so far suggest the cut off scale can't be much smaller than $\sim 1$ TeV.

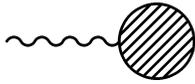
\begin{figure} []
\vskip -.5in
\begin{center}
\begin{fmffile}{fmftadpole}
	\parbox{40mm}{\begin{fmfgraph*}(40,55)
		\fmfleft{g1}
		\fmfright{g2}
		\fmf{photon}{g1,v}
		\fmf{phantom}{v,g2}
		\fmfblob{.25w}{v}
	\end{fmfgraph*}}
\end{fmffile}
\end{center}
\vskip -.6in
\caption[Tadpole diagram corresponding to the cosmological constant term]{\small Feynman diagram of the coupling of a single graviton to the cosmological constant $\Lambda$ in Eq. (\ref{einsteinhilbert}).  The blob may also be thought of as a collection of vacuum-to-vacuum quantum processes.}
\label{fig:tadpole}
\end{figure}

In any case, we get a positive value of $\Lambda$ (the ``cosmological constant'') far, far in excess of what observation allows.  To see qualitatively the effect of large positive $\Lambda$, imagine vacuum energy inside a piston.  Its energy density, $\rho$, is constant. If the piston is pulled out, as shown in Fig. \ref{fig:piston}, the total energy must increase:  \teq{dE = \rho dV}.  By energy conservation, we must have supplied that energy when we pulled on the piston: \teq{dW = F d\ell = p dV = -dE}.  Therefore the piston would {\it resist} being pulled out: Pressure is negative, \teq{p = - \rho}.

\begin{figure} []
\bigskip
\begin{center}
\includegraphics[scale=0.35]{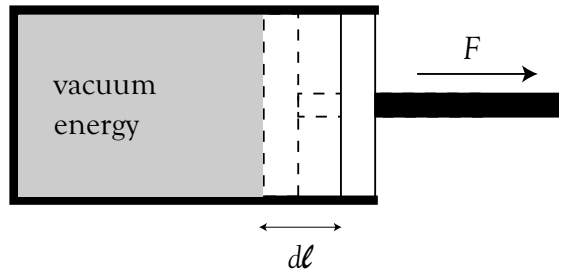}
\end{center}
\caption[Piston filled with vacuum energy]{\small Consider a piston filled with vacuum energy, whose density is constant.  By energy conservation, the piston must resist being pulled out, and therefore the vacuum energy exerts negative pressure.}
\label{fig:piston}
\end{figure}

The Newtonian limit of GR for a test mass on the edge of a uniform sphere of radius $r_0$ gives an acceleration
\beq
g = \frac{4\pi}{3}G \left( \rho + 3p \right)r_0~.
\label{newtoniang}
\eeq
Therefore, the quantum vacuum energy would anti-gravitate.  A value of $\Lambda$ as large as what we would expect on effective theory or quantum mechanical grounds would rip apart the universe, preventing it from developing any structure.  It was long presumed that some unknown symmetry of quantum gravity would forbid the $\Lambda$ term in Eq. (\ref{einsteinhilbert}), thus naturally making the cosmological constant zero.  In Chapter \ref{chap:goldstones} we discussed one such idea: that the graviton was a Goldstone boson of spontaneous Lorentz violation, so that the broken Lorentz invariance protected it from getting any potential at all.

Data on the accelerated expansion of the universe, however, has recently shown that there is a small but non-zero anti-gravitating term.\cite{reiss,perlmutter}  Two possible approaches to this cosmological constant problem that will be of interest to us here are:
\bi
\item to imagine that the true $\Lambda$ is zero, but that the universe contains some other field, coupled only to gravity, which accounts for the accelerated expansion.

\item to imagine that the value of $\Lambda$ varies over some landscape of possible universes, and that we naturally happen to live in one where $\Lambda$ is small enough that structure (and therefore intelligent life) may form.  
\ei
The first line of thought will lead us in Section \ref{sect:nec} to consider whether a cosmological scalar field can have a pressure more negative than $-\rho$.  In Section \ref{sect:ghostdark} we will consider how the ghost condensate of \cite{nima} would behave if it were responsible for the accelerated expansion of the universe.  In Section \ref{sect:anthropic} we will re-examine the second line of thought in light of the proposal that other parameters besides $\Lambda$ vary over the landscape of possible universes.

\section{Gradient instability for scalar models of the dark energy with \teq{w < -1}}
\label{sect:nec}

Matter whose equation of state satisfies $w \equiv p / \rho < -1$
violates a number of conditions, including the weak energy condition,
generally assumed to apply to any reasonable model of physics
\cite{SMC}. However, the  observational data do not exclude the
possibility that the dark energy has $w < -1$
(\cite{Hannestad:2002ur,Melchiorri:2002ux}).  The results reported in
\cite{spergel06} indicate that $-1.26 < w < -0.83$ at 95\%
confidence level. The possibility of $w < -1$ has been explored by
numerous authors (see, for example,
\cite{Caldwell:1999ew}--\cite{Frampton:2002tu}). These models often
contain a field with an unusual kinetic term, which is referred to as
a phantom or ghost field.  In this letter we show that for $w<-1$,
single scalar field models of the dark energy generally have a wrong
sign gradient kinetic term for fluctuations about the homogeneous
background. This result is not dependent on general relativistic
effects and survives in the flat-space limit.  Spatial inhomogeneities
of the dark energy are tightly constrained by observations of the
cosmic microwave background.

In our analysis we will assume a time-dependent but spatially
homogenous scalar background, and show that for $w<-1$ spatial
instabilities inevitably arise.  Consider the low-energy effective
theory of a scalar field coupled to gravity:  \beq
\label{nonminimalcoupling} S = \int d^4x \sqrt{-g}  \left[ M_{Pl}^2
R + P + U \, R + V \, R^{\mu\nu} (\partial_\mu \phi)(\partial_\nu
\phi) + ~\cdots ~\right]~, \eeq where $P$, $U$, and $V$ are functions of
the scalar field $\phi$ and its derivatives.  (Because of the
anti-symmetry of $R^{\mu\nu\rho\sigma}$ in its first two and also in
its last two indices, no non-vanishing invariant can be formed from it
using first derivatives of $\phi$.)  Naively we might expect that the
higher-dimensional couplings of $\phi$ to the Ricci tensor would be
suppressed by powers of the Planck mass $M_{Pl}$, making them
irrelevant for cosmology after the Planck epoch.  However, such terms
are generated by graphs such as that in Figure
\ref{GhostGravitonBlob}.  Writing the metric as $g^{\mu\nu} =
\eta^{\mu\nu} + h^{\mu\nu} / M_{Pl}$, we see that scalar-graviton
interactions in Feynman diagrams are suppressed by the Planck mass,
but when these interactions are reassembled into the Ricci tensor that
suppression is absent.  That is, the higher-dimensional terms in
Eq. (\ref{nonminimalcoupling}) will appear suppressed  only by powers
of the characteristic energy scale of the scalar field, $M$, which may
be much smaller than $M_{Pl}$.

We neglect terms in the action (\ref{nonminimalcoupling}) that
involve higher powers of the Ricci tensor.  The terms we consider are
ones that can generate contributions to the stress-energy tensor
$T_{\mu\nu}$ that are not suppressed by powers of $M_{Pl}$.  Since
$T_{\mu\nu}$ is obtained by varying the action with respect to the
metric, terms with more than one power of $R^{\mu\nu\rho\sigma}$ yield
contributions that are themselves proportional to the Ricci tensor
and therefore vanish in the flat-space limit.

Assuming a spatially homogeneous background, only the time derivatives
of $\phi$ will be non-vanishing in Eq. (\ref{nonminimalcoupling}).  It
may be shown that in the limit $M_{Pl} \rightarrow \infty$, the term
$R^{\mu\nu} \, (\partial_\mu \phi)(\partial_\nu \phi)V$ contributes a
term to the stress-energy tensor, which can be reproduced by an
appropriate change in the function $U$.  Therefore we may restrict
ourselves to $V=0$ and consider the most general $U$ in order to
analyze the flat-space behavior of Eq. (\ref{nonminimalcoupling}).

It is always possible to perform a rescaling of the metric in
Eq. (\ref{nonminimalcoupling}) $g_{\mu\nu} \rightarrow e^{2w}
g_{\mu\nu}$, with $w = \log[1 + U / M_{Pl}^2]$, so that  the $U$ term
in Eq. (\ref{nonminimalcoupling}) disappears, being absorbed into a
redefinition of the $P$ action for the ghost scalar field.  (See, for
example, \cite{Polchinski}.)  The action resulting from this
rescaling, up to terms suppressed by powers of $1 / M_{Pl}$, is then
\beq S = \int d^4 x \sqrt{-g} \left[ M_{Pl}^2 R + P
\right]~. \label{Einsteinframe} \eeq

\begin{figure}[t]
\vskip -.4in
\begin{center}
\begin{fmffile}{fmfGhostGravitonBlob}
	\parbox{40mm}{\begin{fmfgraph*}(40,55)
		\fmfsurround{s1,p1,p2,p3,s2,s3,g1,g2}
		\fmf{photon}{g1,v}
		\fmf{photon}{g2,v}
		\fmf{dashes}{s1,v}
		\fmf{dashes}{s2,v}
		\fmf{dashes}{s3,v}
		\fmf{phantom}{p1,pi1,v}
		\fmf{phantom}{p2,pi2,v}
		\fmf{phantom}{p3,pi3,v}
		\fmfv{decor.shape=circle,decor.filled=full,decor.size=.25thick}{pi1,pi2,pi3}
		\fmfblob{.45w}{v}
	\end{fmfgraph*}}
\end{fmffile}
\caption[Effective coupling of two gravitons to several quanta of the scalar ghost field]{\small The effective couplings of two gravitons to several quanta of the scalar field.  The shaded region represents interactions involving
only scalars.}
\label{GhostGravitonBlob}
\end{center}
\end{figure}
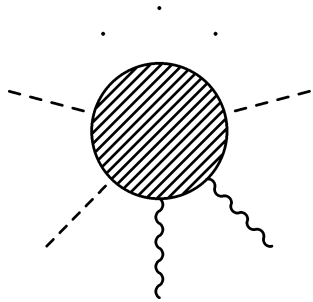

The most general Lorentz-invariant scalar Lagrangian without higher-derivative terms (which we will consider later) is \beq {\cal L} = P(X,\phi)~~, \label{PXX} \eeq where $X =
g^{\mu\nu}\partial_\mu \phi \, \partial_\nu \phi$.  (A potential term
$V$ would be the component of $P(X, \phi)$ that is independent of
$X$.)  Henceforth, $P'(X,\phi)$ will denote differentiation with
respect to $X$.   Since the scalar field $\phi$ is minimally coupled
to gravity in Eq.  (\ref{Einsteinframe}), the stress-energy tensor is
\beq
\label{T} T_{\mu \nu} = - {\cal L} g_{\mu \nu} + 2 P'(X,\phi)
\partial_\mu \phi \partial_\nu \phi~~,\eeq and \beq
\label{WPP}
w = { {P(X,\phi)} \over {T_{00}}} = {{P(X,\phi)} \over {-P(X,\phi) +
2{\dot{\phi}}^2P'(X,\phi)}} = -1 + {{2 {\dot{\phi}}^2P'(X,\phi)} \over
{T_{00}}}~~.  \eeq  For $\phi$ to account for the dark energy, we must
have $T_{00} > 0$.  Then, $w < -1$ requires that $P'(X,\phi) < 0$. Let
$\phi_0=\phi_0(t)$ be a solution to the equations of motion, and
consider the fluctuations about this solution: $\phi = \phi_0 + \pi
(x,t)$. When expanded in $\pi$, the effective Lagrangian will contain
a term \beq
\label{spatial} {\cal L} ~=~  - P'(X,\phi) \vert \nabla \pi \vert^2
~+~ \cdots~, \eeq which implies that for $P'(X,\phi) < 0$ there will
exist field configurations with non-zero spatial gradients that have
lower energy than the homogeneous configuration.\footnote{Here we mean
energy constructed from the Hamiltonian for fluctations about the
background field configuration.}   There is no direct connection
between the sign of $w+1$ and that of the $\dot \pi^2$ term in the
effective Lagrangian.

If $P'(X,\phi)$ is negative, a finite expectation value for the
gradients may be obtained by adding higher powers of $(\nabla \pi)^2$
to the $\pi$ Lagrangian, but this is problematic because it gives rise
to a spatially inhomogeneous ground state for the dark energy and
would lead to inhomogeneities far larger than the limit of $10^{-5}$
imposed by observations of the cosmic microwave background.\footnote{A
condensate of gradients with a preferred magnitude, determined by the
higher-order terms that stabilize Eq. (\ref{spatial}), will
spontaneously break the $O(3)$ rotational symmetry down to $O(2)$.
The homotopy group $\pi_2[O(3)/O(2)]$ is non-trivial, which leads to
the formation of global monopole (hedgehog) configurations.} While a
potential term such as $m^2 \phi^2$ tends to confine the gradients to
regions of size $1/m$, in most models of the dark energy $V'' (\phi)$
must be small enough that these regions are of cosmological size.

In the $w<-1$ case, it is possible, by adding higher-derivative terms
to the Lagrangian, to avoid having finite spatial gradients lower the
energy of the field.  Consider, for example,\
\beq
\label{higherd}
{\cal L} = P(X,\phi) + S(X,\phi)(\bert \phi)^2 \eeq in which case  \beq
\label{higherdrho}
T_{00} = -{\cal L} + 2[P'(X,\phi)\dot{\phi}^2 + S'(X,\phi)
\dot{\phi}^2 (\partial^2 \phi)^2 + 2S(X,\phi)\ddot{\phi}(\partial^2
\phi) -\partial_0(\dot\phi S \partial^2\phi)]~. \eeq  Setting the
spatial gradients of $\phi$ to zero, we have that \beq \dot{\phi}^2
C_{grad} -2 \partial_0(S\ddot\phi)\dot\phi = {(w+1) \over 2} \,
T_{00}~, \eeq where $C_{grad}$ is the coefficient of $-(\nabla \pi)^2$
in the $\pi$ Lagrangian.  If $\partial_0(S\ddot\phi)\dot\phi>0$, then
a model may have both $C_{grad}>0$ and $w<-1$.  But for $w$
significantly less than $-1$, this also requires $\ddot{\phi}^2$ to be
at least of order $M^2\dot\phi^2$, unless $S(X,\phi)$ is made
unnaturally large.  It is not clear how to treat these higher-derivative terms self-consistently beyond perturbation theory, so the
dynamics of such models cannot be analyzed in a straightforward
manner.  The models we consider below have higher powers of first
derivatives, but they satisfy the condition that $\ddot \phi^2 \ll
(\dot \phi M)^2$.

Our analysis shows that $w<-1$ scalar models typically require a wrong
sign $(\nabla \pi)^2$ term in the effective Lagrangian.  Previous
analyses of ghost models (\cite{SMC,MCG}) have focused on the problems
associated with negative energy, particularly with a kinetic term
${\cal L} = -(\partial_{\mu} \phi)^2$ that has the wrong sign for {\it
both} the time and space derivatives.  The classical equations of
motion for such models do not exhibit growing modes of non-zero
spatial gradients, although the energy of the field is unbounded from
below.  Models with $w<-1$ that do not have a wrong sign
time-derivative kinetic term in the effective Lagrangian can result
from a Lorentz-invariant action, as we demonstrate below.   However,
both Lorentz invariance and time translation invariance are
spontaneously broken by a time-dependent condensate.

In \cite{nima} a model with ${\cal L} = P(X)$  was proposed in which
a ghost field has a time-dependent condensate (from now on we take the
Lagrangian to be a function of X only, and therefore invariant under
the shift $\phi \rightarrow \phi + c$).  We use units in which the
dimensional scale $M$ of the model is unity ($M \sim 10^{-3}$ eV if
the ghost comprises the dark energy).  The flat-space equation motion
is \beq \label{EM} \partial_\mu \left[ P^\prime(X) \partial^\mu \phi
\right] = 0~~. \eeq Homogeneous solutions of the equations of motion
with $\dot{\phi}^2 = c^2$ were considered in \cite{nima}.  In
general, the existence of a $\dot{\phi}$ condensate allows for exotic
equations of state, including $w<-1$.  In what follows we let \beq
\label{Pmodel} P(X) = -1 + 2\, (X-1)^2 + \, (X-1)^3~, \eeq which leads
to $w < -1$ with $T_{00} > 0$ when $X < 1$.

The energy density is given by \beq
\label{H}
T_{00} = {\cal H} = {{\partial {\cal L}} \over {\partial \dot{\phi}}}
\dot{\phi} - {\cal L} = 2 \dot{\phi}^2 P^\prime(X) - {\cal L}~~~, \eeq
which is not necessarily minimized by a particular ghost condensate
$\phi = ct$, although it is a solution to the flat-space equations of
motion for any value of $c$.  This is possible  because there is a
conserved charge associated with the shift symmetry, \beq Q = \int
d^3x~ P'(X) \dot{\phi}~~, \eeq so configurations that do not
extremize $T_{00}$ can still be stable.  In fact, the Lagrangian
describing small fluctuations has the correct sign of $\dot{\pi}^2$ if
$P'(X) + 2XP''(X) > 0$. This condition is satisfied in the region
$X<1$ by (\ref{Pmodel}) given above. There is then a local instability
to the formation of gradients, as required by our earlier results.

\section{Time evolution of $w$ for ghost models of the dark energy}
\label{sect:ghostdark}

Ghost models of the dark energy that approach $w=-1$ asymptotically
make potentially interesting predictions for the time evolution of the
equation of state for the dark energy.  In a FRW universe, the
equation of motion for the ghost field is \beq 
\partial_\mu \left[ a^3(t) \, P^\prime(X) \partial^\mu \phi \right] =
0~~, \label{FRWEoM}\eeq where $a(t)$ is the FRW scale factor.  If
there is a value $c_*^2 = \dot\phi^2 = X$ such that $P'(c_*^2)=0$,
then Eq. (\ref{WPP}) implies that $w=-1$ when $X=c_*^2$.  The model
described by Eq. (\ref{Pmodel}) has $c_*^2=1$, and if we apply
Eq. (\ref{FRWEoM}) to it, we see that if we start from $X=c_i^2$ with
$c_i$ close to $c_*$, then we are driven asymptotically towards
$X=c_*^2$ and $w=-1$.

In the model described by Eq. (\ref{Pmodel}), we may be driven towards
$w=-1$ either from above or from below, depending on whether we chose
to start from $c_i^2>1$ or from $c_i^2<1$.  We have argued that $w<-1$
is problematic because of spatial gradient instabilities, so that the
case in which we are driven to $w=-1$ from above is more interesting.

Near the asymptotic value $c_*=1$ we have \beq \dot\pi =
\frac{P'(c_i^2)c_i}{2P''(c_*^2)c_*^2}\left(\frac{a_i}{a}\right)^3~~. \eeq
Thus, in this regime, \beq \label{w} w = -1 -
\frac{4P''(c_*^2)c_*^3\dot\pi}{P(c_*^2)} = -1 - \frac{2P'(c_i^2)
c_*c_i}{P(c_*^2)}\left(\frac{1+z_i}{1+z}\right)^3~.\eeq  Equation
(\ref{w}) offers a prediction for the $w$ parameter of the dark energy
as a function of the redshift $z$, which could be tested by
cosmological data.

In summary, from Eqs. (\ref{WPP}) and (\ref{spatial}) we find that in
single scalar field models of the dark energy with $w<-1$, the kinetic
term for fluctuations about the homogeneous background has a  wrong
sign gradient term.  On the other hand, there is no direct connection
between the sign of the $\dot\pi^2$ kinetic term in the effective
Hamiltonian and the sign of $w+1$.

\section[Anthropic distribution for $\Lambda$ and primordial density perturbations]{Anthropic distribution for cosmological constant and primordial density perturbations}
\label{sect:anthropic}

The anthropic principle has been proposed as a possible solution to
the two cosmological constant problems: why the cosmological constant
$\Lambda$ is orders of magnitude smaller than any theoretical
expectation, and why it is non-zero and comparable today to the energy
density in other forms of matter
(\cite{linde&banks,barrowtipler,weinberg1}).  This anthropic argument,
which predates direct cosmological evidence of the dark energy, is the
only theoretical prediction for a small, non-zero $\Lambda$
(\cite{weinberg1,weinberg2}).  It is based on the observation that the
existence of life capable of measuring $\Lambda$ requires a universe
with cosmological structures such as galaxies or clusters of stars.  A
universe with too large a cosmological constant either doesn't develop
any structure, since perturbations that could lead to clustering have
not gone non-linear before the universe becomes dominated by
$\Lambda$, or else has a very low probability of exhibiting
structure-forming perturbations, because such perturbations would have
to be so large that they would lie in the far tail-end of the cosmic
variance. The existence of the string theory landscape, in which
causally disconnected regions can have different cosmological and
particle physics properties, adds support to the notion of an
anthropic rule for selecting a vacuum.

How well does this principle explain the observed value of $\Lambda$
in our universe? Careful analysis by \cite{weinberg2} finds that 5$\%$
to 12$\%$ of universes would have a cosmological constant smaller than
our own. In everyday experience we encounter events at this level of
confidence,\footnote{For instance, drawing two pairs in a poker hand.}
so as an explanation this is not unreasonable.

If the value of $\Lambda$ is not fixed {\it a priori}, then one might
expect other fundamental parameters to vary between universes as well.
This is the case if one sums over wormhole configurations in the path
integral for quantum gravity (\cite{coleman}), as well as in the string
theory landscape (\cite{stringlandscape,douglas,others,banksdine}). In
\cite{banksdine} it was emphasized that all the parameters of the low
energy theory would vary over the space of vacua (``the landscape'').
Douglas (\cite{douglas}) has initiated a program to quantify the
statistical properties of these vacua, with additional contributions
by others (\cite{others}).

In \cite{aguirre}, Aguirre stressed that life might be possible in
universes for which some of the cosmological parameters are orders of
magnitude different from those of our own universe.  The point is that
large changes in one parameter can be compensated by changes in
another in such a way that life remains possible.  Anthropic
predictions for a {\it particular} parameter value will therefore be
weakened if other parameters are allowed to vary between
universes. One cosmological parameter that may significantly affect
the anthropic argument is $Q$, the standard deviation of the amplitude
of primordial cosmological density perturbations.  Rees in \cite{rees}
and Tegmark and Rees in \cite{tegmark} have pointed out that if the
anthropic argument is applied to universes where $Q$ is not fixed but
randomly distributed, then our own universe becomes less likely
because universes with both $\Lambda$ and $Q$ larger than our own are
anthropically allowed. The purpose of the work in this section is to quantify this
expectation within a broad class of inflationary models.  Restrictions
on the {\it a prori} probability distribution for $Q$ necessary for
obtaining a successful anthropic prediction for $\Lambda$, were
considered in \cite{garrigaliviovilenkin,garrigavilenkin2}.

In our analysis we let both $\Lambda$ and $Q$ vary between universes
and then quantify the anthropic likelihood of a positive cosmological
constant less than or equal to that observed in our own universe.  We
offer a class of toy inflationary models that allow us to restrict the
{\it a priori} probability distribution for $Q$, making only modest
assumptions about the behavior of the {\it a priori} distribution for
the parameter of the inflaton potential in the anthropically allowed
range. Cosmological and particle physics parameters other than
$\Lambda$ and $Q$ are held fixed as initial conditions at
recombination.  We provisionally adopt Tegmark and Rees's anthropic
bound on $Q$: a factor of 10 above and below the value measured in our
universe.  Even though this interval is small, we find that the
likelihood that our universe has a typical cosmological constant is
drastically reduced. The likelihood tends to decrease further if
larger intervals are considered.

Weinberg determined in \cite{weinberg1} that, in order for an
overdense region to go non-linear before the energy density of the
universe becomes dominated by $\Lambda$, the value of the overdensity
$\delta \equiv \delta \rho /\rho$ must satisfy \beq \delta >
\left({729 \Lambda \over 500 \bar{\rho}}\right)^{1/3}
~. \label{deltamin} \eeq In a matter-dominated universe this relation
has no explicit time dependence. Here $\bar{\rho}$ is the energy
density in non-relativistic matter.  Perturbations not satisfying the
bound cease to grow once the universe becomes dominated by the
cosmological constant.  For a fixed amplitude of perturbations, this
observation provides an upper bound on the cosmological constant
compatible with the formation of structure. Throughout our analysis we
assume that at recombination $\Lambda \ll \bar{\rho}$.

To quantify whether our universe is a typical, anthropically allowed
universe, additional assumptions about the distribution of
cosmological parameters and the spectrum of density perturbations
across the ensemble of universes are needed.

A given slow-roll inflationary model with reheating leads to a
Friedman-Robertson-Walker universe with a (late-time) cosmological
constant $\Lambda$ and a spectrum of perturbations that is
approximately scale-invariant and Gaussian with a variance \beq Q^2
\equiv \langle \tilde{\delta}^2 \rangle_{\sl{HC}} ~. \eeq The
expectation value is computed using the ground state in the
inflationary era and perturbations are evaluated at
horizon-crossing. The variance is fixed by the parameters of the
inflationary model together with some initial conditions.  Typically,
for single-field $\phi$ slow-roll inflationary models, \beq Q^2 \sim
\left. { H^4 \over \dot{\phi}^2 }\right\vert _{\sl{HC}}~.  \eeq This
leads to spatially separated over- or underdense regions with an
amplitude $\delta$ that for a scale-invariant spectrum are distributed
(at recombination) according to \beq \label{N} {\cal N}(\sigma,\delta)
= \sqrt{2 \over \pi} {1 \over \sigma} e^{- \delta^2/2 \sigma^2 } ~.
\eeq (The linear relation between $Q$ and the filtered $\sigma$ in Eq.
(\ref{N}) is discussed below.)

By Bayes's theorem, the probability for an anthropically allowed
universe (i.e., the probability that the cosmological parameters should
take certain values, given that life has evolved to measure them) is
proportional to the product of the {\it a priori} probability
distribution $P$ for the cosmological parameters, times the
probability that intelligent life would evolve given that choice of
parameter values.  Following \cite{weinberg2}, we estimate that second
factor as being proportional to the mean fraction ${\cal
F}(\sigma,\Lambda)$ of matter that collapses into galaxies. The latter
is obtained in a universe with cosmological parameters $\Lambda$ and
$\sigma$ by spatially averaging over all over- or underdense regions,
so that (\cite{weinberg2}) \beq {\cal F}(\sigma, \Lambda)=
\int^{\infty}_{\delta_{\sl{min}}} d \delta \, N(\sigma, \delta) {\cal
F}(\delta,\Lambda) ~.  \eeq The lower limit of integration is provided
by the anthropic bound of Eq.  (\ref{deltamin}), which gives
$\delta_{\sl{min}} \equiv (729 \Lambda /500 \bar{\rho})^{1/3} $.  The
anthropic probability distribution is

\beq \label{calP} {\cal P}(\sigma,\Lambda) = P(\Lambda,\sigma) {\cal
F}(\sigma, \Lambda)d\Lambda \,d\sigma ~.  \eeq

Computing the mean fraction of matter collapsed into structures
requires a model for the growth and collapse of inhomogeneities.  The
Gunn-Gott model (\cite{peebles,gunngott}) describes the growth and
collapse of an overdense spherical region surrounded by a compensating
underdense shell. The weighting function ${\cal F}(\delta,\Lambda)$
gives the fraction of mass in the inhomogeneous region of density
contrast $\delta$ that eventually collapses (and then forms
galaxies). To a good approximation it is given by (\cite{weinberg2})
\beq {\cal F}(\delta,\Lambda) =\delta {1 \over \delta +
\delta_{\sl{min}}}~. \eeq Additional model dependence occurs in the
introduction of the parameter $s$ given by the ratio of the volume of
overdense sphere to the volume of the underdense shell surrounding
the sphere. We will set $s=1$ throughout.

Since the anthropically allowed values for $\Lambda$ are so much
smaller than any other mass scale in particle physics, and since we
assume that $\Lambda=0$ is not a special point in the landscape, we
follow \cite{vilenkin,weinberg2} in using the approximation
$P(\Lambda) \simeq P(\Lambda = 0)$ for $\Lambda$ within the
anthropically allowed window.\footnote{Garriga and Vilenkin point to
examples of quintessence models in which the approximation
\mbox{$P(\Lambda)\simeq P(\Lambda=0)$} in the anthropically allowed
range is not valid \cite{garrigavilenkin1}.}  The requirement that the
universe not recollapse before intelligent life has had time to evolve
anthropically rules out large negative $\Lambda$
(\cite{barrowtipler,weinberg3}). We will assume that the anthropic
cutoff for negative $\Lambda$ is close enough to $\Lambda = 0$ that
all $\Lambda < 0$ may be ignored in our calculations.

As an example of a concrete model for the variation in $Q$ between
universes, we consider inflaton potentials of the form (see, for
example, \cite{kolbturner}) \beq \label{inflatonV} V= \Lambda +
\lambda \phi^{2p}~, \eeq where $p$ is a positive
integer.\footnote{Recent analysis of astronomical data disfavors the
$\lambda \phi^4$ inflationary model (\cite{SDSS}), but for generality we
will consider an arbitrary $p$ in Eq.  (\ref{inflatonV}).} We assume
there are additional couplings that provide an efficient reheating
mechanism, but are unimportant for the evolution of $\phi$ during the
inflationary epoch.  The standard deviation of the amplitude of
perturbations gives \beq Q = A \sqrt{\lambda} \,
\frac{\phi^{p+1}_{\sl{HC}}}{M^3_{\sl{Pl}}}~, \label{Qlambda} \eeq
where $A$ is a constant, and $\phi_{\sl{HC}}$ is the value of the
field when the mode of wave number $k$ leaves the horizon.  This
$\phi_{\sl{HC}}$ has logarithmic dependence on $\lambda$ and $k$,
which we neglect.  Randomness in the initial value for $\phi$ affects
only those modes that are (exponentially) well outside our
horizon. Throughout this section, we will set the spectral index to 1
and ignore its running. Equation (\ref{Qlambda}) then gives $\lambda
\propto Q^2$.

Next, suppose that the fundamental parameters of the Lagrangrian are
not fixed, but vary between universes, as might be expected if one
sums over wormhole configurations in the path integral for quantum
gravity (\cite{coleman}) or in the string theory landscape
(\cite{stringlandscape,douglas,banksdine,others}).  To obtain the
correct normalization for the density perturbations observed in our
universe, the self-coupling must be extremely small. As the standard
deviation $Q$ will be allowed to vary by an order of magnitude around
$10^{-5}$, for this model the self-coupling in alternate universes
will be very small as well.

We may then perform an expansion about $\lambda = 0$ for the {\it a
priori} probability distribution of $\lambda$.  The smallness of
$\lambda$ suggests that we may keep only the leading term in that
expansion.  If the {\it a priori} probability distribution extends to
negative values of $\lambda$ (which are anthropically excluded due to
the instability of the resulting action for $\phi$), we expect it to
be smooth near $\lambda=0$, and the leading term in the power series
expansion to be zeroth order in $\lambda$ (i.e., a constant).
Therefore we expect a flat {\it a priori} probability distribution for
$\lambda$. The {\it a priori} probability distribution for $Q $ is
then \beq P(Q) \propto {d \lambda \over d Q} \sim Q~, \eeq where the
normalization constant is determined by the range of integration in
$Q$. Note that this distribution favors large $Q$.  On the other hand,
if the {\it a priori} probability distribution for the coupling
$\lambda$ only has support for $\lambda > 0$ then $\lambda=0$ is a
special point and we cannot argue that $P(Q) \propto Q$.  However,
since the anthropically allowed values of $\lambda$ are very small,
the {\it a priori} distribution for $\lambda$ should be dominated, in
the anthropically allowed window, by a leading term such as
$P(\lambda) \sim \lambda^q$.  Normalizability requires $q>-1$.  Using
$\lambda \propto Q^2$, this gives $P(Q)\sim Q^{2q+1}$.

Before proceeding, it is convenient to transform to the new variables:
\beq y \equiv {\Lambda \over \rho_\ast} ~~;~~ \hat{\sigma} \equiv
\sigma \left( { \bar{\rho} \over \rho_{\ast}} \right)^{1 \over 3}
~. \eeq Here $\bar{\rho}$ is the energy density in non-relativistic
matter at recombination, which we take to be fixed in all universes,
and $\rho_\ast$ is the value for the present-day energy density of
non-relativistic matter in our own universe. For a matter-dominated
universe $\hat{\sigma}$ is time-independent, whereas $y$ is constant
for any era.  Here and throughout this section, a subscript $\ast$
denotes the value that is observed in our universe for the
corresponding quantity.  The only quantities whose variation from
universe to universe we will consider are $y$ and $\hat{\sigma}$.

In terms of these variables and following \cite{weinberg2}, the
probability distribution of Eq.  (\ref{calP}) is found to be \beq
{\cal P} = N d\hat{\sigma} dy \, P(\hat{\sigma}) \int
^{\infty}_{\beta} dx {e^{-x} \over \beta^{1/2} + x^{1/2}}
\label{pdffinal}~, \eeq where \beq \beta \equiv {1 \over 2
\hat{\sigma}^2} \left({729 y \over 500}\right)^{2 /3}~, \eeq and $N$
is the normalization constant.

Notice that, since $x \geq \beta$, large $\beta$ implies that ${\cal
P} \sim e^{-\beta} \ll 1$.  For a fixed $\hat{\sigma}$, large $y$
implies large $\beta$.  Thus, for fixed $\hat{\sigma}$, large
cosmological constants are anthropically disfavored.  But if
$\hat{\sigma}$ is allowed to increase, then $\beta \sim {\cal O}(1)$
may be maintained at larger $y$.  Garriga and Vilenkin have pointed
out that the distribution in Eq.  (\ref{pdffinal}) may be rewritten
using the change of variables $(\hat \sigma, y) \mapsto (\hat \sigma,
\beta)$ (\cite{garrigavilenkin2}).  The Jacobian for that transformation
is a function only of $\hat{\sigma}$.  Equation (\ref{pdffinal}) then
factorizes into two parts: one depending only on $\hat{\sigma}$, the
other only on $\beta$.  Integration over $\hat{\sigma}$ produces an
overall multiplicative factor that cancels out after normalization, so
that any choice of $P(\hat{\sigma})$ will give the same distribution
for the dimensionless parameter $\beta$.  In that sense, even in a
scenario where $\hat{\sigma}$ is randomly distributed, the computation
in \cite{weinberg2} may be seen as an anthropic prediction for
$\beta$.\footnote{We thank Garriga and Vilenkin for explaining this
point to us.} The measured value of $\beta$ is, indeed, typical of
anthropically allowed universes, but an anthropic explanation for
$\beta$ alone does not address the problem of why both $\Lambda$ and
$Q$ should be so small in our universe.

Implementing the anthropic principle requires making an assumption
about the minimum mass of ``stuff'' collapsed into stars, galaxies,
or clusters of galaxies that is needed for the formation of life.  It
is more convenient to express the minimum mass $M_{\sl{min}}$ in terms
of a comoving scale $R$: $M_{\sl{min}}= 4 \pi \bar{\rho} a_{eq}^3
R^3/3 $ (by convention $a=1$ today, so $R$ is a physical scale). We do
not know the precise value of $R$.  A better understanding of biology
would in principle determine its value, which should only depend on
chemistry, the fraction of matter in the form of baryons, and Newton's
constant.  In our analysis these are all fixed initial conditions at
recombination. In particular, we would not expect $M_{\sl{min}}$ to
depend on $\Lambda$ or $Q$.\footnote{Note, however, that requiring
life to last for billions of years (long enough for it to develop
intelligence and the ability to do astronomy) might place bounds on
$Q$.  See \cite{tegmark}.} Therefore, even though the relation between
$M_{\sl{min}}$ and $R$ depends on present-day cosmological parameters,
the value of this threshold will be constant between universes because
it depends only on parameters that we are treating as fixed initial
conditions.  Thus, in computing the probability distribution over
universes, we will fix $R$.  Since we don't know what is the correct
anthropic value for $R$, we will present our results for both $R$=1
and 2 Mpc.  ($R$ on the order of a few Mpc corresponds to requiring
that structures as large as our galaxy be necessary for life.)

We then proceed to filter out perturbations with wavelength smaller
than $R$, leading to a variance $\sigma^2$ that depends on the
filtering scale.  Expressed in terms of the power spectrum evaluated
at recombination, \beq \sigma^2 = {1 \over 2 \pi^2} \int^{\infty}_0 dk
\, k^2 P(k) W^2(kR) \label{sigmaR} \eeq where $W$ is the filter
function, which we take to be a Gaussian $W(x)=e^{-x^2/2}$.  $P(k)$ is
the power spectrum, which we assume to be scale-invariant.  (For
$P(k)$ we use Eq. (39) of \cite{weinberg2}, setting $n=1$).

Evaluating (\ref{sigmaR}) at recombination gives, for our universe,
\beq \hat{\sigma}_\ast = C_\ast Q_{\ast} ~. \eeq The number $C_\ast$
contains the growth factor and transfer function evaluated from
horizon crossing to recombination and only depends on physics from
that era.  We assume $\Lambda$ is small enough so that at
recombination it can be ignored and thus we take the variation in
$\hat{\sigma}$ between universes to come solely from its explicit
dependence on $Q$.

We may then use observations of $Q_\ast$ and $\sigma_\ast$ to
determine $\hat{\sigma} = C_\ast Q$, valid for all universes.  We use
the explicit expression for $C_\ast$ that is obtained from
Eqs. (39)-(43) and (48)-(51) in \cite{weinberg2}.  This takes as
inputs the Hubble parameter $H_0\equiv 100 h_\ast $km/s, the energy
density in non-relativistic matter $\Omega_\ast$, the cosmological
constant $\lambda_\ast=1- \Omega_\ast$, the baryon fraction $\Omega_b
=0.023 h_\ast^{-2}$, the smoothing scale $R$, and the COBE-normalized
amplitude of fluctuations at horizon crossing, $Q_\ast=
1.94\times10^{-5} \Omega_\ast^{-.785 - 0.05\ast\ln\Omega_\ast}$.

As we have argued, the dependence of $C_\ast$ on the cosmological
constant is not relevant for our purposes.  For our calculations we
use $\Omega_\ast=0.134 h_\ast^{-2}$, and $h_\ast=0.73$, consistent
with their observed best-fit values (\cite{pdg}).  The smoothing scale
$R$ will be taken to be either 1 Mpc or 2 Mpc, and the corresponding
values for $C_\ast$ are $5.2 \cdot 10^4$ and $3.8 \cdot 10^4$.

The values chosen for the range of $Q$ are motivated by the discussion
in \cite{tegmark} about anthropic limits on the amplitude of the
primordial density perturbations.  The authors of \cite{tegmark} argue
that $Q$ between $10^{-3}$ and $10^{-1}$ leads to the formation of
numerous supermassive blackholes, which might obstruct the emergence of
life.\footnote{They also note that for $Q> 10^{-4}$ formation of life
is possible, but planetary disruptions caused by flybys may make it
unlikely for planetary life to last billions of years.} They then
claim that universes with $Q$ less than $10^{-6}$ are less likely to
form stars, or if star clusters do form, that they would not be bound
strongly enough to retain supernova remnants.  Since there is
considerable uncertainty in these limits, we carry out calculations
using both the range indicated by \cite{tegmark} as well as a range
that is somewhat broader.\footnote{Notice that we are using the ranges
indicated in \cite{tegmark} as absolute anthropic cutoffs.  Arguments
like those made in \cite{tegmark} introduce some correction to the
approximation made in \cite{weinberg2} that the probability of life is
proportional to the amount of matter that collapses into compact
structures.  Since we are largely ignorant of what the form of this
correction is, we have approximated it as a simple window function.}

Previous work on applying the anthropic principle to variable
$\Lambda$ and $Q$ has assumed {\it a priori} distributions $P(Q)$ that
fall off as $1/Q^k$ for large $Q$, with $k\geq 3$
\cite{garrigaliviovilenkin,garrigavilenkin2}.  Such distributions were
chosen in order to keep the anthropic probability ${\cal P}(y,Q)$
normalizable, and they usually yield anthropic predictions for the
cosmological constant similar to those that were obtained in
\cite{weinberg2} by fixing $Q$ to its observed value, because they
naturally favor a $Q$ as small as its observed value in our universe.
For instance, for $P(Q) \propto 1/Q^3$ in the range $Q_\ast/10 < Q <
10 Q_\ast$, $P(y < y_\ast) = 5\%$ for $R=1$ Mpc, while $P(y < y_\ast)
= 7\%$ for $R=2$ Mpc.)

However, if we accept the argument of Tegmark and Rees in
\cite{tegmark} that there are natural anthropic cutoffs on $Q$, it
follows that the behavior of $P(Q)$ at large $Q$ is irrelevant to the
normalizability of ${\cal P}(y,Q)$.  Furthermore, $P(Q)\sim 1/Q^k$ in
the neighborhood of $Q=0$ for $k\geq1$ leads to an unnormalizable
distribution, since the integral $\int P(Q) dQ$ blows up.  In what
follows we shall consider two {\it a priori} distributions: $P(Q)
\propto Q$, and $P(Q)\propto 1/Q^{0.9}$ inside the anthropic window,
motivated by the inflationary models we have discussed.

The results are summarized in Table \ref{y-table}, where $P(y <
y_\ast)$ is the anthropic probability that the value $y$ be no greater
than what is observed in our own universe, $\langle y \rangle$ is the
anthropically weighed mean value of $y$, and $y_{\sl{5\%}}$ is the
value of $y$ such that the anthropic probability of obtaining a value
no greater than that is 5\%.

\begin{table}[t] \fontsize{10}{12} \selectfont \centerline{
\begin{tabular}{| c || c | c | c | c | c | c | c | c |} \hline
 &\multicolumn{4}{c|}{$P(Q)\propto 1/Q^{0.9}$ in the range}
 &\multicolumn{4}{|c|}{$P(Q)\propto Q$ in the range} \\ \cline{2-9}
 &\multicolumn{2}{c|}{$Q_\ast/10 < Q < 10Q_\ast$} &
 \multicolumn{2}{|c|}{$Q_\ast/15 < Q < 15Q_\ast$}
 &\multicolumn{2}{|c|}{$Q_\ast/10 < Q < 10Q_\ast$} &
 \multicolumn{2}{|c|}{$Q_\ast/15 < Q < 15Q_\ast$}\\ \cline{2-9} &
 {$R=1$ Mpc} & {$R=2$ Mpc} & {$R=1$ Mpc} & {$R=2$ Mpc} & {$R=1$ Mpc} &
 {$R=2$ Mpc} & {$R=1$ Mpc} & {$R=2$ Mpc} \\ \hline \hline $P(y <
 y_\ast)$ & 1 $\cdot 10^{-3}$ & 3 $\cdot 10^{-3}$ & 4$\cdot 10^{-4}$ &
 1 $\cdot 10^{-3}$& 5 $\cdot 10^{-4}$ & 1 $\cdot 10^{-3}$ & 1 $\cdot
 10^{-4}$ & 4 $\cdot 10^{-4}$ \\ \hline $\langle y \rangle/y_\ast$ & 1
 $\cdot 10^4 $ & 4 $\cdot 10^3 $ & 4 $\cdot 10^4 $ & 1 $\cdot 10^4$ &
 1 $\cdot 10^4 $ & 5 $\cdot 10^3 $ & 4 $\cdot 10^4 $ & 2 $\cdot 10^4 $
 \\ \hline $y_{\sl{5\%}}/y_\ast$ & 9 $\cdot 10$ & 4 $\cdot 10$ &
 3$\cdot 10^2$ &1 $ \cdot 10^2$ & 2 $ \cdot 10^2$ & 7$\cdot 10$ &
 6$\cdot 10^2$ & 2 $\cdot 10^2$ \\ \hline
\end{tabular}}
\caption{\small Anthropically determined properties of the cosmological constant.}
\label{y-table}
\end{table}

\begin{table}[t] \fontsize{10}{12} \selectfont \centerline{ 
\begin{tabular}{| c || c | c | c | c | c | c | c | c |} \hline
 &\multicolumn{4}{c|}{$P(Q)\propto 1/Q^{0.9}$ in the range}
 &\multicolumn{4}{|c|}{$P(Q)\propto Q$ in the range} \\ \cline{2-9}
 &\multicolumn{2}{c|}{$Q_\ast/10 < Q < 10Q_\ast$} &
 \multicolumn{2}{|c|}{$Q_\ast/15 < Q < 15Q_\ast$}
 &\multicolumn{2}{|c|}{$Q_\ast/10 < Q < 10Q_\ast$} &
 \multicolumn{2}{|c|}{$Q_\ast/15 < Q < 15Q_\ast$}\\ \cline{2-9} &
 {$R=1$ Mpc} & {$R=2$ Mpc} & {$R=1$ Mpc} & {$R=2$ Mpc} & {$R=1$ Mpc} &
 {$R=2$ Mpc} & {$R=1$ Mpc} & {$R=2$ Mpc} \\ \hline \hline $P(Q <
 Q_\ast)$ & 8 $\cdot 10^{-4}$ & 8 $\cdot 10^{-4}$ & 2 $\cdot 10^{-4}$
 & 2 $\cdot 10^{-4}$ & 1 $\cdot 10^{-5}$ & 1 $\cdot 10^{-5}$ & 1
 $\cdot 10^{-6}$ & 1 $\cdot 10^{-6}$ \\ \hline $\langle Q
 \rangle/Q_\ast$ & 8 & 8 & 11 & 11 & 8 & 8 & 13 & 13 \\ \hline
 $Q_{\sl{5\%}}/Q_\ast$ & 4 & 4 & 6 & 6 & 5 & 5 & 8 & 8 \\ \hline
\end{tabular}} 
\caption{\small Anthropically determined properties of the amplitude for density pertubations.}
\label{Q-table}
\end{table}

By comparison, for this choice of cosmological parameters, the authors
of \cite{weinberg2} find that, for $Q$ fixed (or measured), the
probability of a universe having a cosmological constant no greater
than our own is much higher: $P(y<0.7/0.3)=.05$ and $0.1$, for $R=1$
Mpc and $R=2$ Mpc, respectively.\footnote{These numbers are taken from
Table 1 in the published version of \cite{weinberg2}.}

One can also ask what is the probability of observing a value for $Q$
in the range $Q_\ast/10 < Q < Q_\ast$, after averaging over all
possible cosmological constants.  Table \ref{Q-table} summarizes the
resulting distribution in $Q$.

In summary, inflation and a landscape of anthropically determined
coupling constants provides (in some scenarios) a conceptually clean
framework for variation between universes in the magnitude of
$Q$. Since increasing $Q$ allows the probability of structure to
remain non-negligible for $\Lambda$ considerably larger than in our
own universe, anthropic solutions to the cosmological constant problem
are weakened by allowing $Q$ as well as $\Lambda$ to vary from one
universe to another.
\chapter{The reverse sprinkler}
\label{chap:sprinkler}

\begin{epigraphs}
\qitem{Everything's got a moral, if only you can find it.}
{Lewis Carroll, {\it Alice's Adventures in Wonderland}}
\end{epigraphs}

This chapter is based largely on \cite{sprinkler}.  Some followups that have appeared since the publication of that article include \cite{creutz} and \cite{navy}.  

\section{Introduction}
\label{sect:sprinklerintro}

In 1985, R. P. Feynman, one of most distinguished theoretical
physicists of his time, published a collection of autobiographical
anecdotes that attracted much attention on account of their humor and
outrageousness (\cite{Feynman}). While describing his time at Princeton
as a graduate student (1939--1942), Feynman tells the following
story (\cite{Feynmanquote1}):

\begin{quote}

There was a problem in a hydrodynamics book,\footnote{It has not been possible to identify the book to which Feynman was referring. As we shall discuss, the matter is treated in
Ernst Mach's {\it Mechanik}, first published in 1883 (\cite{Machquote1}). Yet
this book is not a ``hydrodynamics book'' and the reverse sprinkler is
presented as an example, not a problem. In \cite{Wheeler},
John Wheeler suggests that the problem occurred to them while
discussing a different question in the undergraduate mechanics course
that Wheeler was teaching and for which Feynman was the grader.} that was
being discussed by all the physics students. The problem is this: You
have an S-shaped lawn sprinkler \ldots and the water squirts out at
right angles to the axis and makes it spin in a certain
direction. Everybody knows which way it goes around; it backs away
from the outgoing water.  Now the question is this: If you \ldots put
the sprinkler completely under water, and sucked the water in \ldots
which way would it turn?

\end{quote}

Feynman went on to say that many Princeton physicists, when presented
with the problem, judged the solution to be obvious, only to find that
others arrived with equal confidence at the opposite answer, or that
they had changed their minds by the following day. Feynman claims that
after a while he finally decided what the answer should be and
proceeded to test it experimentally by using a very large water
bottle, a piece of copper tubing, a rubber hose, a cork, and the air
pressure supply from the Princeton cyclotron laboratory. Instead of
attaching a vacuum to suck the water, he applied high air pressure inside
of the water bottle to push the water out through the sprinkler. According to 
Feynman's account, the experiment initially
went well, but after he cranked up the setting for the pressure
supply, the bottle exploded, and ``\ldots the whole thing just blew
glass and water in all directions throughout the laboratory
\ldots'' (\cite{Feynmanquote2}).

Feynman (\cite{Feynman}) did not inform the reader what his answer to the
reverse sprinkler problem was or what the experiment revealed before
exploding. Over the years, and particularly after Feynman's
autobiographical recollections appeared in print, many people have
offered their analyses, both theoretical and experimental, of this
reverse sprinkler problem.\footnote{In the literature it is more usual to see this
problem identified as the ``Feynman inverse sprinkler.'' Because the
problem did not originate with Feynman and Feynman never published an
answer to the problem, we have preferred not to attach his name to the
sprinkler. Furthermore, even though it is a pedantic point, a query of
the {\it Oxford English Dictionary} suggests that ``reverse''
(opposite or contrary in character, order, or succession) is a more
appropriate description than ``inverse'' (turned up-side down) for a
sprinkler that sucks water.} The solutions presented often
have been contradictory and the theoretical treatments, even when they
have been correct, have introduced unnecessary conceptual
complications that have obscured the basic physics involved.

All physicists will probably know the frustration of being confronted
by an elementary question to which they cannot give a ready answer in
spite of all the time dedicated to the study of the subject, often at
a much higher level of sophistication than what the problem at hand
would seem to require. Our intention is to offer an elementary
treatment of this problem, which should be accessible to a bright
secondary school student who has learned basic mechanics and fluid
dynamics. We believe that our answer is about as simple as it can be
made, and we discuss it in light of published theoretical and
experimental treatments.

\section{Pressure difference and momentum transfer}
\label{sect:pressuremomentum}

Feynman speaks in his memoirs of ``an S-shaped lawn sprinkler.'' It
should not be difficult, however, to convince yourself that the
problem does not depend on the exact shape of the sprinkler, and for
simplicity we shall refer in our argument to an L-shaped structure.
In Fig.~\ref{sprinklerclosed} the sprinkler is closed: Water cannot
flow into it or out of it. Because the water pressure is equal on
opposite sides of the sprinkler, it will not turn: there is no net
torque around the sprinkler pivot.

Let us imagine that we then remove part of the wall on the right, as
pictured in Fig.~\ref{sprinkleropen}, opening the sprinkler to the
flow of water. If water is flowing in, then the pressure marked $P_2$
must be lower than the pressure $P_1$, because water flows from higher
to lower pressure. In both Fig.~\ref{sprinklerclosed} and
Fig.~\ref{sprinkleropen}, the pressure $P_1$ acts on the left.  But
because a piece of the sprinkler wall is missing in
Fig.~\ref{sprinkleropen}, the relevant pressure on the upper right
part of the open sprinkler will be $P_2$. It would seem then that the
reverse sprinkler should turn toward the water, because if $P_2$ is
less than $P_1$, there would be a net force to the right in the upper
part of the sprinkler, and the resulting torque would make the
sprinkler turn clockwise. If $A$ is the cross section of the sprinkler
intake pipe, this torque-inducing force is $A(P_1-P_2)$.

\begin{figure} []
\bigskip
\begin{center}
\includegraphics{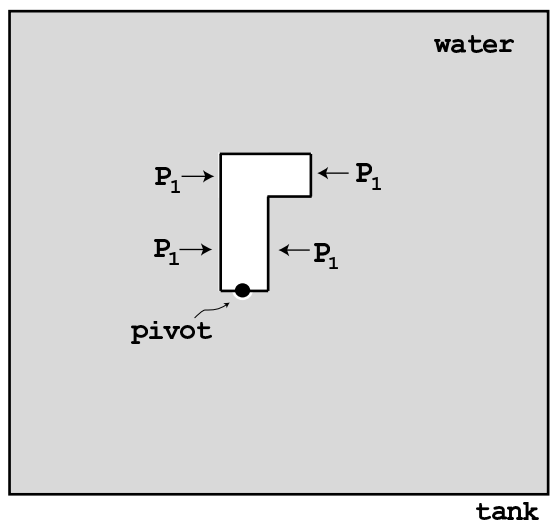}
\caption[Closed sprinkler in a tank]{\small A sprinkler submerged in a tank of water as seen
from above. The L-shaped sprinkler is closed, and the forces
and torques exerted by the water pressure balance each other.}
\label{sprinklerclosed}
\end{center}
\end{figure}

But we have not taken into account that even though the water hitting
the inside wall of the sprinkler in Fig.~\ref{sprinkleropen} has lower
pressure, it also has left-pointing momentum. The incoming water
transfers that momentum to the sprinkler as it hits the inner
wall. This momentum transfer would tend to make the sprinkler turn
counterclockwise. One of the reasons why the reverse sprinkler is a
confusing problem is that there are two effects in play, each of
which, acting on its own, would make the sprinkler turn in opposite
directions. The problem is to figure out the net result of these two
effects.

\begin{figure} []
\bigskip
\begin{center}
\includegraphics{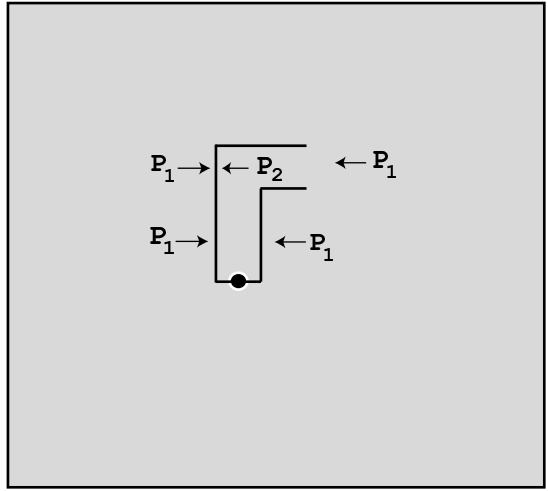}
\caption[Open sprinkler in a tank]{\small The sprinkler is now open. If water is flowing into
it, then the pressures marked $P_1$ and $P_2$ must satisfy $P_1 >
P_2$.}
\label{sprinkleropen}
\end{center}
\end{figure}

How much momentum is being transferred by the incoming water to the
inner sprinkler wall in Fig.~\ref{sprinkleropen}? If water is moving
across a pressure gradient, then over a differential time $dt$, a
given ``chunk'' of water will pass from an area of pressure $P$ to an
area of pressure $P-dP$ as illustrated in
Fig.~\ref{pressuregradient}. If the water travels down a pipe of
cross-section $A$, its momentum gain per unit time is
$A\,dP$. Therefore, over the entire length of the pipe, the water
picks up momentum at a rate $A(P_1-P_2)$, where $P_1$ and $P_2$ are
the values of the pressure at the endpoints of the pipe. (In the
language of calculus, $A(P_1-P_2)$ is the total force that the
pressure gradient across the pipe exerts on the water. We obtain it by
integrating over the differential force $A\,dP$.)\footnote{As some readers of \cite{sprinkler} pointed out to us (\cite{mammel,jeng}), this simplified discussion ignores the fact that the cross-section of a fluid flow is not in general constant when a pressure gradient exists.  For example, for an ideal, incompressible fluid the velocity (and therefore, through Bernoulli's equation, also the pressure) must be constant inside a pipe of fixed cross-section $A$.  In that case all of the acceleration of the fluid would have to occur outside of the sprinkler tube, as the flow narrows down to a cross-section $A$.  However, if $P_1$ is the pressure of the fluid at rest, then $A(P_1-P_2)$ is still the correct expression for the rate at which the flow is gaining momentum.  In fact, the shape of the flow into the reverse sprinkler will not be relevant to our discussion at all, as should become more clear from the discussion of conservation of angular momentum conservation in Section \ref{sect:sprinklerL}.}

For steady flow, the rate $A(P_1-P_2)$ must be the same rate at which the water is transferring momentum to the sprinkler wall in Fig.~\ref{sprinkleropen}, because
otherwise the total amount of momentum contained in the flow of water would not be constant.  Therefore $A(P_1-P_2)$ is the force that the incoming water exerts on the inner
sprinkler wall in Fig.~\ref{sprinkleropen} by virtue of the momentum
it has gained in traveling down the intake pipe.

\begin{figure} []
\bigskip
\begin{center}
\includegraphics{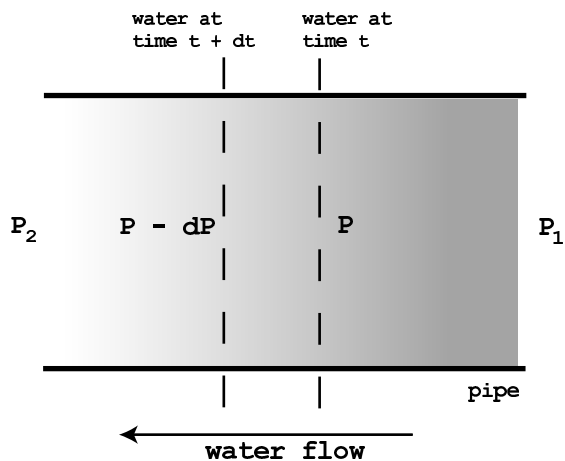}
\caption[Fluid flow in a pressure gradient]{\small As water flows down a tube with a pressure gradient,
it picks up momentum.}
\label{pressuregradient}
\end{center}
\end{figure}

Because the pressure difference and the momentum transfer effects
cancel each other, it would seem that the reverse sprinkler would not
move at all. Notice, however, that we considered the reverse sprinkler
only after water was already flowing continuously into it. In fact,
the sprinkler {\it will} turn toward the water initially, because the
forces will balance only after water has begun to hit the inner wall
of the sprinkler, and by then the sprinkler will have begun to turn
toward the incoming water. That is, initially only the pressure
difference effect and not the momentum transfer effect is
relevant. (As the water flow stops, there will be a brief period
during which only the momentum transfer and not the pressure
difference will be acting on the sprinkler, thus producing a momentary
torque opposite to the one that acted when the water flow was being
established.)

Why can't we similarly ``prove'' the patently false statement that a
non-sucking sprinkler submerged in water will not turn as water flows
steadily out of it? In that case the water is going out and hitting
the upper inner wall, not the left inner wall. It exerts a force, but
that force produces no torque around the pivot.  The pressure
difference, on the other hand, does exert a torque. The pressure in
this case has to be higher inside the sprinkler than outside it, so
the sprinkler turns counterclockwise, as we expect from experience.

\section{Conservation of angular momentum}
\label{sect:sprinklerL}

We have argued that, if we ignore the transient effects from the
switching on and switching off of the fluid flow, we do not expect the
reverse sprinkler to turn at all. A pertinent question is why, for the
case of the regular sprinkler, the sprinkler-water system clearly
exhibits no net angular momentum around the pivot (with the angular
momentum of the outgoing water cancelling the angular momentum of the
rotating sprinkler), while for the reverse sprinkler the system would
appear to have a net angular momentum given by the incoming water. The
answer lies in the simple observation that if the water in a tank is
flowing, then something must be pushing it. In the regular sprinkler,
there is a high-pressure zone near the sprinkler wall next to the
pivot, so it is this lower inner wall that is doing the original
pushing, as shown in Fig.~\ref{twosprinklers}(a).

\begin{figure} []
\bigskip
\centering
\subfigure[]{\includegraphics[scale=1]{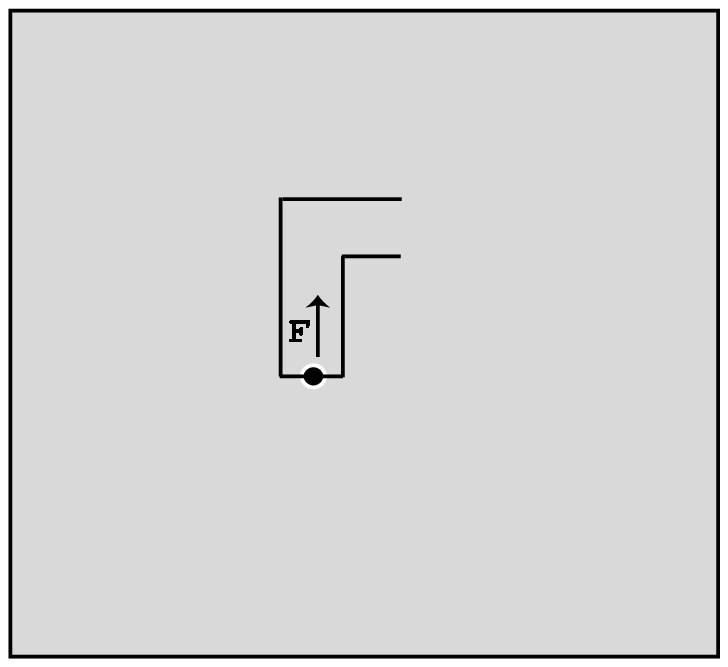}}
\subfigure[]{\includegraphics[scale=1]{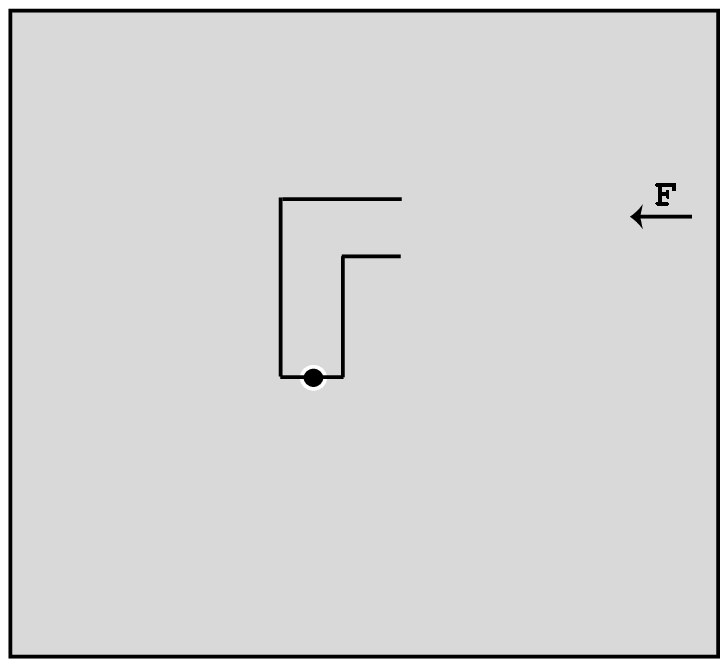}}
\caption[Force creating the flow into the reverse sprinkler]{\small The force that pushes the water must originally come from
a solid wall. The force that causes the water flow is shown for
both the regular and the reverse sprinklers when submerged in a
tank of water.}
\label{twosprinklers}
\end{figure}

For the reverse sprinkler, the highest pressure is outside the
sprinkler, so the pushing originally comes from the right wall of the
tank in which the whole system sits, as shown in
Fig.~\ref{twosprinklers}(b). The force on the regular sprinkler
clearly causes no torque around the pivot, while the force on the
reverse sprinkler does. That the water should acquire a net angular
momentum around the sprinkler pivot in the absence of an external
torque might seem a violation of Newton's laws, but only because we
are neglecting the movement of the tank itself. Consider a water tank
with a hole in its side, such as the one pictured in
Fig.~\ref{puncturedtank}. The water acquires a net angular momentum
with respect to any point on the tank's bottom, but this angular
momentum violates no physical laws because the tank is not inertial:
It recoils as water flows out of it.

But there is one further complication: In the reverse sprinkler shown
in Fig.~\ref{twosprinklers}, the water that has acquired left-pointing
momentum from the pushing of the tank wall will transfer that momentum
back to the tank when it hits the inner sprinkler wall, so that once
water is flowing steadily into the reverse sprinkler, the tank will
stop experiencing a recoil force.  The situation is analogous to that
of a ship inside of which a machine gun is fired, as shown in
Fig.~\ref{ship}. As the bullet is fired, the ship recoils, but when
the bullet hits the ship wall and becomes embedded in it, the bullet's
momentum is transferred to the ship. (We assume that the collision of
the bullets with the wall is completely inelastic.)

\begin{figure}[]
\bigskip
\begin{center}
\includegraphics{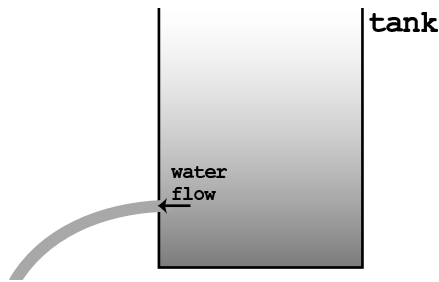}
\caption[Tank recoiling as water rushes out of it]{\small A tank with an opening on its side will exhibit a flow such that the water will have an angular momentum with respect to
the tank's bottom, even though there is no external source of torque
corresponding to the angular momentum. The apparent paradox is
resolved by noting that the tank bottom offers no inertial point of
reference, because the tank is recoiling due to the motion of the
water.}
\label{puncturedtank}
\end{center}
\end{figure}

If the firing rate is very low, the ship periodically acquires a
velocity in a direction opposite to that of the fired bullet, only to
stop when that bullet hits the wall. Thus the ship moves by small
steps in a direction opposite that of the bullets' flight. As the
firing rate is increased, eventually one reaches a rate such that the
interval between successive bullets being fired is equal to the time
it takes for a bullet to travel the length of the ship. If the machine
gun is set for this exact rate from the beginning, then the ship will
move back with a constant velocity from the moment that the first
bullet is fired (when the ship picks up momentum from the recoil) to
the moment the last bullet hits the wall (when the ship comes to a
stop). In between those two events the ship's velocity will not change
because every firing is simultaneous to the previous bullet hitting
the ship wall.

As the firing rate is made still higher, the ship will again move in
steps, because at the time that a bullet is being fired, the previous
bullet will not have quite made it to the ship wall.  Eventually, when
the rate of firing is twice the inverse of the time it takes for a
bullet to travel the length of the ship, the motion of the ship will
be such that it picks up speed upon the first two shots, then moves
uniformly until the penultimate bullet hits the wall, whereupon the
ship loses half its velocity. The ship will finally come to a stop
when the last bullet has hit the wall. At this point it should be
clear how the ship's motion will change as we continue to increase the
firing rate of the gun.\footnote{Two interesting problems for an introductory
university-level physics course suggest themselves. One is to show
that the center of mass of the bullets-and-ship system will not move
in the horizontal direction regardless of the firing rate, as one
expects from momentum conservation. Another would be to analyze this
problem in the light of Einstein's relativity of simultaneity.}

\begin{figure}[]
\bigskip
\begin{center}
\includegraphics{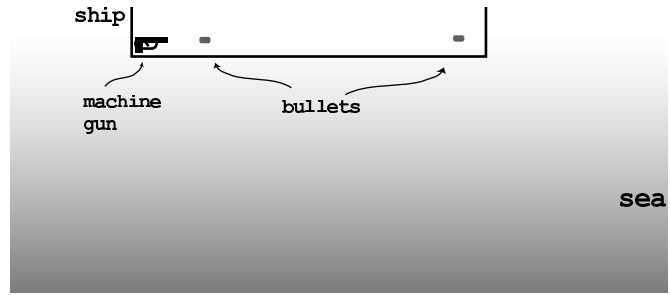}
\caption[Machine gun in a floating ship]{\small In this thought experiment, a ship floats in the ocean
while a machine gun with variable firing rate is placed at one
end. Bullets fired from the gun will travel the length of the ship and
hit the wall on the other side, where they stop.}
\label{ship}
\end{center}
\end{figure}

For the case of continuous flow of water in a tank (rather than a
discrete flow of machine gun bullets in a ship), there clearly will be
no intermediate steps, regardless of the rate of flow. Figure
\ref{shower} shows a water tank connected to a shower head. Water
flows (with a consequent linear and angular momentum) between the
points marked A and B, before exiting via the shower head. When the
faucet valve is opened, the tank will experience a recoil from the
outgoing water, until the water reaches B and begins exiting through
the shower head, at which point the forces on the tank will balance.
By then the tank will have acquired a left-pointing momentum. It will
lose that momentum as the valve is closed or the water tank becomes
empty, when there is no longer water flowing away from A but a flow is
still impinging on B.

A.~K.~Schultz (\cite{letters1}) argues that, at each instant, the water
flowing into the reverse sprinkler's intake carries a constant angular
momentum around the sprinkler pivot, and if the sprinkler could turn
without any resistance (either from the friction of the pivot or the
viscosity of the fluid) this angular momentum would be counterbalanced
by the angular momentum that the sprinkler picked up as the water flow
was being switched on. As the fluid flow is switched off, such an
ideal sprinkler would then lose its angular momentum and come to a
halt. At every instant, the angular momentum of the sprinkler plus the
incoming water would be zero.

\begin{figure}[]
\bigskip
\begin{center}
\includegraphics{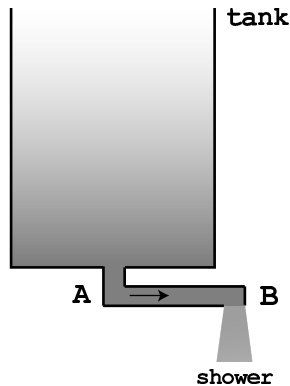}
\caption[Water flowing out of a shower head]{\small A water tank is connected to a shower head, so that
water flows out. Water in the pipe that connects the points marked A
and B has a right-pointing momentum, but as long as that pipe is
completely filled with water there is no net horizontal force on the
tank.}
\label{shower}
\end{center}
\end{figure}

Schultz's discussion is correct: In the absence of any resistance, the
sprinkler arm itself moves so as to cancel the momentum of the
incoming water, in the same way that the ship in Fig.~\ref{ship} moves
to cancel the momentum of the flying bullets. Resistance, on the other
hand, would imply that some of that momentum is picked up not just by
the sprinkler, but by the tank as a whole. If we cement the pivot to
prevent the sprinkler from turning at all, then the tank will pick up
all of the momentum that cancels that of the incoming water.

How does non-ideal fluid behavior affect this analysis? Viscosity,
turbulence, and other such phenomena all dissipate mechanical
energy. Therefore, a non-ideal fluid rushing into the reverse
sprinkler would acquire less momentum with respect to the pivot, for a
given pressure difference, than predicted by the analysis we carried
out in Section \ref{sect:pressuremomentum}. Thus the pressure-difference effect would outweigh the
momentum-transfer effect even in the steady state, leading to a small
torque on the sprinkler even after the fluid has begun to hit the
inside wall of the sprinkler. Total angular momentum is conserved
because the ``missing'' momentum of the incoming fluid is being
transmitted to the surrounding fluid, and finally to the tank.

\section{History of the reverse sprinkler problem}
\label{sect:sprinklerhistory}

The literature on the subject of the reverse sprinkler is abundant and
confusing. Ernst Mach speaks of ``reaction wheels'' blowing or sucking
air where we have spoken of regular or reverse sprinklers
respectively (\cite{Machquote1}):

\begin{quote}

It might be supposed that sucking on
the reaction wheels would produce the opposite motion to that
resulting from blowing. Yet this does not usually take place, and the
reason is obvious \ldots Generally, no perceptible rotation takes place
on the sucking in of the air \ldots If \ldots an elastic ball, which
has one escape-tube, be attached to the reaction-wheel, in the manner
represented in [Fig.~\ref{Machfigs}(a)], and be alternately squeezed
so that the same quantity of air is by turns blown out and sucked in,
the wheel will continue to revolve rapidly in the same direction as it
did in the case in which we blew into it. This is partly due to the
fact that the air sucked into the spokes must participate in the
motion of the latter and therefore can produce no reactional rotation,
but it also results partly from the difference of the motion which the
air outside the tube assumes in the two cases. In blowing, the air
flows out in jets, and performs rotations. In sucking, the air comes
in from all sides, and has no distinct rotation \ldots

\end{quote}

\begin{figure}[]
\bigskip
\centering
\subfigure[]{\includegraphics[scale=1]{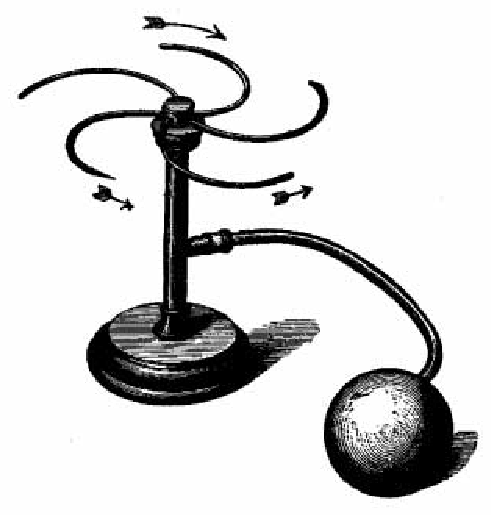}} \quad
\subfigure[]{\includegraphics[scale=1]{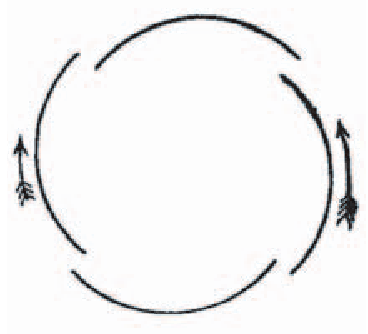}}
\caption[Illustrations from Ernst Mach's {\it Mechanik}]{\small Illustrations from Ernst Mach's {\it Mechanik}  (\cite{Machquote1}):
(a) Figure 153 a in the original. (b) Figure 154 in the
original. (Images in the public domain, copied from the English
edition of 1893.)}
\label{Machfigs}
\end{figure}

Mach appears to base his treatment on the observation that a
``reaction wheel'' is not seen to turn when sucked on. He then sought
a theoretical rationale for this observation without arriving at one
that satisfied him. Thus the bluster about the explanation being
``obvious,'' accompanied by the tentative language about how
``generally, no perceptible rotation takes place'' and by the
equivocation about how the lack of turning is ``partly due'' to the
air ``participating in the motion'' of the wheel and partly to the air
sucked ``coming in from all sides.''

Mach goes on to say that

\begin{quote}

if we perforate the bottom
of a hollow cylinder \ldots and place the cylinder on [a pivot], after
the side has been slit and bent in the manner indicated in
[Fig.~\ref{Machfigs}(b)], the [cylinder] will turn in the direction of
the long arrow when blown into and in the direction of the short arrow
when sucked on.  The air, here, on entering the cylinder can continue
its rotation {\it unimpeded}, and this motion is accordingly
compensated for by a rotation in the opposite direction (\cite{Machquote2}).

\end{quote}

This observation is correct and interesting:  It shows that if the incoming
water did not give up all its angular momentum upon hitting the inner
wall of the reverse sprinkler, then the device would turn toward the
incoming water, as we discussed at the beginning of
Section \ref{sect:sprinklerL}.\footnote{In \cite{Hewitt}, P. Hewitt proposes a
physical setup identical to the one shown in Fig.~\ref{Machfigs}(b),
and observes that the device turns in opposite directions depending on
whether the fluid pours out of or into it. Hewitt's discussion seems
to ignore the important difference between such a setup and the
reverse sprinkler.  The issue has recently been investigated in \cite{navy}.}

In his introduction to Mach's {\it Mechanik}, mathematician Karl
Menger describes it as ``one of the great scientific achievements of
the [nineteenth] century'' (\cite{Menger}), but it seems that the passage
we have quoted was not well known to the twentieth-century scientists
who commented publicly on the reverse sprinkler.
Feynman (\cite{Feynman}) gave no answer to the problem and wrote as if he
expected and observed rotation.  Some have pointed out, however, that the
fact that he cranked up the pressure until the bottle exploded
suggests another explanation: that he expected rotation and didn't see
it.  This interpretation seems to be supported by a recent letter published by E.~Creutz, who claims to have been the only other person at the Princeton cyclotron when Feynman carried out his experiment (\cite{creutz}).  Creutz, however, explicitly disclaims any knowledge of what Feynman's own theoretical understanding of the problem was.

In \cite{Kirkpatrick} and \cite{Belson}, the
authors discuss the problem and claim that no rotation is observed,
but they pursue the matter no further. In \cite{NSF}, it is
suggested that students demonstrate as an exercise that ``the
direction of rotation is the same whether the flow is supplied through
the hub [of a submerged sprinkler] or withdrawn from the hub,'' a
result that is discounted by almost all the rest of the literature.

Shortly after Feynman's memoirs appeared, A.~T.~Forrester published
a paper in which he concluded that if water is sucked out of a tank by
a vacuum attached to a sprinkler then the sprinkler will not
rotate (\cite{Forrester}). But he also made the strange claim that
Feynman's original experiment at the Princeton cyclotron, in which he
had high air pressure in the bottle push the water out, would actually
cause the sprinkler to rotate in the direction of the incoming
water (\cite{Forrester}). An exchange on the issue of conservation of
angular momentum between Shultz and Forrester appeared shortly
thereafter (\cite{letters1, letters2}). The following year L.~Hsu, a
high school student, published an experimental analysis that found no
rotation of the reverse sprinkler and questioned (quite sensibly)
Forrester's claim that pushing the water out of the bottle was not
equivalent to sucking it out (\cite{Hsu}). E.~R.~Lindgren also
published an experimental result that supported the claim that the
reverse sprinkler did not turn (\cite{Lindgren}).

After Feynman's death, his graduate research advisor, J.~A.~Wheeler,
published some reminiscences of Feynman's Princeton days from which it
would appear that Feynman observed no motion in the sprinkler before
the bottle exploded (``a little tremor as the pressure was first
applied \ldots but as the flow continued there was no
reaction'') (\cite{Wheeler}). In 1992 the journalist James Gleick
published a biography of Feynman in which he states that both Feynman
and Wheeler ``were scrupulous about never revealing the answer to the
original question'' and then claims that Feynman's answer all along
was that the sprinkler would not turn (\cite{Gleick}). The physical
justification that Gleick offers for this answer is misleading:  Gleick echoes one of Mach's comments in \cite{Machquote1} that the
water entering the reverse sprinkler comes in from many directions,
unlike the water leaving a regular sprinkler, which forms a narrow
jet. Although this observation is correct, it is not very
relevant to the question at hand.

The most detailed and pertinent work on the subject, both theoretical
and experimental, was published by Berg, Collier, and Ferrell, who
claimed that the reverse sprinkler turns toward the incoming
water (\cite{CollegePark1, CollegePark2}). Guided by Schultz's arguments
about conservation of angular momentum (\cite{letters1}), the authors
offered a somewhat complicated statement of the correct observation
that the sprinkler picks up a bit of angular momentum before reaching
a steady state of zero torque once the water is flowing steadily into
the sprinkler. When the water stops flowing, the sprinkler comes to a
halt.\footnote{There are other references in the literature to the
reverse sprinkler. For a rather humorous exchange, see \cite{Kuzyk} and \cite{CPreply}.  Already in 1990
the {\it American Journal of Physics} had received so many conflicting
analyses of the problem that the editor  proposed ``a moratorium on
publications on Feynman's sprinkler'' (\cite{Mironer}). In one of her
1996 columns for {\it Parade Magazine}, Marilyn vos Savant, who bills
herself as having the highest recorded IQ, offered an account of
Feynman's experiment that, she claimed, settled that the reverse
sprinkler does not move (\cite{Savant}).  Vos Savant's column emphasized
the confusion of Feynman and others when faced with the problem,
leading a reader to respond with a letter to his local newspaper in
which he questioned the credibility of physicists who address matters
more complicated than lawn sprinklers, such as the origin of the
universe (\cite{deGruyter}).}

The air-sucking reverse sprinkler at the Edgerton Center at MIT shows
no movement at all (\cite{MIT}). As in the setups used by Feynman and
others, this sprinkler arm is not mounted on a true pivot, but rather
turns by twisting or bending a flexible tube. Any transient torque
will therefore cause, at most, a brief shaking of such a device. The
University of Maryland's Physics Lecture Demonstration Facility offers
video evidence of a reverse sprinkler, mounted on a true pivot of very
low friction, turning slowly toward the incoming water (\cite{UMD}).
According to R.~E.~Berg, in this particular setup 

\begin{quote}
while the water is flowing the nozzle rotates at a constant angular speed. This would be
consistent with conservation of angular momentum except for one thing:
while the water is flowing into the nozzle, if you reach and stop the
nozzle rotation it should remain still after you release it. [But, in
practice,] after [the nozzle] is released it starts to rotate
again'' (\cite{Bergemail}).
\end{quote}

This behavior is consistent with non-zero dissipation of kinetic
energy in the fluid flow, as we have discussed. Angular momentum is
conserved, but only after the motion of the tank is taken into
account.\footnote{In the late 1950's and early 1960's, there was some
interest in the related physics problem of the so-called putt-putt
(or pop-pop) boat, a fascinating toy boat that propels itself by
heating (usually with a candle) an inner tank connected to a submerged
double exhaust. Steam bubbles cause water to be alternately blown out
of and sucked into the tank (\cite{toyboat1,toyboat2,toyboat3}). The ship
moves forward, much like Mach described the ``reaction wheel'' turning
vigorously in one direction as air was alternately blown out and
sucked in.}  An earlier, unpublished treatment of how dissipation causes a steady-state torque on the reverse sprinkler is due to Titcomb, Rueckner, and Sokol (\cite{wolf}).  Rueckner also reports that the behavior of a sprinkler made to suck argon gas whose viscosity is adjusted by changing its temperature seems to corroborate that higher viscosity leads to a larger steady-state torque.  This experiment, however, would need to be carried out more carefully to fully confirm this effect experimentally (\cite{wolf2}).

\section{Conclusions}
\label{sprinklerconclusion}

We have offered an elementary theoretical treatment of the behavior of
a reverse sprinkler, and concluded that, under idealized conditions,
it should experience no torque while fluid flows steadily into it, but
as the flow commences, it will pick up an angular momentum opposite to
that of the incoming fluid, which it will give up as the flow
ends. However, in the presence of viscosity or turbulence, the reverse
sprinkler will experience a small torque even in steady state, which
would cause it to accelerate toward the incoming water. This torque is
balanced by an opposite torque acting on the surrounding fluid and
finally on the tank itself.

Throughout our discussion, our foremost concern was to emphasize
physical intuition and to make our treatment as simple as it could be
made (but not simpler). A question about what L.\ A.\ Delsasso 
called, according to Feynman's recollection, ``a freshman
experiment'' (\cite{Feynmanquote2}) deserves an answer presented in a 
language at the corresponding level of complication. More important is the
principle, famously put forward by Feynman himself when discussing the
spin statistics theorem, that if we can't ``reduce it to the freshman
level,'' we don't really understand it (\cite{SixEasy}).

We also have commented on the perplexing history of the reverse
sprinkler problem, a history that is interesting not only because
physicists of the stature of Mach, Wheeler, and Feynman enter into it,
but also because it offers a startling illustration of the fallibility
of great scientists faced with a question about ``a freshman
experiment.''

\end{document}